\documentclass{emulateapj}
\usepackage{apjfonts}
\usepackage{mathrsfs}
\usepackage{amsmath}
\usepackage{graphicx}

\begin{document}

\newcommand{\RRab}{RR{\em ab}}
\newcommand{\RRc}{RR{\em c}}
\slugcomment{AJ, in press}
\shorttitle{NGC~5286. II. Variable Stars}
\shortauthors{M. Zorotovic et al.}

\title{The Globular Cluster NGC~5286. II. Variable Stars\footnote{Based on observations obtained in Chile with the 1.3m Warsaw telescope at the Las Campanas Observatory, and the SOAR 4.1m telescope.}}

\author{
M. Zorotovic,\altaffilmark{1,2} M. Catelan,\altaffilmark{1,3,4} H. A. Smith,\altaffilmark{5} B. J. Pritzl,\altaffilmark{6} \\
P. Aguirre,\altaffilmark{1} R. E. Angulo,\altaffilmark{1} M. Aravena,\altaffilmark{1} R. J. Assef,\altaffilmark{1} C. Contreras,\altaffilmark{1} \\ C. Cort\'{e}s,\altaffilmark{1} G. De Martini,\altaffilmark{1} M. E. Escobar,\altaffilmark{1} D. Gonz\'{a}lez,\altaffilmark{1} P. Jofr\'{e},\altaffilmark{1} I. Lacerna,\altaffilmark{1} \\ C. Navarro,\altaffilmark{1} 
O. Palma,\altaffilmark{1} G. E. Prieto,\altaffilmark{1} E. Recabarren,\altaffilmark{1}
J. Trivi\~{n}o,\altaffilmark{1} E. Vidal\altaffilmark{1}
}

\altaffiltext{1}{Departamento de Astronom\'ia y Astrof\'isica, Pontificia Universidad Cat\'olica de Chile, Av. Vicu\~na Mackena 4860, 782-0436 Macul, Santiago, Chile; 
e-mail: mzorotov, mcatelan@astro.puc.cl}
\altaffiltext{2}{European Southern Observatory, Alonso de Cordova 3107, Santiago, Chile}
\altaffiltext{3}{John Simon Guggenheim Memorial Foundation Fellow} 
\altaffiltext{4}{On sabbatical leave at Michigan State University, Department of Physics and Astronomy, East Lansing, MI 48824} 
\altaffiltext{5}{Department of Physics and Astronomy, Michigan State University, East Lansing, MI 48824}
\altaffiltext{6}{Department of Physics and Astronomy, University of Wisconsin, Oshkosh, WI 54901}

\begin{abstract}

 We present the results of a search for variable stars in the globular cluster \objectname{NGC~5286}, which has recently been suggested to be associated with the Canis Major dwarf spheroidal galaxy. 57 variable stars were detected, only 19 of which had previously been known. Among our detections one finds 52 RR Lyrae (22 \RRc\ and 30 \RRab), 4 LPV's, and 1 type II Cepheid of the BL Herculis type. Periods are derived for all of the RR Lyrae as well as the Cepheid, and {\it{BV}} light curves are provided for all the variables.
 
The mean period of the \RRab\ variables is $\left\langle P_{ab} \right\rangle = 0.656$~days, and the number fraction of \RRc\ stars is $N_c/N_{RR} = 0.42$, both consistent with an Oosterhoff II (OoII) type~-- thus making NGC~5286 one of the most metal-rich (${\rm [Fe/H]} = -1.67$; \citeauthor{har1996} \citeyear{har1996}) OoII globulars known to date. The minimum period of the \RRab's, namely $P_{\rm ab,min} = 0.513$~d, while still consistent with an OoII classification, falls towards the short end of the observed  $P_{\rm ab,min}$ distribution for OoII globular clusters. As was recently found in the case of the prototypical OoII globular cluster M15 (NGC~7078), the distribution of stars in the Bailey diagram does not strictly conform to the previously reported locus for OoII stars.  

We provide Fourier decomposition parameters for all of the RR Lyrae stars detected in our survey, and discuss the physical parameters derived therefrom. The values derived for the \RRc's are not consistent with those typically found for OoII clusters, which may be due to the cluster's relatively high metallicity~-- the latter being confirmed by our Fourier analysis of the ab-type RR Lyrae light curves. Using the recent recalibration of the RR Lyrae luminosity scale by \citeauthor{cc08}, we derive for the cluster a revised distance modulus of $(m-M)_V = 16.04$~mag. 
\end{abstract}

\keywords{stars: Hertzsprung-Russell diagram --- stars: variables: other --- Galaxy: globular clusters: individual (NGC~5286) --- galaxies: dwarf --- galaxies: star clusters}

\section{Introduction} \label{sec:intro}
NGC~5286 (C1343-511) is a fairly bright ($M_V = -8.26$) and dense globular cluster (GC), with a central luminosity density $\rho_0 \approx 14,\!800 \, L_{\odot}/{\rm pc}^3$~-- which is more than a factor of six higher than in the case of $\omega$~Centauri (NGC~5139), according to the entries in the \citet{har1996} catalog. In \citet[][hereafter Paper~I]{mzea08} we presented a color-magnitude diagram (CMD) study of the cluster that reveals an unusual horizontal branch (HB) morphology in that it does not contain a prominent red HB component, contrary to what is normally found in GCs with comparable metallicity (${\rm [Fe/H]} = -1.67$; \citeauthor{har1996} \citeyear{har1996}), such as M3 (NGC~5272) or M5 (NGC~5904). As a matter of fact, NGC~5286 contains blue HB stars reaching down all the way to at least the main sequence turnoff level in $V$. Yet, unlike most blue HB GCs, NGC 5286 is known to contain a sizeable population of RR Lyrae variable stars, with at least 15 such variables being known in the field of the cluster \citep{cle2001}. In this sense, NGC~5286 resembles the case of M62 \citep[NGC~6266;][]{rcea05}, thus possibly being yet another member of a new group of GCs with HB types intermediate between M13 (NGC~6205)-like (a very blue HB with relatively few RR Lyrae variables) and that of the Oosterhoff I (Oo I) cluster M3 (a redder HB, with a well-populated instability strip). NGC~5286 thus constitutes an example of the ``missing link'' between M3- and M13-like GCs \citep*{vcea87}.

Previous surveys for variable stars in NGC~5286 \citep[e.g.,][]{ll78,agea97} have turned up relatively large numbers of RR Lyrae stars. However, such studies were carried out either by photographic methods, used comparatively few observations, or utilized reduction methods that have subsequently been superseded by improved techniques, including robust multiple-frame photometry \citep[e.g., ALLFRAME;][]{pbs94} and image subtraction \citep[e.g., ISIS;][]{Al2000}. This, together with the large central surface brightness of the cluster, strongly suggests that a large population of variable stars remains unknown in NGC~5286, especially towards its crowded inner regions. In addition, for the known or suspected variables, it should be possible to obtain light curves of much superior quality to those available, thus leading to better defined periods, amplitudes, and Fourier decomposition parameters. 

Indeed, to our knowledge, no modern variability study has ever been carried out for this cluster. A study of its variable star population appears especially interesting in view of its suggested association with the Canis Major dwarf spheroidal galaxy \citep*{crane2003,forbes2004}, and the constraints that the ancient RR Lyrae variable stars are able to pose on the early formation history of galaxies \citep*[e.g.,][]{mc04,mc07,cat2005,tkea04,cmea09}. Therefore, the time seems ripe for a reassessment of the variable star content of NGC~5286~-- and this is precisely the main subject of the present paper.

In \S\ref{sec:obs}, we describe the variable stars search techniques and the conversion from ISIS relative fluxes to standard magnitudes. In \S\ref{sec:var}, we show the results of our variability search, giving the positions, periods, amplitudes, magnitudes, and colors for the detected variables. We show the positions of the variables in the cluster CMD in \S\ref{sec:cmd}. In \S\ref{sec:four}, we provide the results of a Fourier decomposition of the RR Lyrae light curves, obtaining several useful physical parameters. We analyze the cluster's Oosterhoff type in \S\ref{sec:oos}, whereas \S\ref{sec:cep} is dedicated to the type II Cepheid that we found in NGC~5286. \S\ref{sec:summ} summarizes the main results of our investigation. All of the derived light curves are provided in an Appendix.

\begin{center}
\begin{deluxetable*}{lccccccccccccl}
\tablewidth{0pc} 
\tabletypesize{\scriptsize}
\tablecaption{Photometric Parameters for NGC~5286 Variables}
\tablehead{\colhead{ID} & \colhead{RA (J2000)} & \colhead{DEC (J2000)} & \colhead{$P$} & \colhead{$A_B$} & \colhead{$A_V$} & \colhead{$(B)_{\rm mag}$} & \colhead{$(V)_{\rm mag}$} & \colhead{$\langle B \rangle$} & \colhead{$\langle V \rangle$} &  \colhead{$(\bv)_{\rm mag}$} &  \colhead{$\langle B \rangle - \langle V \rangle$} &  \colhead{$(\bv)_{\rm st}$} & \colhead{Comments} \\
\colhead{} & \colhead{(h:m:s)} & \colhead{(deg:m:s)} & \colhead{(days)} & \colhead{(mag)} & \colhead{(mag)} & \colhead{(mag)} & \colhead{(mag)} & \colhead{(mag)} & \colhead{(mag)} & \colhead{(mag)} & \colhead{(mag)} & \colhead{(mag)} & \colhead{} }
\startdata
V01...........  	& $13:46:21.5$& $-51:20:03.8$ & $0.635$ & $1.26$ & $0.99$ & $17.435$ & $16.743$  & $17.539$ & $16.818$ & $0.692$ & $0.721$ & $0.680$ &\RRab\ \\
V02...........  	& $13:46:35.1$& $-51:23:14.8$ & $0.611$ & $0.82$ & $0.71$ & $17.517$ & $17.002$ & $17.578$ & $17.047$ & $0.515$ & $0.530$ & $0.507$ & \RRab\ \\
V03...........  	& $13:46:54.1$& $-51:23:08.9$ & $0.685$ & $0.92$ & $0.73$ & $17.216$ & $16.697$ & $17.299$ & $16.744$ & $0.520$ & $0.555$ & $0.510$ & \RRab\ \\
V04...........  	& $13:46:19.2$& $-51:23:47.2$ & $0.352$ & $0.60$ & $0.46$ & $17.067$ & $16.639$ & $17.114$ & $16.668$ & $0.428$ & $0.446$ & $0.423$ &\RRc\ \\
V05...........  	& $13:46:33.4$& $-51:22:03.0$ & $0.5873$ & $...$  & $1.31$ & $...$ & $17.383$ & $...   $ & $17.518$ & $...  $ & $...  $ & $...  $ &\RRab\  \\
V06...........  	& $13:46:32.9$& $-51:23:04.3$ & $0.646$ & $1.38$ & $1.08$ & $16.850$ & $16.363$ & $16.990$ & $16.485$ & $0.487$ & $0.504$ & $0.476$ &\RRab\ \\
V07...........  	& $13:46:29.4$& $-51:23:38.4$ & $0.512$ & $...$  & $...$  & $17.08:$ & $16.58:$ & $....  $ & $....  $ & $0.50:$ & $.... $ & $.... $ &\RRab\ \\
V08......  	& $13:46:28.5$& $-51:23:10.2$ & $2.33$ & $1.24$ & $1.15$ & $15.569$ & $15.099$ & $15.730$ & $15.259$ & $0.471$ & $0.471$ & $...  $ &BL Her \\ 
V09...........  	& $13:46:39.0$& $-51:22:00.0$ & $0.3003$ & $0.63$ & $0.48$ & $17.266$ & $16.873$ & $17.314$ & $16.904$ & $0.393$ & $0.411$ & $0.387$ &\RRc\ \\
V10.........  	& $13:46:24.0$& $-51:22:46.4$ & $0.569$ & $1.29$ & $1.18$ & $17.529$ & $17.141$ & $17.635$ & $17.241$ & $0.388$ & $0.394$ & $0.376$ &\RRab\ \\
V11.........  	& $13:46:26.2$& $-51:23:35.7$ & $0.652$ & $0.94$ & $0.69$ & $17.120$ & $16.530$ & $17.194$ & $16.519$ & $0.590$ & $0.676$ & $0.580$ &\RRab\ \\
V12.........  	& $13:46:09.1$& $-51:22:38.9$ & $0.356$ & $0.65$ & $0.48$ & $17.305$ & $16.770$ & $17.352$ & $16.803$ & $0.535$ & $0.550$ & $0.528$ &\RRc\ \\
V13.........  	& $13:46:33.4$& $-51:23:33.8$ & $0.294$ & $0.65$ & $0.49$ & $16.871$ & $16.518$ & $16.919$ & $16.548$ & $0.353$ & $0.371$ & $0.346$ &\RRc\ \\
V14.........  	& $13:46:23.2$& $-51:23:38.9$ & $0.415$ & $0.60$ & $0.47$ & $16.994$ & $16.535$ & $17.037$ & $16.562$ & $0.458$ & $0.474$ & $0.454$ &\RRc\ \\
V15.........  	& $13:46:24.1$& $-51:22:56.8$ & $0.585$ & $1.78$ & $1.19$ & $17.539$ & $16.825$ & $17.768$ & $16.919$ & $0.713$ & $0.849$ & $0.689$ &\RRab\ \\
V17.........  	& $13:46:34.6$& $-51:23:29.0$ & $0.733$ & $0.83$ & $0.23$ & $17.172$ & $16.581$ & $17.182$ & $16.584$ & $0.591$ & $0.598$ & $0.582$ &\RRab\ \\
V18.........  	& $13:46:33.9$& $-51:23:16.0$ & $0.781$ & $0.20$ & $...$  & $17.265$ & $...$  & $17.280$ & $...$  & $...$  & $...$  & $...$  &\RRab\ \\
V20.........	& $13:46:25.2$& $-51:21:38.6$ & $0.319$ & $0.39$ & $0.35$ & $17.077$ & $16.614$ & $17.095$ & $16.628$ & $0.463$ & $0.467$ & $0.467$ &\RRc\ \\
V21.........	& $13:46:25.8$& $-51:24:02.7$ & $0.646$ & $0.89$ & $0.73$ & $17.176$ & $16.673$ & $17.246$ & $16.718$ & $0.503$ & $0.528$ & $0.493$ &\RRab\ \\
NV1........	& $13:46:27.4$& $-51:25:46.1$ & $0.366$ & $0.54$ & $0.43$ & $17.156$ & $16.695$ & $17.199$ & $16.722$ & $0.461$ & $0.476$ & $0.459$ &\RRc\ \\
NV2........  	& $13:46:17.7$& $-51:23:56.8$ & $0.354$ & $0.60$ & $0.50$ & $17.284$ & $16.758$ & $17.336$ & $16.789$ & $0.526$ & $0.547$ & $0.521$ &\RRc\ \\
NV3........  	& $13:46:30.5$& $-51:23:32.5$ & $0.755$ & $0.55$ & $0.30$ & $17.292$ & $16.501$ & $17.321$ & $16.503$ & $0.791$ & $0.819$ & $0.788$ &\RRab\ \\
NV4........  	& $13:46:15.9$& $-51:23:31.7$ & $0.786$ & $0.28$ & $0.21$ & $17.099$ & $16.450$ & $17.108$ & $16.455$ & $0.649$ & $0.653$ & $0.660$ &\RRab\ \\
NV5........  	& $13:46:17.6$& $-51:20:33.0$ & $0.357$ & $0.60$ & $0.45$ & $17.166$ & $16.659$ & $17.207$ & $16.685$ & $0.507$ & $0.522$ & $0.502$ &\RRc\ \\
NV6........  	& $13:46:26.4$& $-51:23:27.6$ & $0.566$ & $1.45$ & $1.12$ & $16.888$ & $16.511$ & $17.028$ & $16.604$ & $0.377$ & $0.425$ & $0.366$ &\RRab\ \\
NV7........  	& $13:46:25.8$& $-51:21:34.6$ & $0.339$ & $0.48$ & $0.41$ & $16.986$ & $16.560$ & $17.017$ & $16.585$ & $0.426$ & $0.431$ & $0.427$ &\RRc\ \\
NV8........  	& $13:46:26.2$& $-51:23:12.4$ & $0.80$ & $0.30$ & $0.30$ & $17.103$ & $16.463$ & $17.113$ & $16.473$ & $0.640$ & $0.640$ & $0.649$ &\RRab\ \\
NV9........  	& $13:46:24.6$& $-51:23:11.6$ & $0.745$ & $... $ & $... $ & $17.24:$ & $16.58:$ & $...   $ & $... $ & $0.66:$ & $... $ & $...$ &\RRab\ \\
NV10......  	& $13:46:25.1$& $-51:23:02.2$ & $0.339$ & $0.68$ & $0.54$ & $16.922$ & $16.436$ & $16.984$ & $16.454$ & $0.485$ & $0.529$ & $0.477$ &\RRc\ \\
NV11......  	& $13:46:26.0$& $-51:23:02.7$ & $0.536$ & $0.85$ & $0.71$ & $16.813$ & $16.347$ & $16.889$ & $16.402$ & $0.466$ & $0.487$ & $0.457$ &\RRab\ \\
NV12......  	& $13:46:26.7$& $-51:23:02.5$ & $0.905$ & $0.52$ & $0.41$ & $16.876$ & $16.251$ & $16.906$ & $16.269$ & $0.625$ & $0.637$ & $0.622$ &\RRab\ \\
NV13......  	& $13:46:27.3$& $-51:22:58.2$ & $0.583$ & $0.95$ & $1.35$ & $16.514$ & $16.083$ & $16.585$ & $16.215$ & $0.431$ & $0.370$ & $0.421$ &\RRab\ \\
NV14......  	& $13:46:29.2$& $-51:22:07.8$ & $0.284$ & $0.64$ & $0.44$ & $17.319$ & $16.803$ & $17.366$ & $16.826$ & $0.516$ & $0.540$ & $0.509$ &\RRc\ \\
NV15......  	& $13:46:25.5$& $-51:22:53.1$ & $0.742$ & $0.34$ & $0.52$ & $17.128$ & $16.477$ & $17.138$ & $16.505$ & $0.651$ & $0.633$ & $0.657$ &\RRab\ \\
NV16......  	& $13:46:27.0$& $-51:22:50.7$ & $0.366$ & $0.54$ & $0.32$ & $17.125$ & $16.618$ & $17.158$ & $16.629$ & $0.507$ & $0.529$ & $0.505$ &\RRc\ \\
NV17......  	& $13:46:26.6$& $-51:22:49.3$ & $0.322$ & $0.49$ & $0.50$ & $16.923$ & $16.490$ & $16.946$ & $16.431$ & $0.433$ & $0.415$ & $0.433$ &\RRc\ \\
NV18......  	& $13:46:29.1$& $-51:22:46.9$ & $0.362$ & $0.57$ & $0.44$ & $17.257$ & $16.410$ & $17.295$ & $16.431$ & $0.846$ & $0.864$ & $0.843$ &\RRc\ \\
NV19......  	& $13:46:26.8$& $-51:22:44.2$ & $0.658$ & $0.94$ & $1.70$ & $16.902$ & $16.624$ & $16.924$ & $16.706$ & $0.278$ & $0.218$ & $0.268$ &\RRab\ \\
NV20......  	& $13:46:22.5$& $-51:22:43.2$ & $0.3103$ & $1.90$ & $0.65$ & $17.555$ & $16.753$ & $18.023$ & $16.791$ & $0.802$ & $1.232$ & $0.736$ &\RRc\ \\
NV21......  	& $13:46:29.4$& $-51:22:41.0$ & $0.570$ & $0.95$ & $1.06$ & $16.830$ & $16.583$ & $16.902$ & $16.669$ & $0.247$ & $0.233$ & $0.237$ &\RRab\ \\
NV22......  	& $13:46:27.5$& $-51:22:39.9$ & $0.68$ & $1.15$ & $1.20$ & $16.829$ & $16.273$ & $17.022$ & $16.374$ & $0.556$ & $0.648$ & $0.544$ &\RRab\ \\
NV23......  	& $13:46:26.7$& $-51:22:16.1$ & $0.598$ & $1.14$ & $1.30$ & $17.090$ & $16.609$ & $17.182$ & $16.729$ & $0.481$ & $0.453$ & $0.469$ &\RRab\ \\
NV24......  	& $13:46:29.1$& $-51:22:39.1$ & $0.60$ & $0.75$ & $0.70$ & $16.927$ & $16.346$ & $16.965$ & $16.382$ & $0.581$ & $0.583$ & $0.574$ &\RRab\ \\
NV25......  	& $13:46:24.2$& $-51:22:17.2$ & $0.550$ & $1.44$ & $1.20$ & $17.328$ & $16.724$ & $17.048$ & $16.664$ & $0.605$ & $0.384$ & $0.593$ &\RRab\ \\
NV26......  	& $13:46:26.0$& $-51:22:36.5$ & $0.364$ & $0.46$ & $0.36$ & $17.011$ & $16.523$ & $17.040$ & $16.536$ & $0.489$ & $0.505$ & $0.490$ &\RRc\ \\
NV27......  	& $13:46:30.1$& $-51:22:22.0$ & $0.706$ & $1.17$ & $0.80$ & $16.952$ & $16.599$ & $17.489$ & $16.811$ & $0.490$ & $0.530$ & $0.340$ &\RRab\ \\
NV28......  	& $13:46:29.8$& $-51:22:35.4$ & $0.540$ & $0.97$ & $0.70$ & $16.880$ & $16.390$ & $16.954$ & $16.424$ & $0.490$ & $0.530$ & $0.479$ &\RRab\ \\
NV29......  	& $13:46:28.8$& $-51:22:33.8$ & $0.301$ & $0.97$ & $0.63$ & $17.458$ & $16.963$ & $17.565$ & $17.010$ & $0.495$ & $0.555$ & $0.473$ &\RRc\ \\
NV30......  	& $13:46:29.1$& $-51:22:24.1$ & $0.72$ & $0.65$ & $0.55$ & $17.251$ & $16.641$ & $17.292$ & $16.672$ & $0.610$ & $0.620$ & $0.605$ &\RRab\ \\
NV31......  	& $13:46:25.5$& $-51:22:31.6$ & $0.289$ & $0.41$ & $0.40$ & $17.163$ & $16.519$ & $17.189$ & $16.535$ & $0.645$ & $0.654$ & $0.648$ &\RRc\ \\
NV32......  	& $13:46:25.2$& $-51:22:30.0$ & $0.283$ & $0.34$ & $0.24$ & $16.804$ & $16.339$ & $16.817$ & $16.344$ & $0.465$ & $0.473$ & $0.471$ &\RRc\ \\
NV33......  	& $13:46:27.9$& $-51:22:26.6$ & $0.294$ & $0.34$ & $0.50$ & $16.801$ & $16.457$ & $16.815$ & $16.483$ & $0.345$ & $0.333$ & $0.351$ &\RRc\ \\
NV34......  	& $13:46:26.7$& $-51:22:24.9$ & $0.367$ & $0.42$ & $0.48$ & $16.569$ & $16.073$ & $16.591$ & $16.096$ & $0.496$ & $0.495$ & $0.499$ &\RRc\ \\
NV35......  	& $13:46:24.9$& $-51:23:03.8$ & $...$ & $...$ & $...$ & $14.70:$ & $13.38:$ & $...$ & $...$ & $1.32:$ & $...$ & $...$ & LPV \\ 
NV36......  	& $13:46:29.1$& $-51:22:58.4$ & $...$ & $...$ & $...$ & $15.30:$ & $13.40:$ & $...$ & $...$ & $1.90:$ & $...$ & $...$ & LPV \\ 
NV37......  	& $13:46:28.8$& $-51:22:12.9$ & $...$ & $...$ & $...$ & $15.70:$ & $14.60:$ & $...$ & $...$ & $1.10:$ & $...$ & $...$ & LPV \\ 
NV38......  	& $13:46:28.8$& $-51:20:37.0$ & $...$ & $...$ & $...$ & $15.30:$ & $13.74:$ & $...$ & $...$ & $1.56:$ & $...$ & $...$ & LPV 
\enddata
\label{tab1}
\end{deluxetable*}
\end{center}

\section{Observations and Data Reduction} \label{sec:obs}
The images used in this paper are the same as described in Paper~I, constituting a set of 128 frames in $V$ and 133 in $B$, acquired with the 1.3m Warsaw University Telescope at Las Campanas Observatory, Chile, in the course of a one-week run in April 2003. Further details can be found in Paper~I. In addition, a few images were taken in Feb. 2008 using the 4.1m Southern Astrophysical Research (SOAR) Telescope, located in Cerro Pach\'on, Chile, to further check the positions of the variables in the crowded regions around the cluster center.

The variable stars search was made using the image subtraction package ISIS v2.2 \citep{Al2000}. In order to convert the ISIS differential fluxes to standard magnitudes, we used DAOPHOT II/ALLFRAME \citep{stet1987,pbs94} to obtain instrumental magnitudes for each of the variables in the $B$ and $V$ reference images of the ISIS reductions.
First we obtained the flux of the variable star in the reference image, given by

\begin{equation}
 F_{\rm ref} = 10^{\left(\frac{C_0-m_{\rm ref}}{2.5}\right)},
\end{equation}

\noindent where $m_{\rm ref}$ is the instrumental magnitude of the star in the reference image and $C_0$ is a constant which depends on the photometric reduction package (for DAOPHOT II/ALLFRAME it is $C_0$ = 25). 
Then we derived instrumental magnitudes for each epoch from the differential fluxes $\Delta F_i = F_{\rm ref} - F_i$ given by ISIS using the equation

\begin{equation}
m_i= C_0 - 2.5\log\left(F_{\rm ref} - \Delta F_i \right).
\end{equation}

\noindent Finally, the equations to obtain the calibrated magnitudes ($M_i$) from the instrumental magnitudes are of the following form:

\begin{equation}
M_i = m_i + m_{\rm std}-m_{\rm ref}, 
\end{equation}

\noindent where $m_{\rm std}$ is the calibrated magnitude of the star in the reference image (we used the standard magnitude data from Paper~I).

\begin{deluxetable}{lccccc}
\tablewidth{0pc} 
\tabletypesize{\scriptsize}
\tablecaption{Photometry of the variable stars}
\tablehead{\colhead{Name} & \colhead{Filter} & \colhead{JD} & \colhead{Phase} & \colhead{Mag} & \colhead{e\_Mag}\\
\colhead{} & \colhead{} & \colhead{(d)} & \colhead{} & \colhead{(mag)} & \colhead{(mag)}}
\startdata
 V01 & $V$ & $2,452,736.54861$ & $0.0000$ & $16.2351$ & $0.0031$ \\ 
 V01 & $V$ & $2,452,736.55367$ & $0.0080$ & $16.2779$ & $0.0042$ \\ 
 V01 & $V$ & $2,452,736.55977$ & $0.0176$ & $16.2787$ & $0.0051$ \\ 
 V01 & $V$ & $2,452,736.57490$ & $0.0414$ & $16.3361$ & $0.0054$ \\ 
 V01 & $V$ & $2,452,736.58749$ & $0.0612$ & $16.3878$ & $0.0059$ \\
 V01 & $V$ & $2,452,736.59664$ & $0.0756$ & $16.3996$ & $0.0059$ \\
\enddata
\tablecomments{This table is published in its entirety in the 
electronic edition of the {\it Astronomical Journal}.  A portion is 
shown here for guidance regarding its form and content.}
\label{longtab}
\end{deluxetable}

\section{Variable Stars} \label{sec:var}
\begin{figure*}[t]
	  \begin{center}
      \includegraphics[scale=0.70]{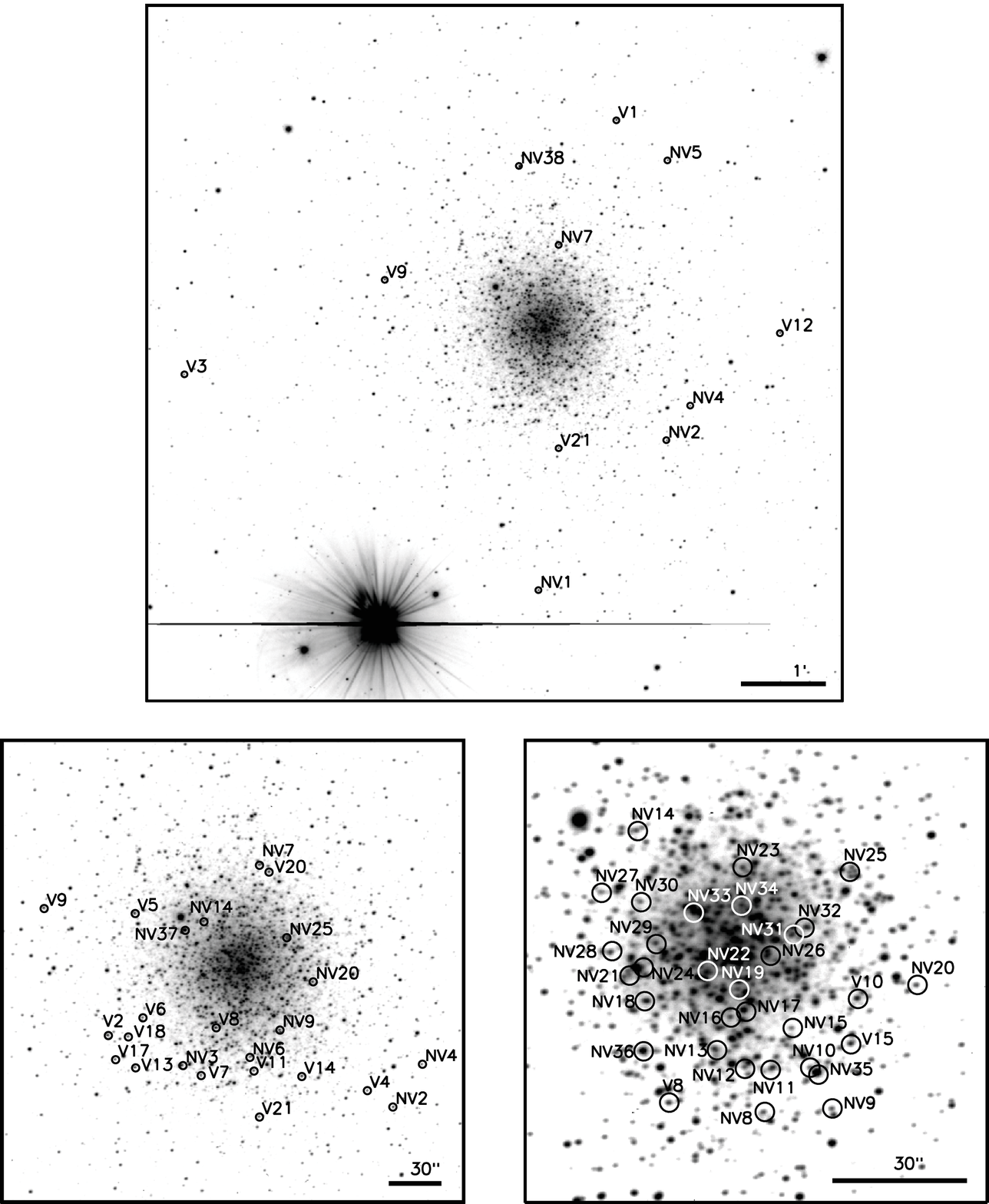}
	  \end{center}
      \caption{\footnotesize{Finding chart for the NGC~5286 variables. {\em Top}: The outer variables in the $V$ image obtained with the 0.9m CTIO telescope (note scale at the bottom right of the image). {\em Bottom left}: the same image as before, but zoomed in slightly. {\em Bottom right}: finding chart for the variables located closest to the cluster center.}}
      \label{ubica}
      \end{figure*}

In our variability search, we found 57 variable stars: 52 RR Lyrae (22 \RRc, 30 \RRab), 4 LPV's, and 1 type II Cepheid (more specifically, a BL Herculis star). We identified 19 of the 24 previously catalogued variables, and discovered 38 new variables. A finding chart is provided in Figure~\ref{ubica}.
Of the 16 previously catalogued variables with known periods (V1-V16) we were able to find 15. V16 is the only one not present in our data, because it is not in the chip of the CCD that we have analyzed for variability. The other 8 previously catalogued variables (V17-V24) were suggested by \citet{agea97} based just on their position on the CMD. We found that only 4 of these stars (V17, V18, V20, and V21) are real variables in our survey. We do not detect any variable sources at the coordinates that they provide for the remaining 4 candidate variables in their study (V19, V22, V23 and V24).
The recent images taken with a better spatial resolution at the 4.1m SOAR Telescope reveal that V19 is very close to two other stars, and probably is not resolved in the images used by \citet{agea97}. V22 and V23 are close to the instability strip but they still fall in the blue part of the HB, so they are not variable stars. V24 is in the instability strip but very close to the blue part of the HB. It is possible that this star belongs to the blue HB and is contaminated by a redder star.

\subsection{Periods and Light Curves} \label{subsec:per} 
Periods were determined using the phase dispersion minimization (PDM; \citealp{stel1978}) program in IRAF. Periods, along with the coordinates and several important photometric parameters, are provided in Table~\ref{tab1}. In this table, column~1 indicates the star's name. Columns~2 and 3 provide the right ascension and declination (J2000 epoch), respectively, whereas column~4 shows our derived period. Columns~5 and 6 list the derived amplitudes in the $B$ and $V$ bands, respectively, whereas columns~7 and 8 show the magnitude-weighted mean $B$ and $V$ magnitudes, corrected for differential reddening (see Paper I for details). The corresponding intensity-mean averages are provided in columns 9 and 10 (also corrected for differential reddening). The average $\bv$ color in magnitude units and the intensity-mean color $\langle B\rangle - \langle V\rangle$ are given in columns 11 and 12, respectively, whereas column~13 lists the $\bv$ color corresponding to the equivalent static star. Finally, the last column indicates the star's variability type.

To derive the color of the equivalent static star (i.e., the color the star would have if it were not pulsating), we first derived the magnitude-weighted mean color, and then applied an amplitude-dependent correction by interpolating on Table~4 from \citet*{bcs1995}. 

We calculate the HB level, $V_{\rm HB} = 16.63 \pm 0.04$, as the average $\langle V \rangle$ magnitude of all the RR Lyrae detected.

Magnitude data as a function of Julian Date and phase for the variable stars detected in our study  are given in Table~\ref{longtab}. In this table, column~1 indicates the star's name, following the \citet{cle2001} designation (when available). Column~2 indicates the filter used. Column~3 provides the Julian Date of the observation, whereas column~4 shows the phase according to our derived period (from Table~\ref{tab1}). Columns~5 and 6 list the observed magnitude in the corresponding filter and the associated error, respectively. Light curves based on our derived periods (when available) are shown in the Appendix.

	 \section{Color-Magnitude Diagram} \label{sec:cmd}
\begin{figure}[t]
     \plotone{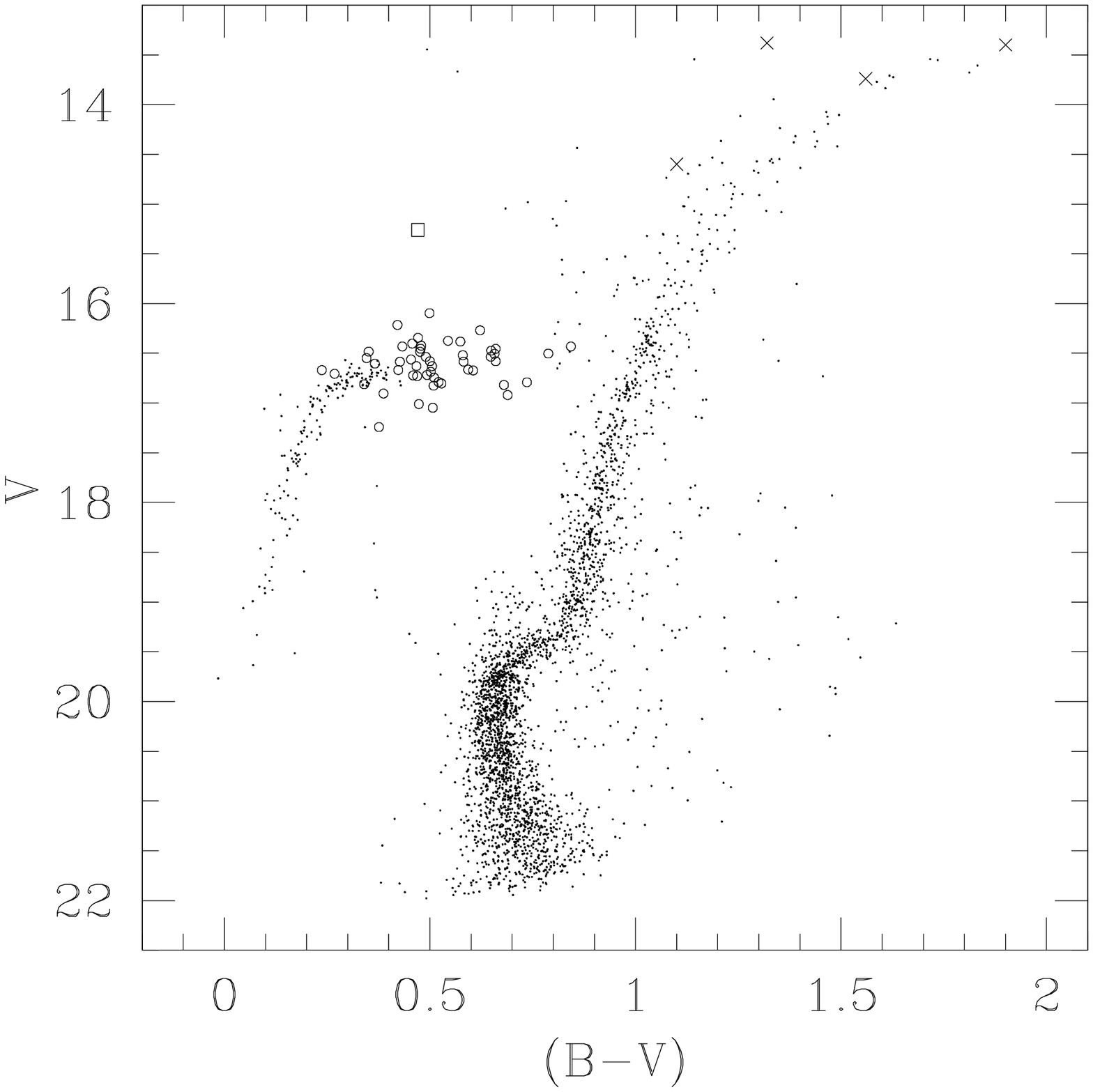}
     \caption{\footnotesize{CMD for NGC~5286 (from Paper~I) including the variables analyzed in this paper. {\em Crosses} indicate the LPV's, {\em circles} the RR Lyrae, and the {\em open square} the type II Cepheid.} The positions of the variables are based on the intensity-mean magnitudes $\langle V\rangle$ and on the colors of the equivalent static stars $(\bv)_{\rm st}$, as given in Table~\ref{tab1}.}
     \label{cmdcv}
     \end{figure}

Figure~\ref{cmdcv} shows the variable stars in the \objectname{NGC~5286} CMD, decontaminated from field stars as described in Paper~I. We can see that all RR Lyrae stars fall around the HB region, whereas the type II Cepheid is brighter than the HB. The four detected LPVs all fall close to the top of the RGB. These trends are precisely as expected if all the detected variables are cluster members. Note, however, that the CMD positions of the 4 LPVs are not precisely defined, since we lack adequate phase coverage for these stars.

\begin{figure}[t]
     \plotone{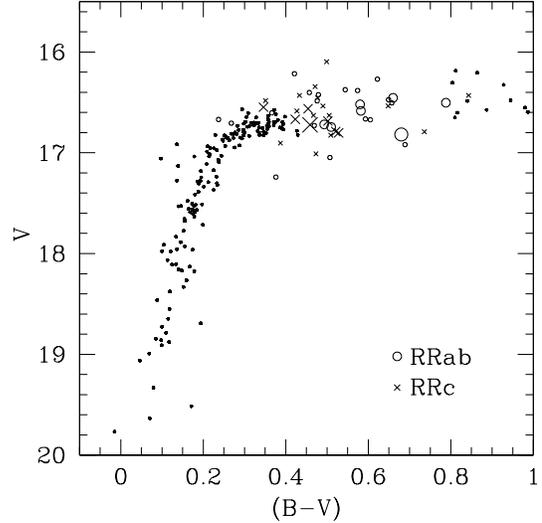}
     \caption{\footnotesize{As in Figure~\ref{cmdcv}, but focusing on the HB region of the CMD. Here we use different symbol sizes for the variables in different radial annuli from the cluster center: Small symbols are related to stars inside the core radius ($r \leq 0.29'$); medium-sized symbols refer to stars with $0.29' < r \leq 0.58'$; and the larger symbols refer to stars in the outermost cluster regions ($r > 0.58'$).}}
     \label{hb}
     \end{figure}

Figure~\ref{hb} is a magnified CMD showing only the HB region. To assess the effects of crowding, we use different symbol sizes for the variables in different radial annuli from the cluster center: small sizes for stars in the innermost cluster regions ($r \leq 0.29'$) which are badly affected by crowding, medium sizes for stars with $0.29' < r \leq 0.58'$, and large sizes for stars in the outermost cluster regions ($r > 0.58'$). As expected, the variables in the innermost cluster region present more scatter. Apart from this, in general the detected RR Lyrae fall inside a reasonably well-defined instability strip (IS). However, we can see that there is not a clear-cut separation in color between \RRc's and \RRab's. Although the \RRc's tend to be found preferentially towards the blue side of the IS, as expected, the \RRab's are more homogeneously distributed. This can be again a scatter effect, because the RRab stars that are in the less crowded regions of the cluster (large and medium size circles) are more concentrated at the red part.

\subsection{Notes on Individual Stars}
{\bf V5}: This star was only detected in our $V$ images, so that we were unable to determine its color. It remains unclear to us why ISIS was unable to detect this variable in the $B$ data~-- note, from Figure~\ref{ubica}, that it is not located in an especially crowded region of the cluster.

{\bf V7}: Our derived period, 0.512~d, is slightly longer than the previously reported period of
0.50667~d in \citet{ll78}.  As a matter of fact, V7 remains one of the
shortest-period RRab stars among all known OoII GCs.  Because the period
is so close to half a day, there is a considerable gap in the phased light curve for V7
(as was also the case, though to a somewhat lesser degree, for the light curve of \citeauthor{ll78}).  
We carefully checked the PDM periodogram of the star in search of acceptable
longer periods, but could find none that fit our data, nor could we find a period
that reduced significantly the spread seen in the phased light curve close to minimum 
light (which is particularly obvious in the $V$-band light curve). The \citeauthor{ll78}
light curve also shows substantial scatter close to minimum light, although in their case 
much of the scatter is clearly due to photometric error.

{\bf V8}: \citet{ll78} found a period of 0.7~d for this star and suggested that it is an RR Lyrae-type variable. In our analysis we found an alias at 0.7~d for the period, but the best fit is obtained with a period of 2.33~d, which corresponds to a type II Cepheid of the BL Her type.

{\bf V18}: As for V5, we only detected this star in one of the filters. In this case ISIS only detected it in our $B$ images, so that we were unable to determine its color and also to perform Fourier decomposition.

{\bf V7} and {\bf NV9}: We do not see the minimum and maximum, respectively, for these two variables. For that reason, 
we could not perform a Fourier analysis, and the mean $V$ and $B$ values provided in Table~\ref{tab1} are just approximate. 

\section{Fourier Decomposition} \label{sec:four} 
Light curves for RR Lyrae variables were analyzed by Fourier decomposition 
using the same equations as in \citet{cor2003}, namely 

\begin{equation}
\ mag = A_0 + \sum_{j=1}^{N} A_j \sin(2 \pi j \, t/P + \phi_j)
\label{corwin1}
\end{equation}

\noindent (for RR{\em ab} variables) and 

\begin{equation}
\ mag = A_0 + \sum_{j=1}^{N} A_j \cos(2 \pi j \, t/P + \phi_j)
\label{corwin2}
\end{equation}

\noindent (for RR{\em c} variables), where again $N = 10$ was usually adopted. 

\subsection{RR{\em c} Variables} 
Amplitude ratios $A_{j1} \equiv A_j/A_1$ and phase differences $\phi_{j1} \equiv \phi_j - j \phi_1$ for the lower-order terms are provided in Table~\ref{tab2}. In this table, a colon symbol (``:'') indicates an uncertain value, whereas a double colon (``::'') indicates a very uncertain value. 

\begin{deluxetable}{lllllll}
\tablewidth{0pc} 
\tabletypesize{\scriptsize}
\tablecaption{Fourier Coefficients: RR{\em c} Stars}
\tablehead{\colhead{ID} & \colhead{$A_{21}$} & \colhead{$A_{31}$} & \colhead{$A_{41}$} & \colhead{$\phi_{21}$} & \colhead{$\phi_{31}$} & \colhead{$\phi_{41}$} }
\startdata
 V4  	& $	0.113	$ & $	0.059	$ & $	0.037	$ & $	4.503	$ & $	3.667	$ & $	2.103	$ \\
 V9  	& $	0.226	$ & $	0.056	$ & $	0.053	$ & $	4.735	$ & $	2.905	$ & $	1.283	$ \\
 V12 	& $	0.159	$ & $	0.096	$ & $	0.032	$ & $	5.064	$ & $	3.454	$ & $	2.635	$ \\
 V13 	& $	0.194	$ & $	0.063	$ & $	0.058	$ & $	4.748	$ & $	2.451	$ & $	1.248	$ \\
 V14 	& $	0.072	$ & $	0.064	$ & $	0.062	$ & $	5.411	$ & $	4.522	$ & $	3.401	$ \\
 V20 	& $	0.059	$ & $	0.016	$ & $	0.026	$ & $	4.723	$ & $	3.832	$ & $	3.128	$ \\
 NV1 	& $	0.117	$ & $	0.091	$ & $	0.053	$ & $	4.768	$ & $	3.635	$ & $	3.295	$ \\
 NV2 	& $	0.123	$ & $	0.051	$ & $	0.036	$ & $	4.621	$ & $	3.406	$ & $	2.433	$ \\
 NV5 	& $	0.083	$ & $	0.059	$ & $	0.043	$ & $	4.766	$ & $	4.134	$ & $	2.693	$ \\
 NV7 	& $	0.151::	$ & $	0.064::	$ & $	0.012::	$ & $	5.443::	$ & $	4.886::	$ & $	5.665::	$ \\
NV10 	& $	0.144	$ & $	0.056	$ & $	0.020	$ & $	4.936	$ & $	3.237	$ & $	2.660	$ \\
NV14 	& $	0.193	$ & $	0.053	$ & $	0.044	$ & $	4.645	$ & $	2.881	$ & $	1.954	$ \\
NV16 	& $	0.069::	$ & $	0.094::	$ & $	0.058::	$ & $	4.448::	$ & $	4.293::	$ & $	2.468::	$ \\
NV17 	& $	0.115::	$ & $	0.052::	$ & $	0.085::	$ & $	5.515::	$ & $	3.764::	$ & $	2.698::	$ \\
NV18 	& $	0.087:	$ & $	0.013:	$ & $	0.041:	$ & $	4.484:	$ & $	5.416:	$ & $	3.313:	$ \\
NV20 	& $	0.164	$ & $	0.098	$ & $	0.072	$ & $	4.844	$ & $	3.104	$ & $	1.823	$ \\
NV26 	& $	0.068:	$ & $	0.072:	$ & $	0.035:	$ & $	5.041:	$ & $	4.154:	$ & $	3.058:	$ \\
NV29 	& $	0.233:	$ & $	0.091:	$ & $	0.029:	$ & $	4.929:	$ & $	3.344:	$ & $	1.328:	$ \\
NV31 	& $	0.152:	$ & $	0.002:	$ & $	0.046:	$ & $	4.557:	$ & $	1.086:	$ & $	0.590:	$ \\
NV32 	& $	0.074:	$ & $	0.017:	$ & $	0.050:	$ & $	5.495:	$ & $	3.484:	$ & $	3.811:	$ \\
NV33 	& $	0.168:	$ & $	0.075:	$ & $	0.073:	$ & $	4.704:	$ & $	2.230:	$ & $	0.747:	$ \\
NV34 	& $	0.043:	$ & $	0.103:	$ & $	0.017:	$ & $	5.336:	$ & $	3.616:	$ & $	2.641:	$ 
\enddata
\label{tab2}
\end{deluxetable}

\begin{deluxetable*}{lcccccc}
\tablewidth{0pc} 
\tabletypesize{\scriptsize}
\tablecaption{Fourier-Based Physical Parameters: RR{\em c} Stars}
\tablehead{\colhead{ID} & \colhead{$M/M_{\odot}$} & \colhead{$\log{(L/L_{\odot})}$} & \colhead{$T_e ({\rm K})$} & \colhead{{\it{y}}} & \colhead{[Fe/H]} & \colhead{$ \left\langle M_V \right\rangle $}}
\startdata
 V4  	& $0.563$ & $1.725$ & $7264$ & $0.271$ & $-1.690$ & $0.747$\\
 V9  	& $0.629$ & $1.698$ & $7358$ & $0.275$ & $-1.891$ & $0.811$\\
 V12 	& $0.598$ & $1.744$ & $7229$ & $0.264$ & $-1.681$ & $0.737$\\
 V13 	& $0.625$ & $1.619$ & $7568$ & $0.298$ & $-1.898$ & $0.822$\\
 V14 	& $0.495$ & $1.751$ & $7167$ & $0.268$ & $-1.069$ & $0.711$\\
 V20 	& $0.513$ & $1.672$ & $7382$ & $0.289$ & $-1.837$ & $0.763$\\
 NV1 	& $0.579$ & $1.745$ & $7218$ & $0.265$ & $-1.605$ & $0.745$\\
 NV2 	& $0.604$ & $1.744$ & $7231$ & $0.264$ & $-1.673$ & $0.742$\\
 NV5 	& $0.504$ & $1.705$ & $7290$ & $0.280$ & $-1.656$ & $0.748$\\
NV10 	& $0.617$ & $1.734$ & $7259$ & $0.266$ & $-1.766$ & $0.764$\\
NV14 	& $0.614$ & $1.673$ & $7418$ & $0.282$ & $-1.896$ & $0.816$\\
NV20 	& $0.609$ & $1.701$ & $7342$ & $0.275$ & $-1.887$ & $0.830$\\
Mean 	& $0.601 \pm 0.049$ &$1.715 \pm 0.040$ & $7276 \pm 110$ & $0.273 \pm 0.011$  & $-1.713 \pm 0.230$ & $0.770 \pm 0.039$ 
\enddata
\label{tab3}
\end{deluxetable*}

 \citet{sicl93} used light curves of \RRc\ variables, as provided by their hydrodynamical pulsation models, to derive equations to calculate mass, luminosity, effective temperature, and a ``helium parameter'' for \RRc\ variables~-- all as a function of the Fourier phase difference $\phi_{31} \equiv \phi_3 - 3 \phi_1$ and the period. However, \citet[][see his \S4]{mc04} pointed out that the \citeauthor{sicl93} set of equations cannot, in their current form, provide physically correct values for luminosities and masses, since they violate the constraints imposed by the Ritter (period-mean density) relation. We still provide the derived values for the NGC~5286 \RRc\ stars in this paper though, chiefly for comparison with similar work for other GCs.
 
Accordingly, we use \citeauthor{sicl93}'s (\citeyear{sicl93}) equations~(2), (3), (6), and (7) to compute $M/M_{\odot}$, $\log{(L/L_{\odot})}$, $T_e$, and {\it{y}}, respectively, for 12 of our \RRc\ variables (i.e., those with the best defined Fourier coefficients). We also use equation~(3) in \citet{smea07} to compute ${\rm [Fe/H]}$, and equation~(10) in \citet{kov1998} to compute $M_V$. The results are given in Table~\ref{tab3}. 
 
 The unweighted mean values and corresponding standard errors of the derived mean mass, $\log$ luminosity (in solar units), effective temperature, ``helium parameter'', metallicity (in the \citeauthor{zw1984} \citeyear{zw1984} scale), and mean absolute magnitude in $V$ are $(0.601 \pm 0.049) \, M_{\odot}$, $(1.715 \pm 0.040)$, $(7276 \pm 110)$~K, $(0.273 \pm 0.011)$, $(-1.71  \pm 0.23)$, and $(0.770 \pm 0.039)$~mag, respectively.

\subsection{RR{\em ab} Variables}\label{sec:subab}
Amplitude ratios $A_{j1}$ and phase differences $\phi_{j1}$ for the lower-order terms are provided, in the case of the \RRab's, in Table~\ref{tab4}. We also give the Jurcsik-Kov\'acs $D_m$ value (\citealp{jurko96}, computed on the basis of their eq.~[6] and Table~6), which is intended to differentiate \RRab\ stars with ``regular'' light curves from those with ``anomalous'' light curves (e.g., presenting the Blazhko effect~-- but see also \citeauthor*{cac05} \citeyear{cac05} for a critical discussion of $D_m$ as an indicator of the occurrence of the Blazhko phenomenon). As before, a colon symbol (``:'') indicates an uncertain value, whereas a double colon (``::'') indicates a very uncertain value.  

\begin{deluxetable*}{lllllllr}
\tablewidth{0pc} 
\tabletypesize{\scriptsize}
\tablecaption{Fourier Coefficients: RR{\em ab} Stars}
\tablehead{\colhead{ID} & \colhead{$A_{21}$} & \colhead{$A_{31}$} & \colhead{$A_{41}$} & \colhead{$\phi_{21}$} & \colhead{$\phi_{31}$} & \colhead{$\phi_{41}$} & \colhead{$D_m$} }
\startdata
 V1  	& $	0.524	$ & $	0.359	$ & $	0.246	$ & $	2.406	$ & $	5.113	$ & $	1.638	$ & $	0.751	$ \\
 V2  	& $	0.379	$ & $	0.257	$ & $	0.108	$ & $	2.293	$ & $	4.754	$ & $	1.181	$ & $	2.534	$ \\
 V3  	& $	0.412	$ & $	0.258	$ & $	0.126	$ & $	2.623	$ & $	5.379	$ & $	2.263	$ & $	1.211	$ \\
 V5  	& $	0.472	$ & $	0.303	$ & $	0.231	$ & $	2.329	$ & $	5.040	$ & $	1.459	$ & $	5.957	$ \\
 V6  	& $	0.563	$ & $	0.330	$ & $	0.199	$ & $	2.419	$ & $	5.198	$ & $	1.683	$ & $	14.276	$ \\
 V7  	& $	...	$ & $	...	$ & $	...	$ & $	...	$ & $	...	$ & $	...	$ & $	...	$ \\
 V10 	& $	0.488	$ & $	0.325	$ & $	0.193	$ & $	2.330	$ & $	4.948	$ & $	1.148	$ & $	2.117	$ \\
 V11 	& $	0.491	$ & $	0.226	$ & $	0.176	$ & $	2.484	$ & $	5.228	$ & $	1.844	$ & $	16.137	$ \\
 V15 	& $	0.448	$ & $	0.362	$ & $	0.222	$ & $	2.363	$ & $	5.235	$ & $	1.517	$ & $	4.592	$ \\
 V17 	& $	0.267	$ & $	0.078	$ & $	0.019	$ & $	2.653	$ & $	5.840	$ & $	3.345	$ & $	1.502	$ \\
 V21 	& $	0.494	$ & $	0.315	$ & $	0.142	$ & $	2.585	$ & $	5.403	$ & $	2.227	$ & $	1.956	$ \\
 NV3 	& $	0.346	$ & $	0.204	$ & $	0.023	$ & $	2.647	$ & $	5.167	$ & $	1.894	$ & $	6.447	$ \\
 NV4 	& $	0.246:	$ & $	0.087:	$ & $	0.037:	$ & $	2.712:	$ & $	6.177:	$ & $	3.153:	$ & $	3.197:	$ \\
 NV6 	& $	0.458	$ & $	0.352	$ & $	0.237	$ & $	2.260	$ & $	4.839	$ & $	1.164	$ & $	2.117	$ \\
 NV8 	& $	0.330	$ & $	0.135	$ & $	0.021	$ & $	2.775	$ & $	5.838	$ & $	3.085	$ & $	0.379	$ \\
 NV9 	& $	...	$ & $	...	$ & $	...	$ & $	...	$ & $	...	$ & $	...	$ & $	...	$ \\
NV11 	& $	0.334	$ & $	0.210	$ & $	0.107	$ & $	2.176	$ & $	4.616	$ & $	1.198	$ & $	3.550	$ \\
NV12 	& $	0.295:	$ & $	0.080:	$ & $	0.106:	$ & $	3.511:	$ & $	0.525:	$ & $	4.805:	$ & $	3.143:	$ \\
NV13 	& $	0.517	$ & $	0.353	$ & $	0.217	$ & $	2.438	$ & $	5.241	$ & $	1.588	$ & $	0.324	$ \\
NV15 	& $	0.439	$ & $	0.194	$ & $	0.078	$ & $	2.619	$ & $	5.645	$ & $	2.320	$ & $	4.390	$ \\
NV19 	& $	0.475	$ & $	0.231	$ & $	0.000	$ & $	2.450	$ & $	5.234	$ & $	1.749	$ & $	5.938	$ \\
NV21 	& $	0.487	$ & $	0.341	$ & $	0.245	$ & $	2.357	$ & $	4.812	$ & $	1.232	$ & $	0.313	$ \\
NV22 	& $	0.566	$ & $	0.339	$ & $	0.217	$ & $	2.670	$ & $	5.588	$ & $	2.607	$ & $	10.859	$ \\
NV23 	& $	0.463	$ & $	0.321	$ & $	0.221	$ & $	2.208	$ & $	5.015	$ & $	1.187	$ & $	2.600	$ \\
NV24 	& $	0.631	$ & $	0.282	$ & $	0.136	$ & $	1.964	$ & $	4.458	$ & $	0.793	$ & $	4.763	$ \\
NV25 	& $	0.448	$ & $	0.344	$ & $	0.241	$ & $	2.277	$ & $	4.859	$ & $	1.108	$ & $	4.151	$ \\
NV27 	& $	0.481	$ & $	0.296	$ & $	0.156	$ & $	2.627	$ & $	5.540	$ & $	2.191	$ & $	0.686	$ \\
NV28 	& $	0.497	$ & $	0.277	$ & $	0.158	$ & $	2.278	$ & $	4.901	$ & $	0.799	$ & $	3.703	$ \\
NV30 	& $	0.444	$ & $	0.262	$ & $	0.119	$ & $	2.474	$ & $	5.587	$ & $	2.174	$ & $	3.312	$ 
\enddata
\label{tab4}
\end{deluxetable*}

\begin{deluxetable*}{lcccccccc}
\tablewidth{0pc}
\tabletypesize{\scriptsize}
\tablecaption{Fourier-Based Physical Parameters: RR{\em ab} Stars}
\tablehead{\colhead{ID} & \colhead{[Fe/H]} & \colhead{$ \left\langle M_V \right\rangle $} & \colhead{$ \left\langle V-K \right\rangle $} & \colhead{$ \log{T_e}^{ \left\langle V-K \right\rangle } $ } & \colhead{$ \left\langle B-V \right\rangle $} & \colhead{$ \log{T_e}^{ \left\langle B-V \right\rangle } $} & \colhead{$ \left\langle V-I \right\rangle $} & \colhead{$ \log{T_e}^{ \left\langle V-I \right\rangle } $}  }
\startdata
V1 	& $-1.587$ & $0.707$ & $1.200$ & $3.801$ & $0.355$ & $3.804$ & $0.450$ & $3.837$ \\
V2 	& $-1.938$ & $0.728$ & $1.239$ & $3.798$ & $0.355$ & $3.802$ & $0.443$ & $3.843$ \\
V3 	& $-1.501$ & $0.692$ & $1.256$ & $3.794$ & $0.366$ & $3.801$ & $0.475$ & $3.829$ \\
V10 	& $-1.452$ & $0.747$ & $1.105$ & $3.811$ & $0.327$ & $3.815$ & $0.398$ & $3.848$ \\
V17	& $-1.139$ & $0.752$ & $1.321$ & $3.786$ & $0.406$ & $3.790$ & $0.534$ & $3.810$ \\
V21 	& $-1.529$ & $0.687$ & $1.267$ & $3.793$ & $0.376$ & $3.798$ & $0.489$ & $3.826$ \\
NV6 	& $-1.584$ & $0.752$ & $1.135$ & $3.808$ & $0.335$ & $3.812$ & $0.407$ & $3.848$ \\
NV8 	& $-1.509$ & $0.645$ & $1.385$ & $3.780$ & $0.409$ & $3.785$ & $0.551$ & $3.810$ \\
NV13 	& $-1.521$ & $0.662$ & $1.292$ & $3.790$ & $0.391$ & $3.794$ & $0.518$ & $3.818$ \\
NV21 	& $-1.140$ & $0.732$ & $1.069$ & $3.814$ & $0.322$ & $3.819$ & $0.396$ & $3.845$ \\
NV23 	& $-1.642$ & $0.752$ & $1.156$ & $3.806$ & $0.339$ & $3.809$ & $0.413$ & $3.847$ \\
NV27 	& $-1.397$ & $0.677$ & $1.239$ & $3.796$ & $0.370$ & $3.800$ & $0.484$ & $3.825$ \\
Mean & $-1.515 \pm 0.213$ & $0.717 \pm 0.038$ & $1.239 \pm 0.093$ & $3.797 \pm 0.010$ & $0.361 \pm 0.029$ & $3.802 \pm 0.010$ & $0.462 \pm 0.054$ & $3.833 \pm 0.015$ 
\enddata
\label{tab5}
\end{deluxetable*}

 \citet{jurko96}, \citet{kovjur1996,kovjur1997}, \citet{jur98}, \citet{koka98}, and \citet{kowa99,kowa2001} derived empirical expressions that relate metallicity, absolute magnitude, and temperature with the Fourier parameters of \RRab\ stars, in the case of sufficiently ``regular'' light curves ($D_m < 3$). We accordingly use equations~(1), (2), (5), and (11) in \citet{jur98} to compute ${\rm [Fe/H]}$, $M_V$, $V\!-\!K$, and $ \log{T_e}^{ \left\langle V-K \right\rangle }$, respectively, for the 12 \RRab\ variables in our sample with $D_m < 3$. The color indices \bv  and $V\!-\!I$ come from equations~(6) and (9) of \citet{kowa2001}; we then use equation~(12) of \citet{kowa99}, assuming a mass of $0.7 \, M_{\odot}$, to derive temperature values from equation~(11) (for \bv) and equation~(12) (for $V\!-\!I$) in \citet{kowa2001}. These results are given in Table~\ref{tab5}.  

Fourier parameters suggest a metallicity of ${\rm [Fe/H]} = -1.52  \pm 0.21$ for NGC~5286 in the \citet{jur95} scale; this corresponds to a value of ${\rm [Fe/H]} =  -1.68$ in the \citet{zw1984} scale. This is consistent with the value derived using the \RRc\ variables and is in excellent agreement with \citet[][${\rm [Fe/H]} = -1.67$]{har1996}, and also with the value obtained in Paper~I (${\rm [Fe/H]} = -1.70\pm 0.05$, again in the \citeauthor{zw1984} scale) on the basis of several different photometric parameters describing the shape and position of the cluster's red giant branch in the $V$, \bv diagram. 

 We find a mean absolute magnitude of $\left\langle M_V \right\rangle = 0.717 \pm 0.038$~mag for the \RRab\ stars. For the same set of 12 \RRab\ stars used to derive this value,  we also find $\left\langle V \right\rangle = 16.64 \pm 0.08$~mag. This implies a distance modulus of $(m-M)_V = 15.92 \pm 0.12$ for NGC~5286, which is in excellent agreement with the value provided in the \citet{har1996} catalog, namely, $(m-M)_V = 15.95$~mag.
If one adopts instead for the HB an average absolute magnitude of $M_V = 0.60$~mag at the NGC~5286 metallicity, as implied by equation~(4a) in \citet{cc08}~-- which is based on a calibration of the RR Lyrae distance scale that uses the latest Hipparcos and {\em Hubble Space Telescope} trigonometric parallaxes for RR Lyr, and takes explicitly into account the evolutionary status of this star~-- one finds for the cluster a distance modulus of $(m-M)_V = 16.04$~mag for NGC~5286. We caution the reader that the NGC~5286 RR Lyrae stars could in principle be somewhat overluminous 
for the cluster's metallicity (in view of the cluster's predominantly blue HB), in which case the correct distance modulus could be even larger, by an amount that could be of the order of $\sim 0.1$~mag \citep[e.g.,][]{lc99,pdea00}. 
Finally, and as also pointed out by \citet{cac05}, we also warn the reader that intrinsic colors and temperatures estimated from Fourier decomposition are not particularly reliable, and should accordingly be used with due caution. The reader is also referred to \citet{kov1998} and \citet{mc04} for caveats regarding the validity of the results obtained based on the \citet{sicl93} relations for \RRc\ stars.

\section{The Oosterhoff Type of NGC~5286}\label{sec:oos} 
\begin{figure}[t]
      \plotone{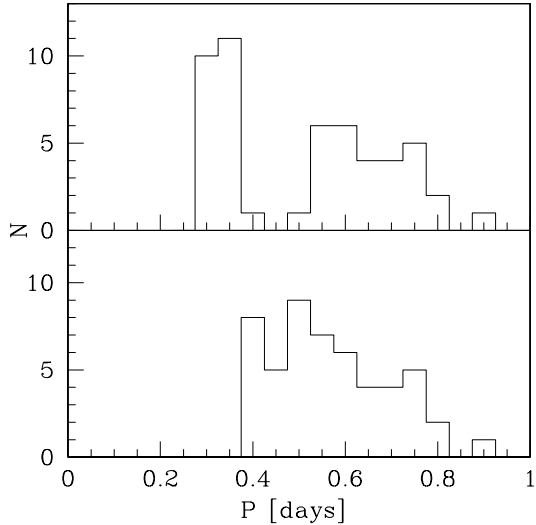}
      \caption{\footnotesize{{\em Top}: period histogram for the NGC~5286 RR Lyrae stars. {\em Bottom}: Same as in the top, but fundamentalized the periods of the \RRc's.}}
      \label{cuantas}
      \end{figure}

Figure~\ref{cuantas} shows a histogram with our derived periods in \objectname{NGC~5286}. The bottom panel is similar to the upper panel, but with the \RRc\ periods ``fundamentalized'' using the equation $\log P_f = \log P_c + 0.128$ \citep[e.g.,][and references therein]{cat2005}. Note the lack of a sharply peaked distribution, contrary to what is seen in several GCs of both Oosterhoff types, but most notably in M3 \citep[e.g.,][and references therein]{mc04a,dc08}. 

In order to assign an Oosterhoff type to NGC~5286, we must compare its RR Lyrae properties with those found in other \citet{o39,o44} type I and II GCs. In this sense, \citet{cle2001} found the mean \RRab\ and \RRc\ (plus RR{\em d}) periods for RR Lyrae stars in Galactic GCs to be $0.559$~days and $0.326$~days, respectively, for OoI clusters, and $0.659$~days and $0.368$~days, respectively, in the case of OoII clusters. In addition, Catelan et al. (2009, in preparation) have recently shown that the minimum period of the ab-type pulsators $P_{\rm ab,min}$, when used in conjunction with $\langle P_{ab}\rangle$, provides a particularly reliable diagnostics of Oosterhoff status. 

In this sense, the key quantities for the cluster can be summarized as follows: 

\begin{align}
\left\langle P_{ab} \right\rangle  & = 0.656 \, {\rm d}, \\
\left\langle P_{c} \right\rangle   & = 0.333 \, {\rm d}, \\
P_{\rm ab,min} & = 0.512 \, {\rm d},\\
N_c/N_{c+ab}   & = 0.42.
\end{align}

\begin{figure}[t]
      \plotone{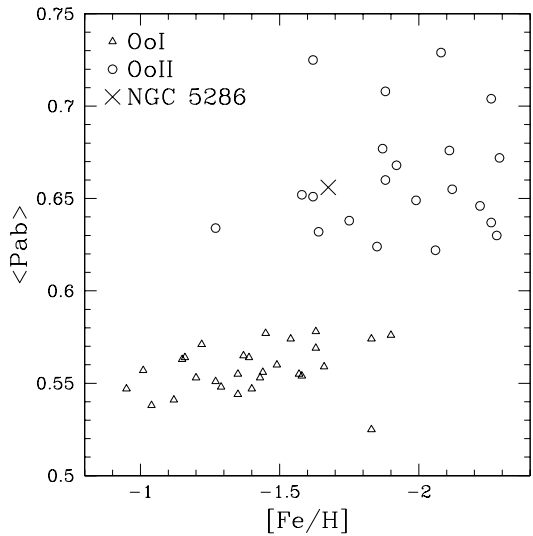}
      \caption{\footnotesize{Distribution of Galactic GCs in the mean \RRab\ period vs. metallicity plane. The position of NGC~5286, as derived on the basis of our new measurements, is highlighted.}}
      \label{pf}
      \end{figure}

One immediately finds that the value of $\langle P_{ab}\rangle$ for the cluster points to an OoII status (see also Fig.~\ref{pf}, which is based on the compilation presented in \citeauthor{cat2005} \citeyear{cat2005})~-- which is also favored by its relatively high c-type number fraction. On the other hand, the average period of the \RRc's is lower than typically found among OoII globulars, being more typical of OoI objects. However, as shown by Catelan et al. (2009), there is a large overlap in $\left\langle P_{c} \right\rangle$ values between OoI and OoII globulars, thus making this quantity a poorer indicator of Oosterhoff status than is often realized. 

The situation regarding $P_{\rm ab,min}$ is rather interesting, for the value derived for NGC~5286 makes it one of the OoII clusters with the shortest $P_{\rm ab,min}$ values to date~-- though 0.512~d would still clearly be too long for an OoI cluster, which generally have $P_{\rm ab,min} < 0.5$~d (Catelan et al. 2009)~-- as opposed to OoII clusters, which typically have instead $P_{\rm ab,min} > 0.5$~d. As can be seen from Table~\ref{tab1} and Figure~\ref{cuantas} ({\em top}), after V7 (the star with $P = 0.512$~d), the next shortest-period star in NGC~5286 is NV11, with $P = 0.536$~d; this is indeed less atypical for an OoII object. 

      \begin{figure}[t]
      \plotone{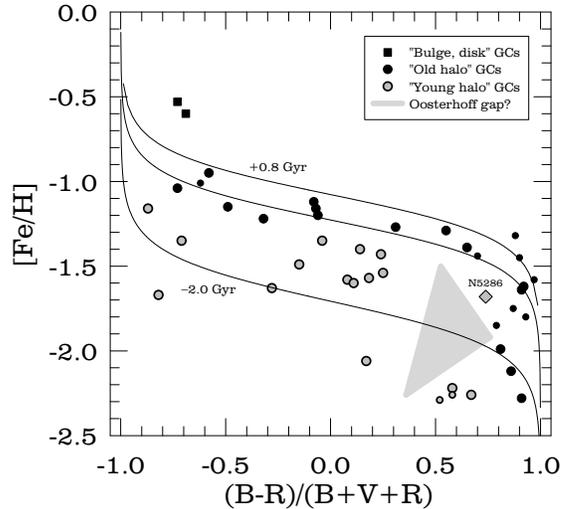}
      \caption{\footnotesize{Distribution of Galactic GCs in the metallicity-HB type $\mathcal{L}$ plane. The region marked as a triangle represents the ``Oosterhoff Gap,'' a seemingly forbidden region for bona-fide Galactic GCs. OoI clusters tend to sit to the left of the Oosterhoff Gap, whereas OoII clusters are mostly found to its right. The overplotted lines are isochrones from \citet{cf1993}. Adapted from \citet{cat2005}.}}
      \label{fl}
      \end{figure} 

      \begin{figure}[t]
      \plotone{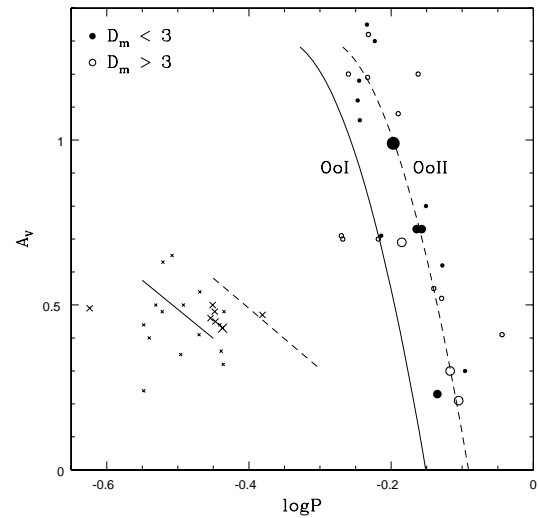}
      \caption{\footnotesize{Position of RR Lyrae stars on the Bailey (period-amplitude) diagram for $V$. {\em Filled circles} show RR{\em ab}'s with $D_m < 3$, {\em open circles} those with $D_m > 3$, and {\em crosses} show the RR{\em c}'s. {\em Solid lines} are the typical lines for OoI clusters and {\em Dashed lines} for OoII clusters, according to \citealp{cac05}. As in Figure~\ref{hb}, different symbol sizes are related to the radial distance from the cluster center.}}
      \label{baiv}
      \end{figure} 
      
      \begin{figure}[t]
      \plotone{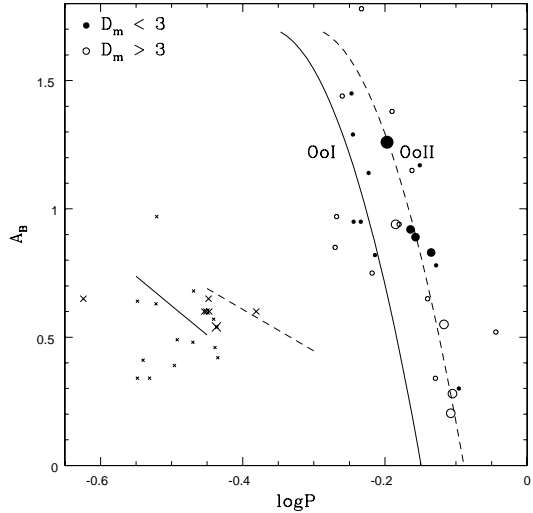}
      \caption{\footnotesize{Same as in Figure~\ref{baiv}, but for $B$.}}
      \label{baib}
      \end{figure}

The position of the cluster in a metallicity versus HB type diagram is displayed in Figure~\ref{fl}. Here HB type ${\mathcal{L \equiv (B-R)/(B+V+R)}}$, where $\mathcal{B, \, R, \, V}$ are the numbers of blue, red, and variable (RR Lyrae-type) HB stars, respectively; this quantity was derived for NGC~5286 in Paper~I. As discussed by \citet{cat2005}, Oosterhoff-intermediate GCs, such as found in several GCs associated with the dwarf satellite galaxies of the Milky Way, tend to cluster inside the triangular-shaped region marked in this diagram~-- whereas Galactic GCs somehow are not found in this same region, thus giving rise to the Oosterhoff dichotomy in the Galaxy. In this same plane, OoI clusters tend to fall to the left (i.e., redder HB types) of the triangular-shaped region, whereas OoII objects are more commonly found to its right. NGC~5286 falls rather close to the Oosterhoff-intermediate region in this plane, but its position is indeed still consistent with OoII status (see also Fig.~8 in \citeauthor{cat2005} \citeyear{cat2005}).   

Figures~\ref{baiv} and \ref{baib} show the positions of the RR Lyrae stars on the Bailey (period-amplitude) diagram for $V$ and $B$ magnitudes, respectively. Circles indicate the \RRab's, whereas crosses are used for the \RRc's. As in Figure~\ref{hb}, we use different symbol sizes for the variables in different radial annuli from the cluster center. Also shown in this figure are typical lines for OoI and OoII clusters, which read as follows (see \citealp{cac05}):

\begin{equation}
A_B^{ab} = -3.123 - 26.331 \,\log{P} - 35.853 \,\log{P^2},
\end{equation}

\begin{equation}
A_V^{ab} = -2.627 - 22.046 \,\log{P} - 30.876 \,\log{P^2}, 
\end{equation}

\noindent for ab-type RR Lyrae stars in OoI clusters. For \RRab's in OoII clusters, in turn, the same lines can be used, but shifted in periods by $\Delta \log{P} = +0.06$.\footnote{\citet{cac05} actually derived their OoII curves based on what appeared to be highly evolved stars in M3, and then verified that the same curves do provide a good description of RR Lyrae stars in several bona-fide OoII GCs.} In the case of c-type stars, we derive reference lines on the basis of Figures~2 and 4 of \citet{cac05}; these read as follows: 

\begin{equation}
A_B^{c} = -0.522 - 2.290 \,\log{P},
\end{equation}

\begin{equation}
A_V^{c} = -0.395 - 1.764 \,\log{P},
\end{equation}

\noindent for c-type RR Lyrae stars in OoI clusters, and 

\begin{equation}
A_B^{c} = -0.039 - 1.619 \,\log{P},
\end{equation}

\begin{equation}
A_V^{c} = -0.244 - 1.834 \,\log{P},
\end{equation}

\noindent for c-type RR Lyrae stars in OoII clusters (again based on presumably ``evolved'' RR Lyrae stars in M3). 

This kind of diagram can be used as a diagnostic tool to investigate the Oosterhoff classification of RR{\em ab} stars. However, the position of a star in this diagram can be strongly affected by the presence of the Blazhko effect. In order to minimize this problem, we make a distinction between variables with a Jurcsik-Kov\'acs compatibility parameter value $D_m < 3$ ({\em filled circles}) and $D_m > 3$ ({\em open circles}). Even considering only the RR Lyrae stars with small $D_m$ and hence presumably ``regular'' light curves (according to the Jurcsik-Kov\'acs criterion), we see that there is still a wide scatter among the \RRab's in the Bailey diagrams, with no clear-cut tendency for stars to cluster tightly around either Oosterhoff reference line, particularly in the $B$ case. It is possible that at least some of the dispersion is caused by unidentified blends in the heavily crowded inner regions of the cluster, where most of the variables studied in this paper can be found (see Fig.~\ref{ubica}). In fact, the variable stars in the innermost cluster regions (small circles) show more scatter. If we only look at the variables that lie outside the core radius (the large and medium-sized circles, respectively, with $r > 0.29'$), we can see that they cluster much more tightly around the Oosterhoff II line. It should be noted, in any case, that very recently \citet{mcea08} have shown that the ab-type RR Lyrae stars in the prototypical OoII cluster M15 (NGC~7078) similarly do not cluster around the OoII reference line derived on the basis of more metal-rich clusters, thus casting some doubt on the validity of these lines as indicators of Oosterhoff status, at least at the more metal-poor end of the RR Lyrae metallicity distribution. 

As far as the positions of the c-type RR Lyrae stars in the Bailey diagrams are concerned, we find that there is also a wide scatter, without any clearly defined tendency for the data to clump tightly around either of the Oosterhoff reference lines~-- although the distribution does seem skewed towards shorter amplitudes at a given period, compared to the typical situation in OoII clusters.
 
Finally, we can also check how the average Fourier-based physical parameters derived for the NGC~5286 variables rank the cluster in terms of Oosterhoff status. This exercise is enabled by a comparison with the data for several clusters of different Oosterhoff types, as compiled in Tables~6 and 7 of \citet{cor2003}. For the \RRc's, we find that the mean masses, luminosities and temperatures are in fact more similar to those found for M3 (a prototypical OoI cluster) than they are for OoII globulars. Part of the problem may be due to the fact that NGC~5286 is significantly more metal-rich than all OoII globulars used in the analysis; recall that $\phi_{31}$ is the {\em only} Fourier parameter used in the \citet{sicl93} calibration of masses, luminosities, and temperatures, and that the impact of metallicity on the \citeauthor{sicl93} relations has still not been comprehensively addressed (see \S5 in \citeauthor*{ccea92} \citeyear{ccea92}, and also \S4 in \citeauthor{cat2005} \citeyear{cat2005} for general caveats regarding the validity of those relations). As a matter of fact, the recent study by \citet*{smea07} clearly shows that, in the case of \RRc\ variables, $\phi_{31}$ depends strongly on the metallicity. 

For the \RRab's, in turn, both the derived temperatures and absolute magnitudes are fully consistent with an OoII classification for the cluster.

\section{The Type II Cepheid}\label{sec:cep}
We find one type II Cepheid (V8) with  a period of $2.33$~days and a visual amplitude of $A_V=1.15$~mag, typical for a BL Herculis star. We use equation~(3) in \citet{pri99} to obtain $M_V = -0.55 \pm 0.07$~mag for V8. Using the intensity-weighted mean magnitude for V8 from Table~\ref{tab1}, we obtain for the cluster a distance modulus $(m-M)_V = 15.81 \pm 0.07$~mag, slightly shorter than the values discussed in \S\ref{sec:subab}. However, as we can see from Figure~9 in \citeauthor{pri99}, a large dispersion in $M_V$ is indeed present for short-period type II Cepheids, thus possibly explaining the small discrepancy.

\section{Summary}\label{sec:summ}
In this paper, we present the results of time-series photometry for NGC~5286, a GC which has been tentatively associated with the Canis Major dwarf spheroidal galaxy. 38 new variables were discovered in the cluster, and 19 previously known ones were recovered in our study (including one BL Her star that was previously catalogued as an RR Lyrae). The population of variable stars consists of 52 RR Lyrae (22 \RRc\ and 30 \RRab), 4 LPV's, and 1 type II Cepheid.
 
From Fourier decomposition of the \RRab\ light curves, we obtained a value for the metalicity of the cluster of ${\rm [Fe/H]} = -1.68  \pm 0.21$~dex in the \citet{zw1984} scale. We also derive a distance modulus of $(m-M)_V = 16.04$~mag for NGC~5286, based on the recent RR Lyrae $M_V - {\rm [Fe/H]}$ calibration of \citet{cc08}. 
 
Using a variety of indicators, we discuss in detail the Oosterhoff type of the cluster, concluding in favor of an OoII classification. The cluster's fairly high metallicity places it among the most metal-rich OoII clusters known, which may help account for what appears to be a fairly unusual behavior for a cluster of this type, including relatively short values of $P_{\rm ab,min}$ and $\langle P_c\rangle$, and unusual physical parameters, as derived for its c-type RR Lyrae stars on the basis of Fourier decomposition of their light curves. 

 In regard to the cluster's suggested association to the Canis Major dwarf spheroidal galaxy, we note that the metallicity and distance modulus derived in this work are very similar to the values previously accepted for the cluster \citep{har1996}, and thus the conclusions reached by previous authors \citep{crane2003,forbes2004} regarding its possible association with this dwarf galaxy are not significantly affected by our new metallicity and distance estimates. In addition, the position of the cluster in the HB morphology-metallicity plane is fairly similar to that found in several nearby extragalactic systems. As far as NGC~5286's RR Lyrae pulsation properties are concerned, the present study shows them to be somewhat unusual compared with bona-fide Galactic globular clusters, but still do not classify the cluster as an Oosterhoff-intermediate system, as frequently found among the Galaxy's dwarf satellites \citep[e.g.,][and references therein]{cat2005}. It is interesting to note, in any case, that the Canis Major field, unlike what is found among other dwarf galaxies, appears to be basically devoid of RR Lyrae stars \citep{tkea04,cmea09}, and also to be chiefly comprised of fairly high-metallicity (${\rm [M/H]} \geq -0.7$), young ($t \lesssim 10$~Gyr) stars \citep[e.g.,][]{mbea04}. The present paper, along with Paper~I, show instead that NGC~5286 is an RR Lyrae-rich, metal-poor globular cluster that is at least as old as the oldest globular clusters in the Galactic halo. It is not immediately clear that such an object as NGC~5286 would be easily formed within a galaxy with the properties observed for the Canis Major main body~-- and this should be taken into account when investigating the physical origin and formation mechanism for the Canis Major overdensity and its associated tidal ring. 

\acknowledgments
We warmly thank I. C. Le\~ao for his help producing the finding chart, and an anonymous referee for several comments that helped improve the presentation of our results. MZ and MC acknowledge financial support by Proyecto Fondecyt Regular \#1071002. Support for MC is also provided by Proyecto Basal PFB-06/2007, by FONDAP Centro de Astrof\'{i}sica 15010003, and by a John Simon Guggenheim Memorial Foundation Fellowship. HAS is supported by CSCE and NSF grant AST 0607249.

\vfill\eject

\appendix 
\section{LIGHT CURVES}

  \begin{figure}[!hb]
  \begin{center}

\includegraphics[width=.28\textwidth]{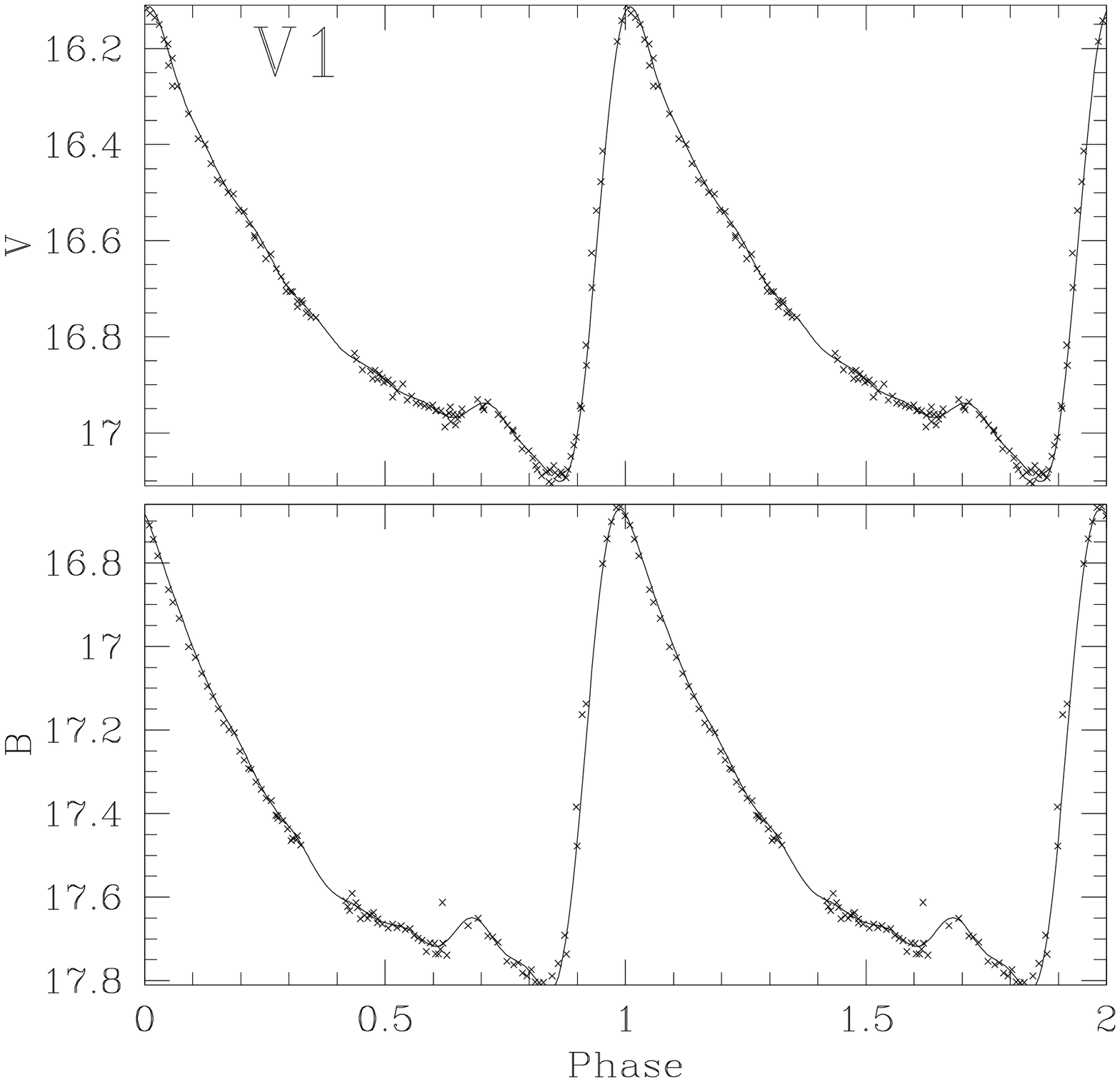}
\includegraphics[width=.28\textwidth]{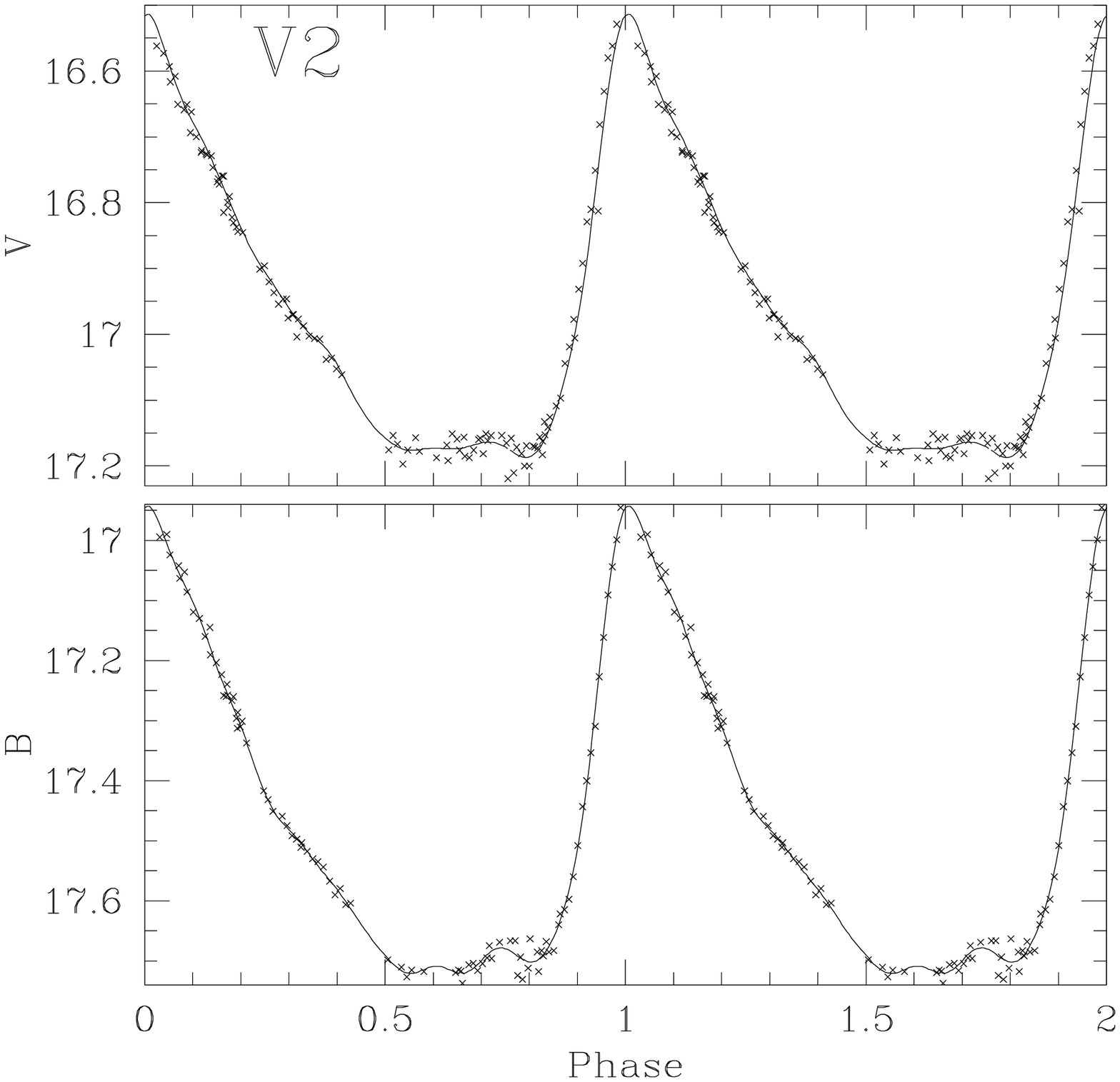}
\includegraphics[width=.28\textwidth]{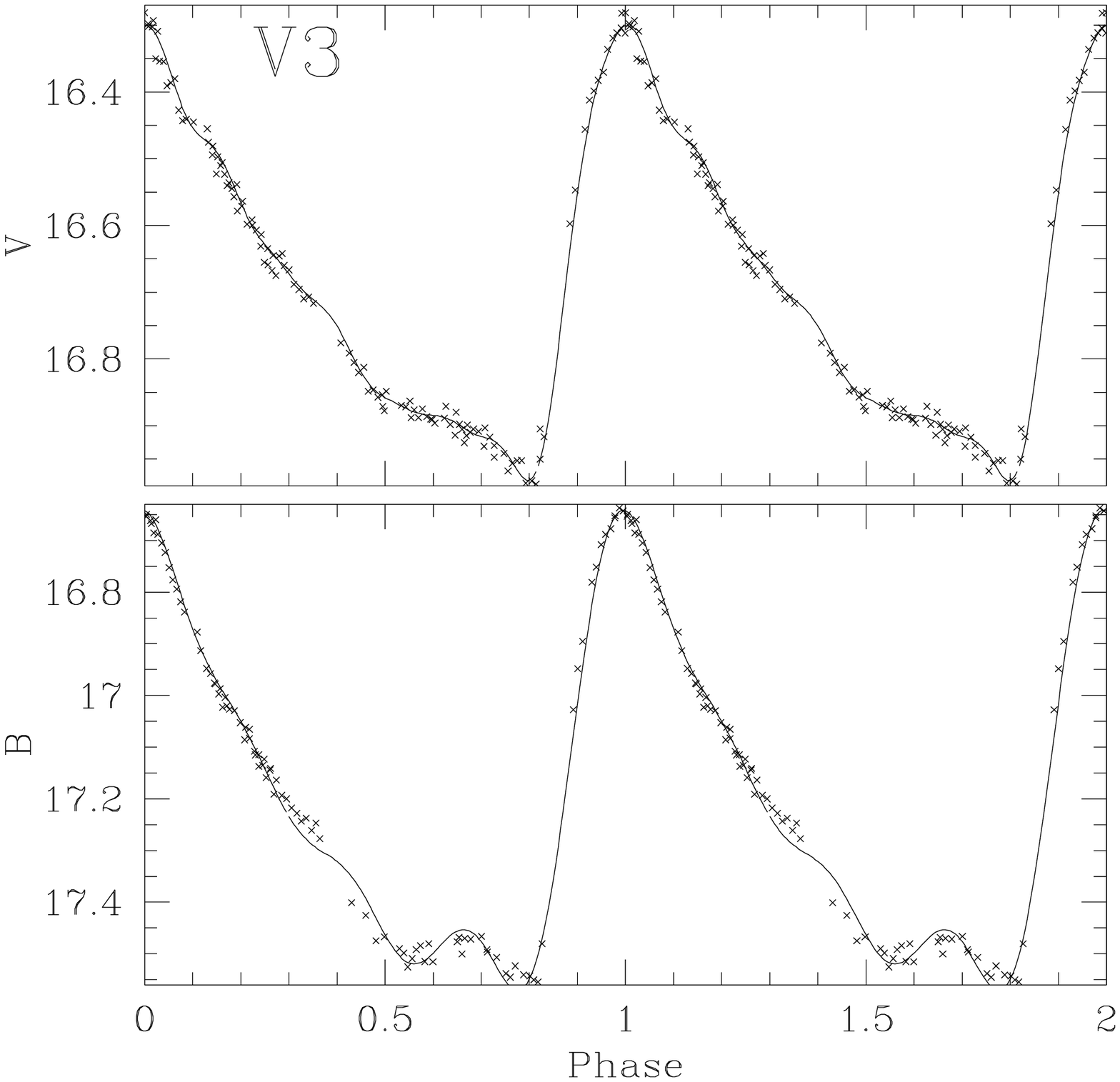}
\includegraphics[width=.28\textwidth]{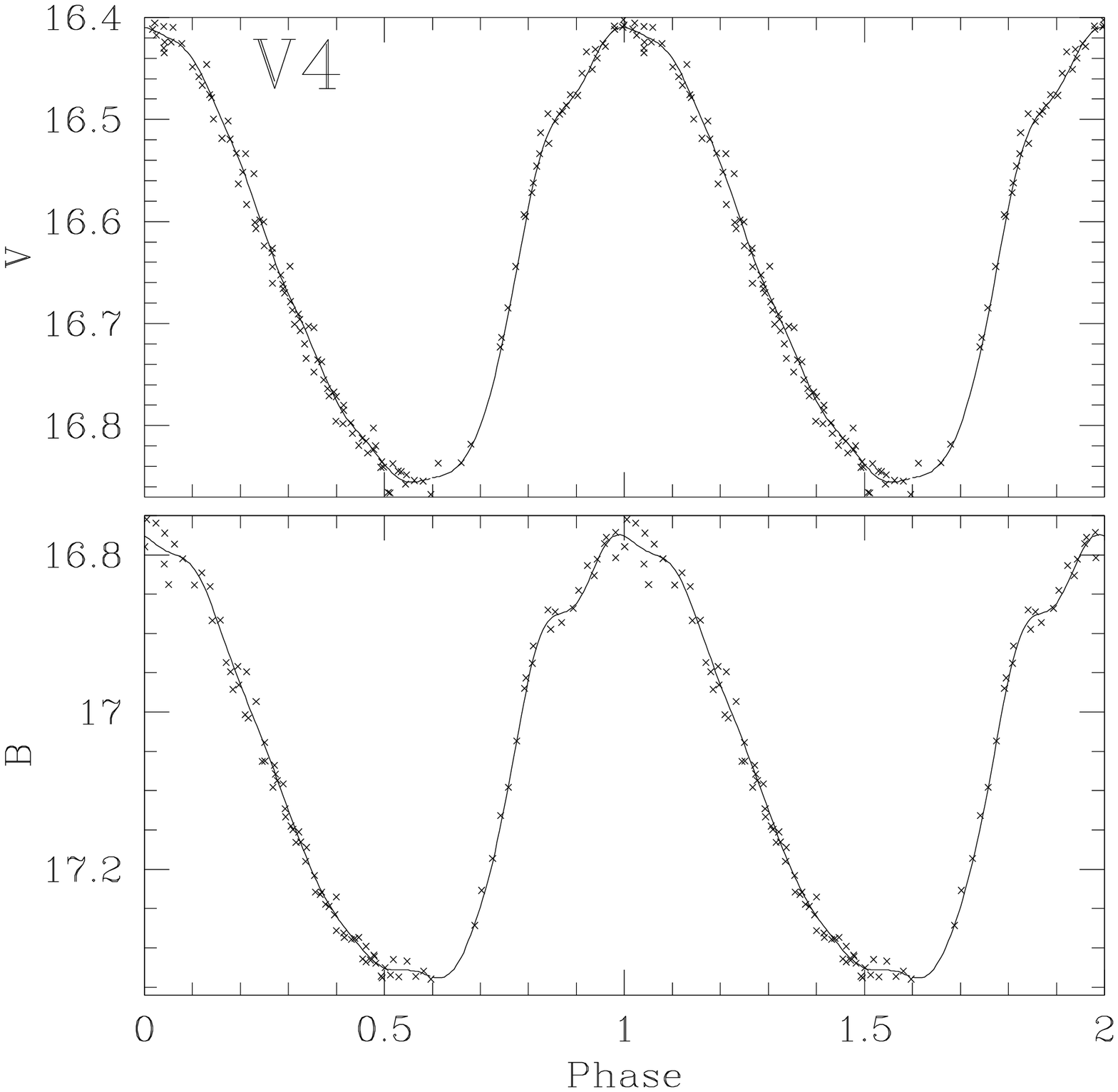}
\includegraphics[width=.28\textwidth]{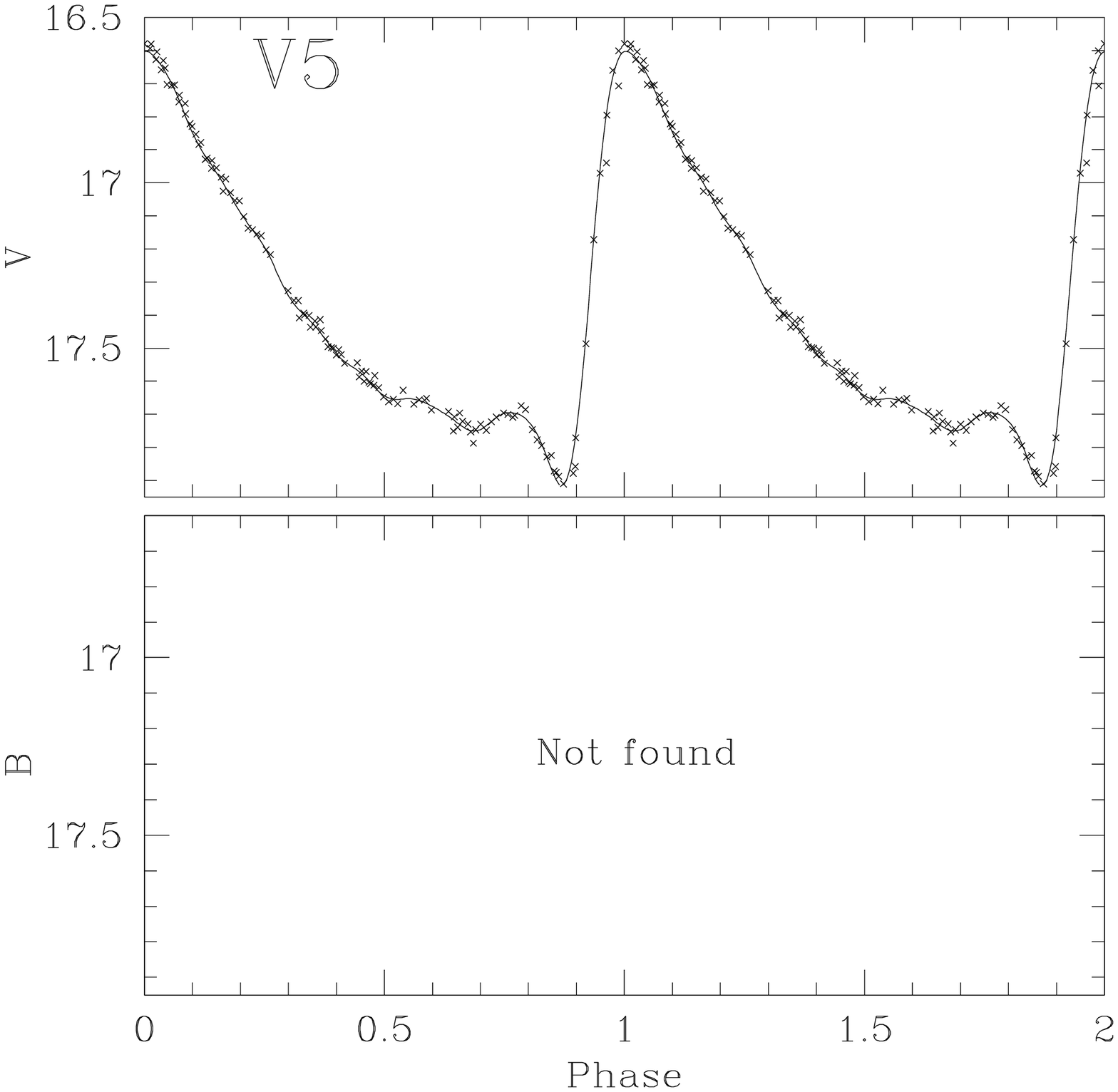}
\includegraphics[width=.28\textwidth]{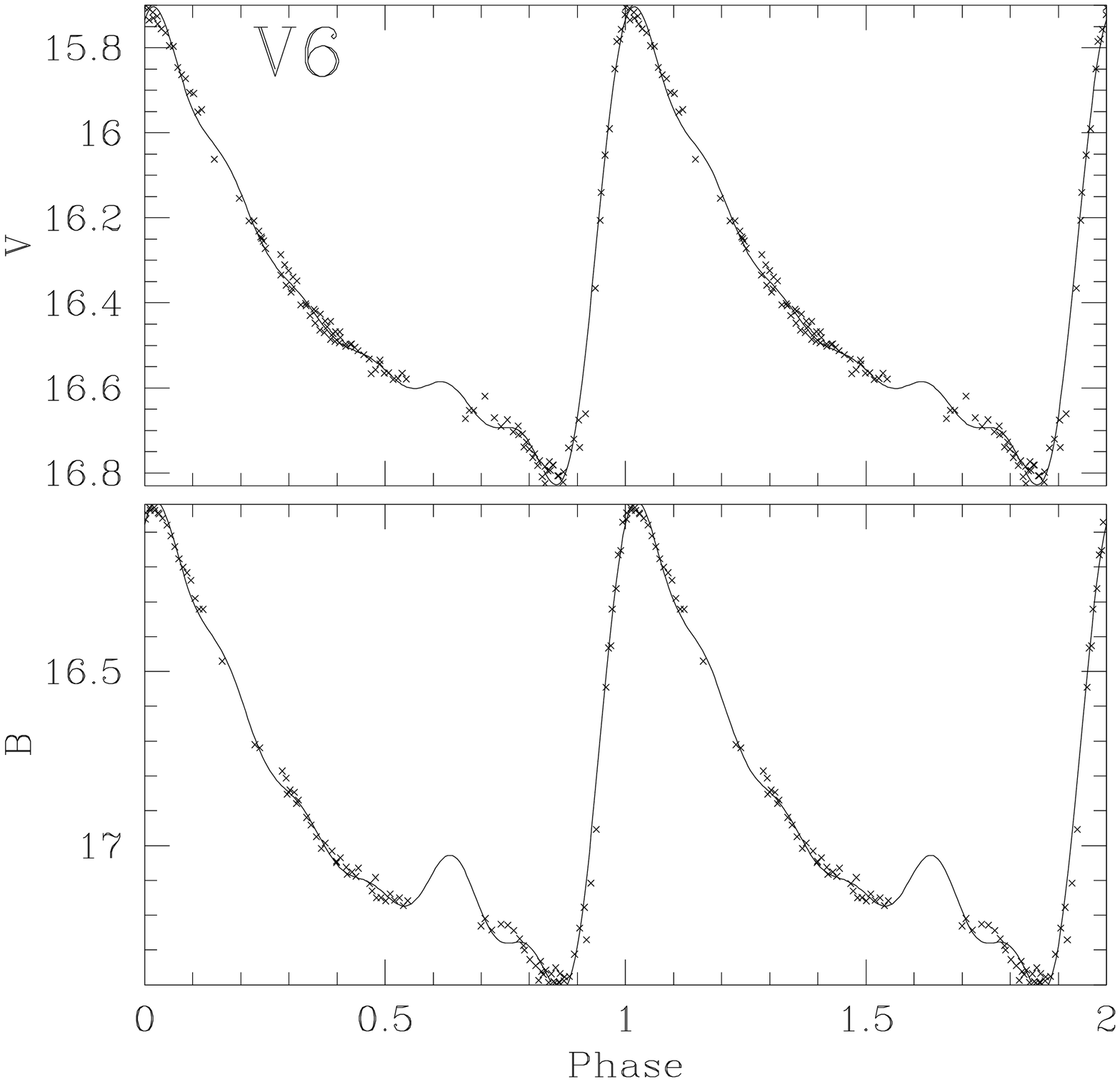}
\includegraphics[width=.28\textwidth]{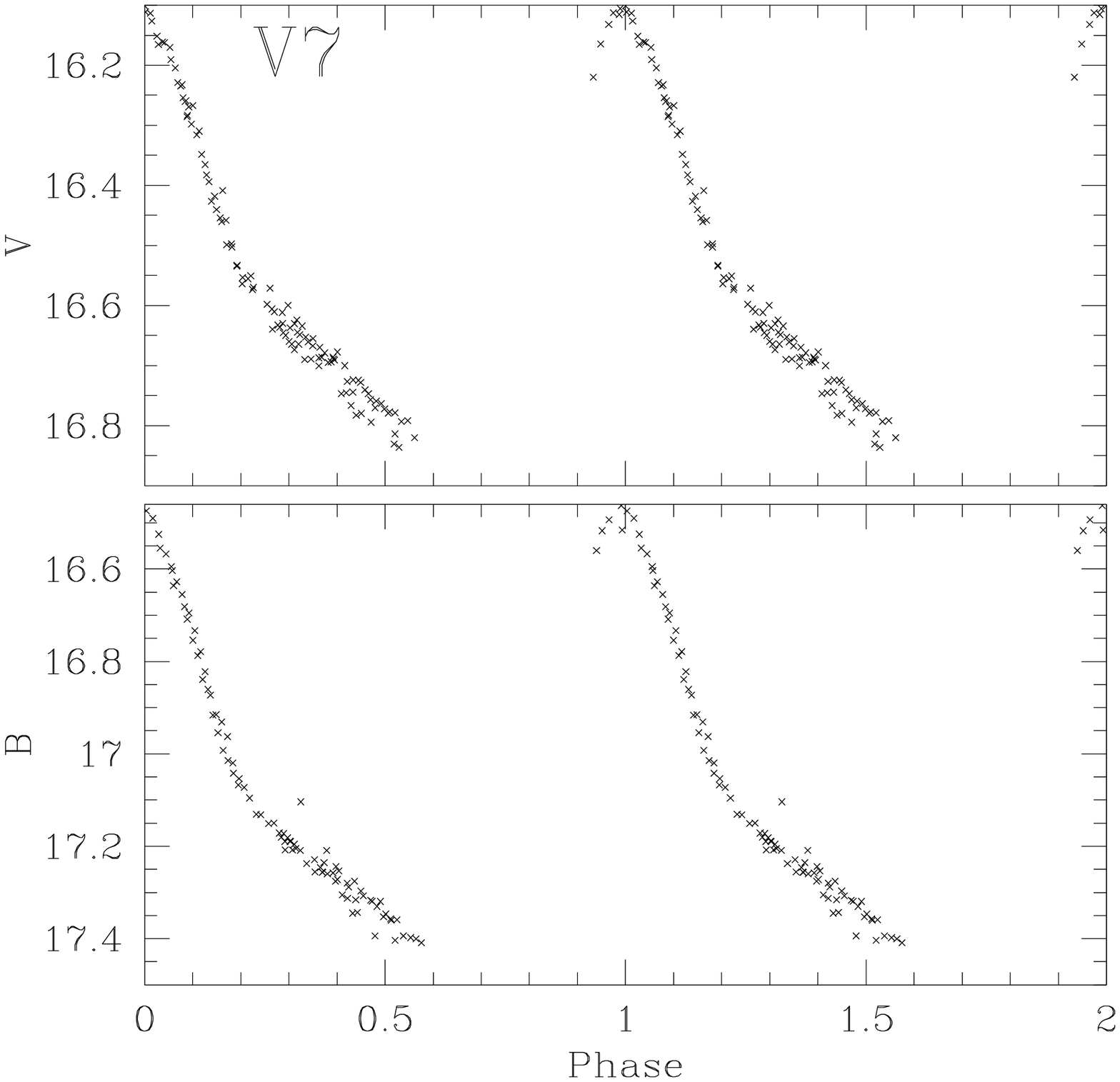}
\includegraphics[width=.28\textwidth]{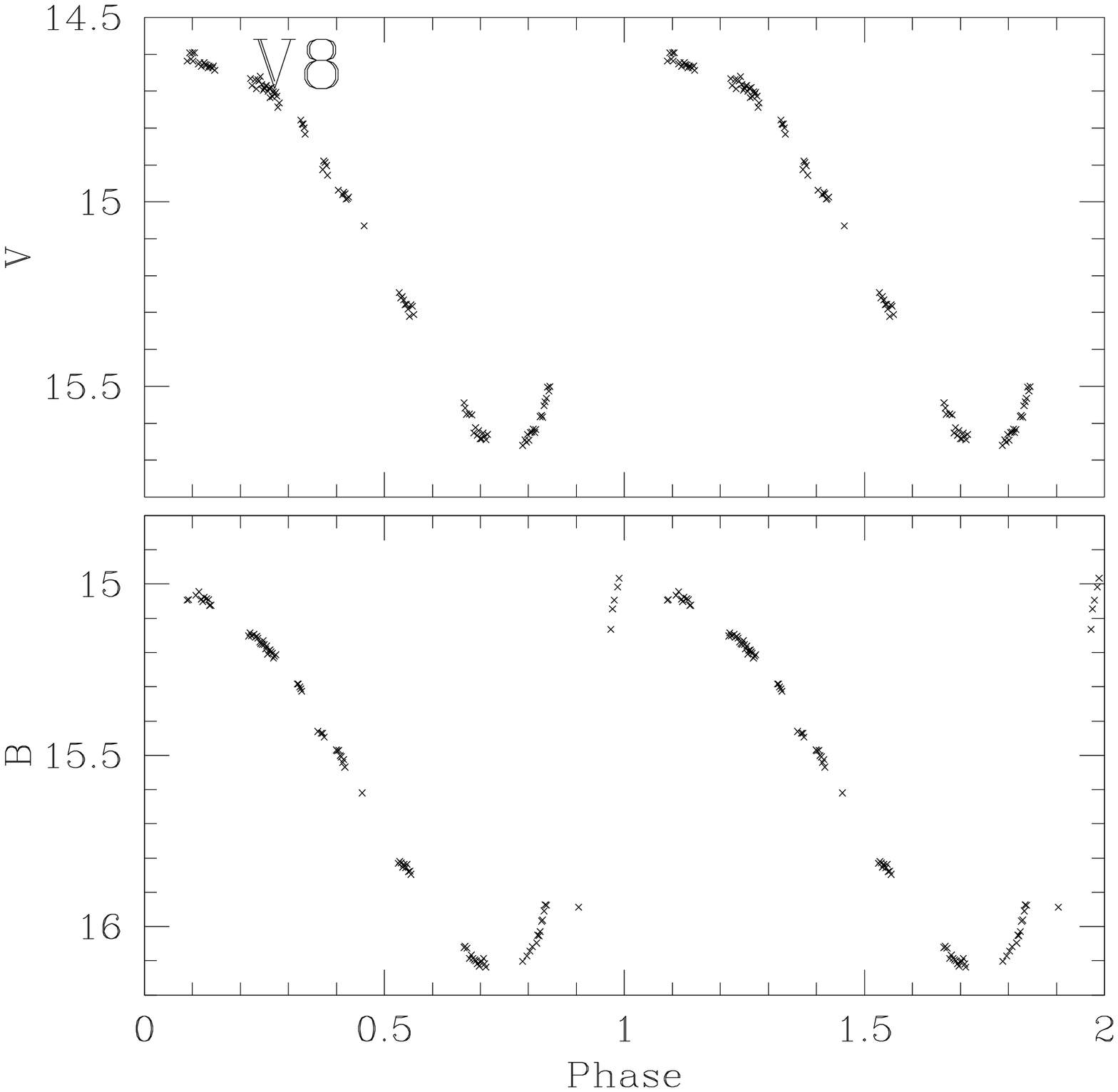}
\includegraphics[width=.28\textwidth]{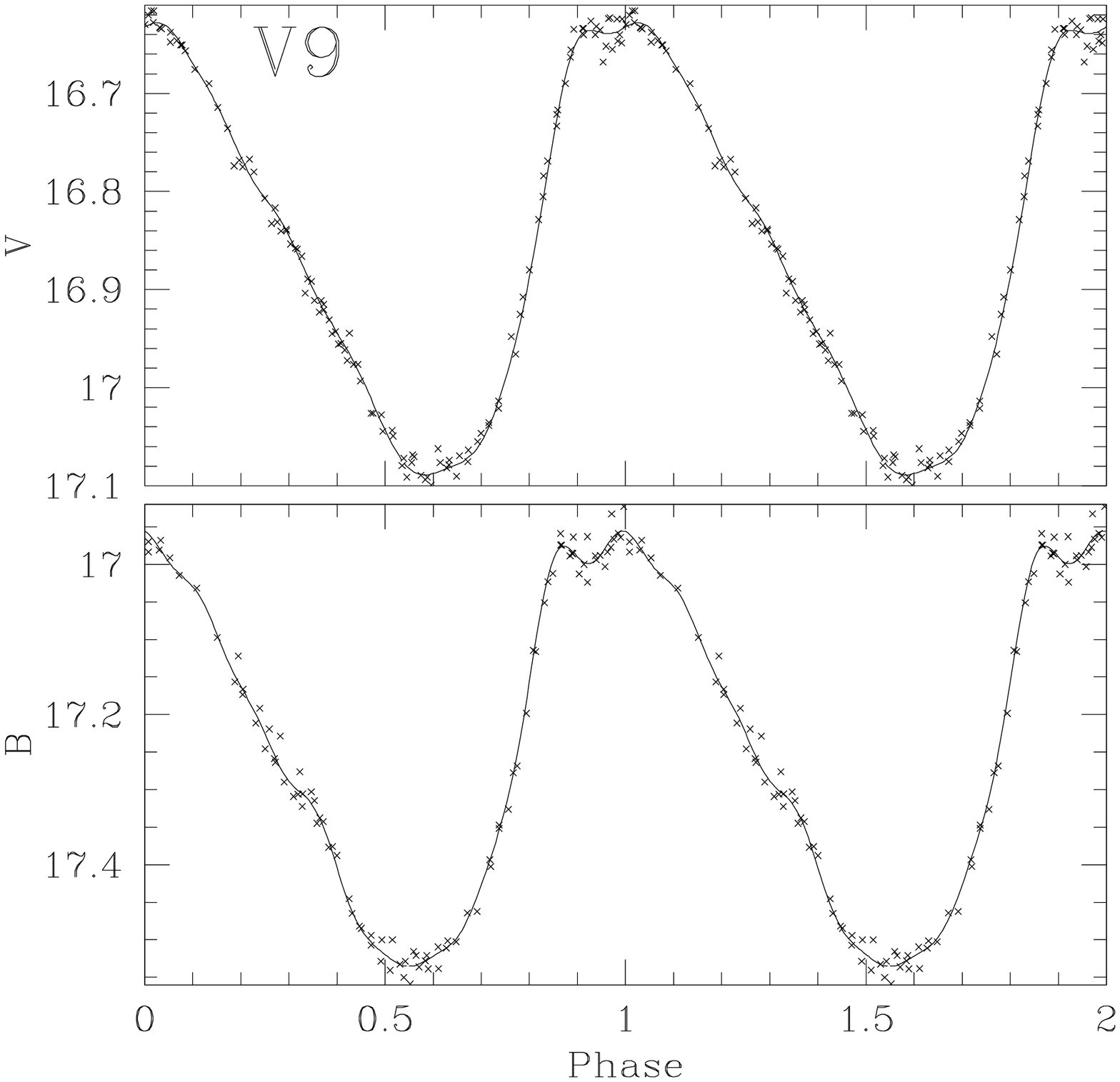}
\includegraphics[width=.28\textwidth]{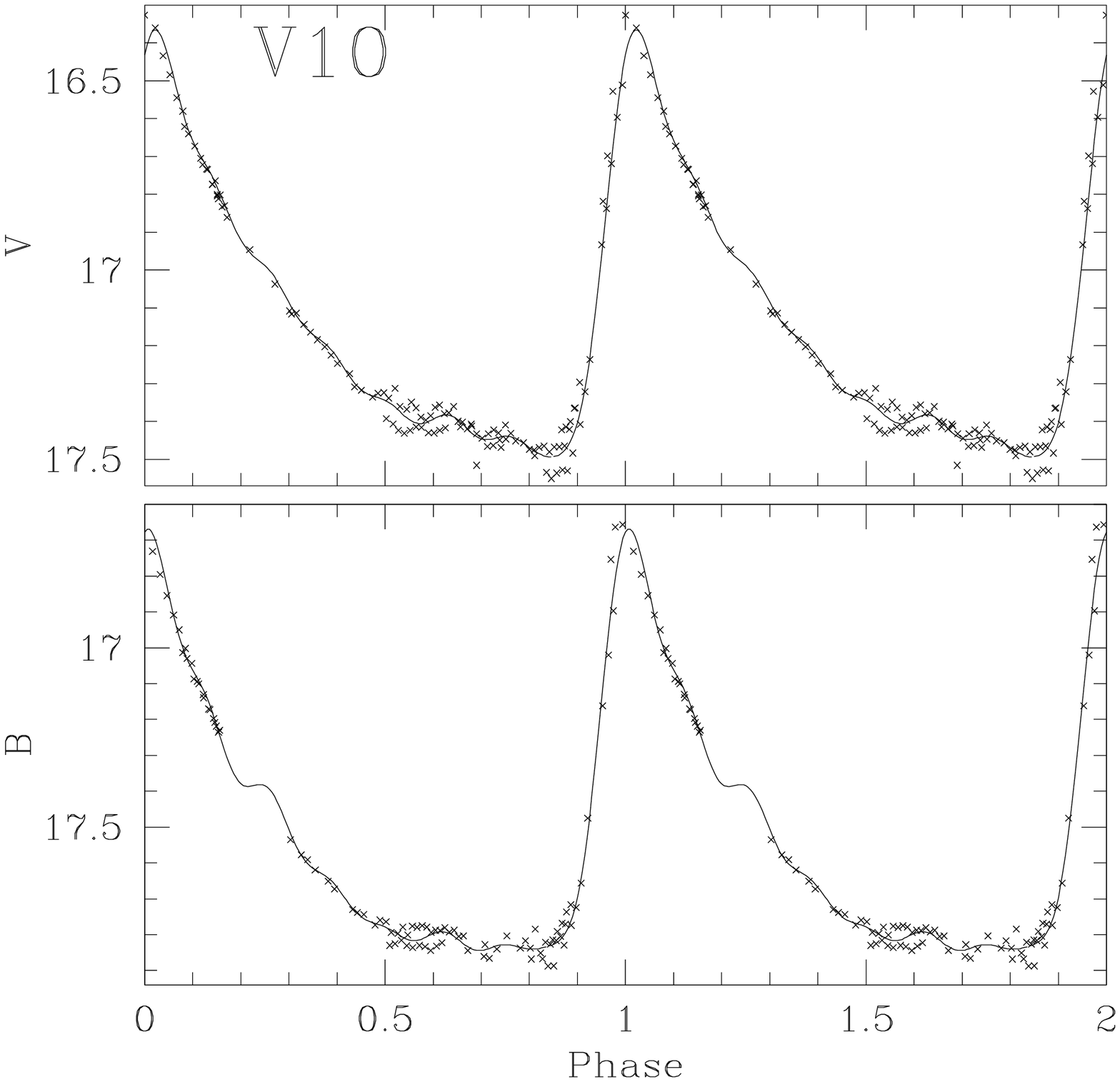}
\includegraphics[width=.28\textwidth]{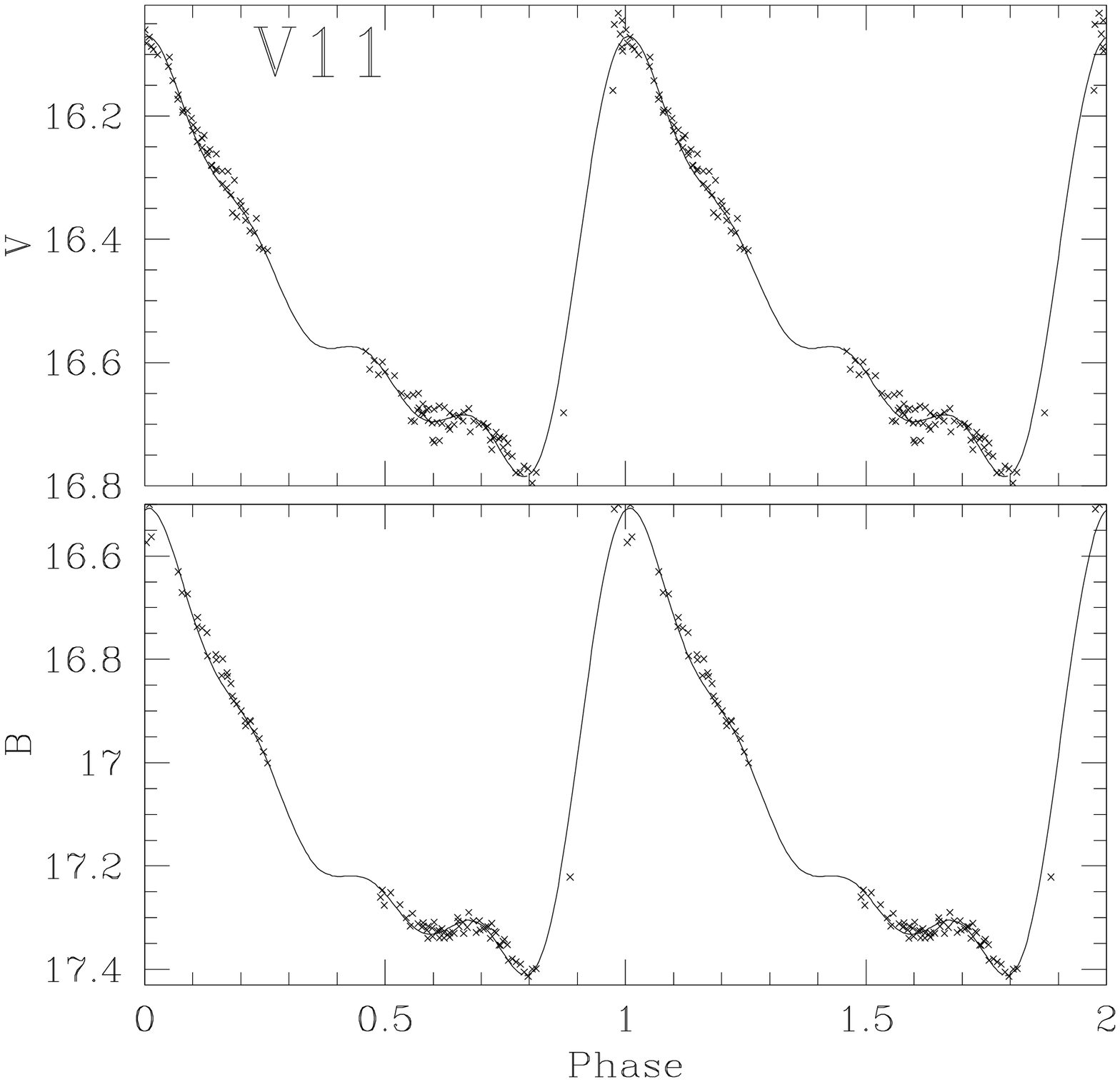}
\includegraphics[width=.28\textwidth]{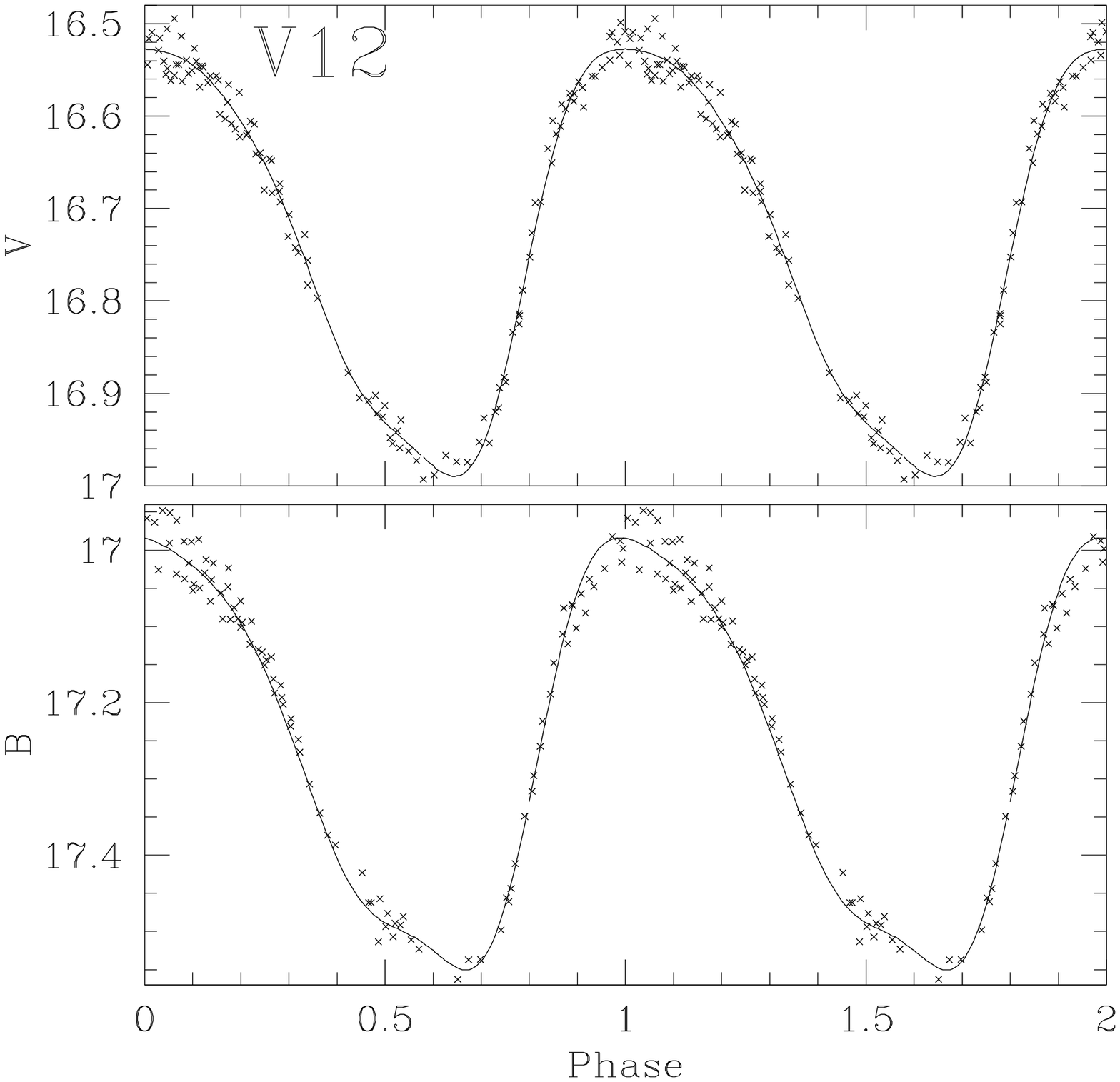}
  \end{center}
  \end{figure}

  \begin{figure}[t]
  \begin{center}
\includegraphics[width=.28\textwidth]{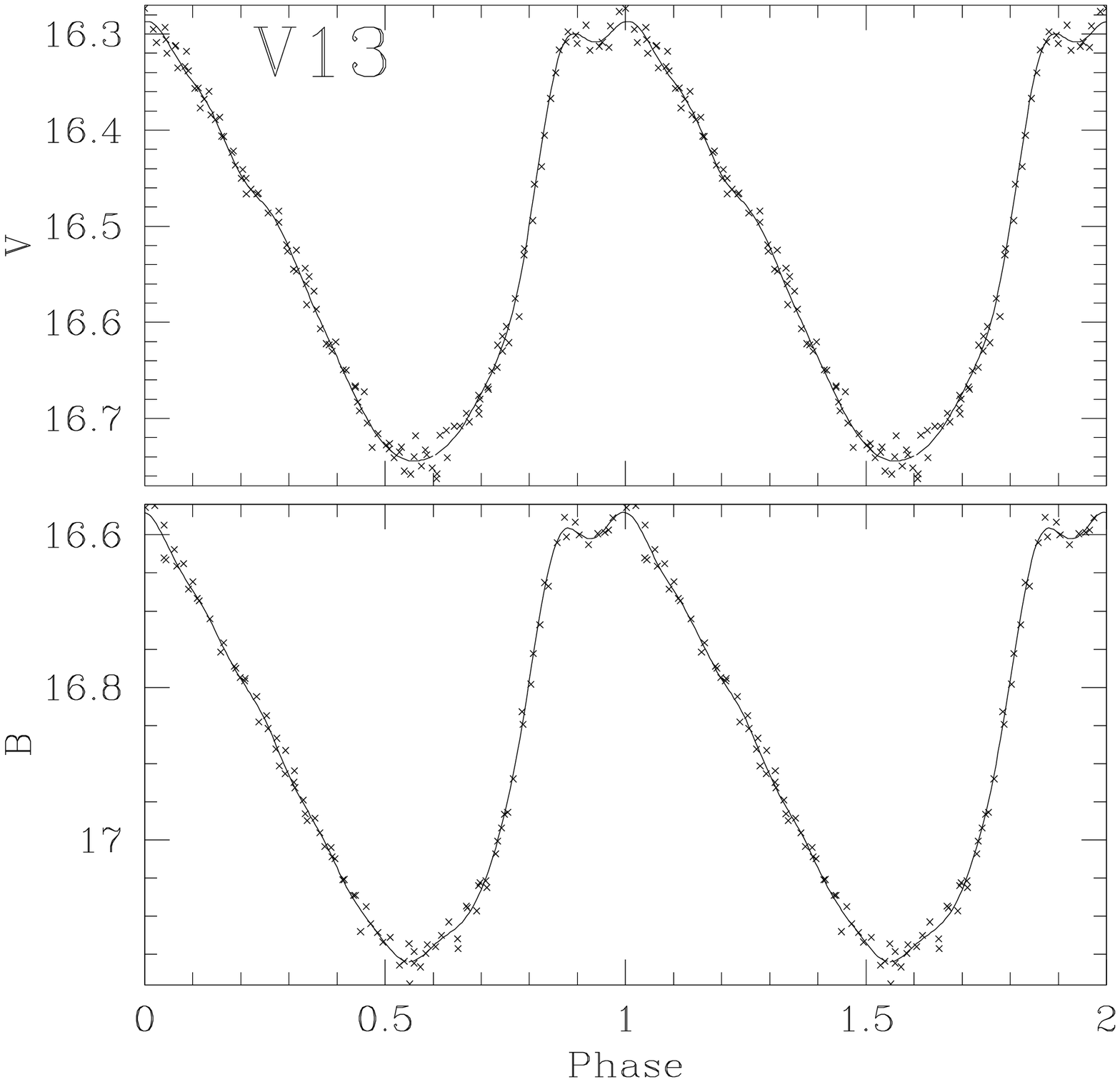}
\includegraphics[width=.28\textwidth]{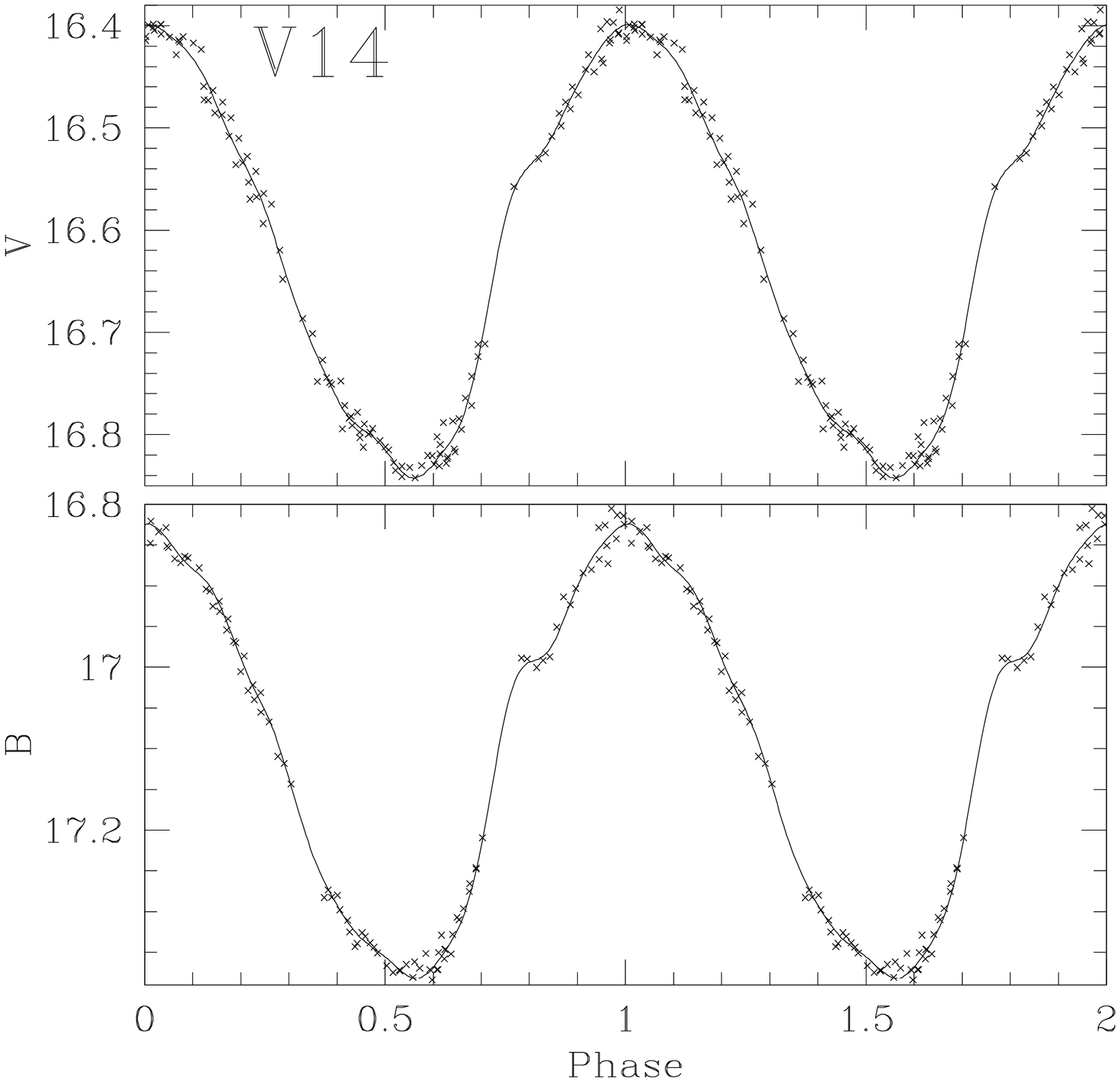}
\includegraphics[width=.28\textwidth]{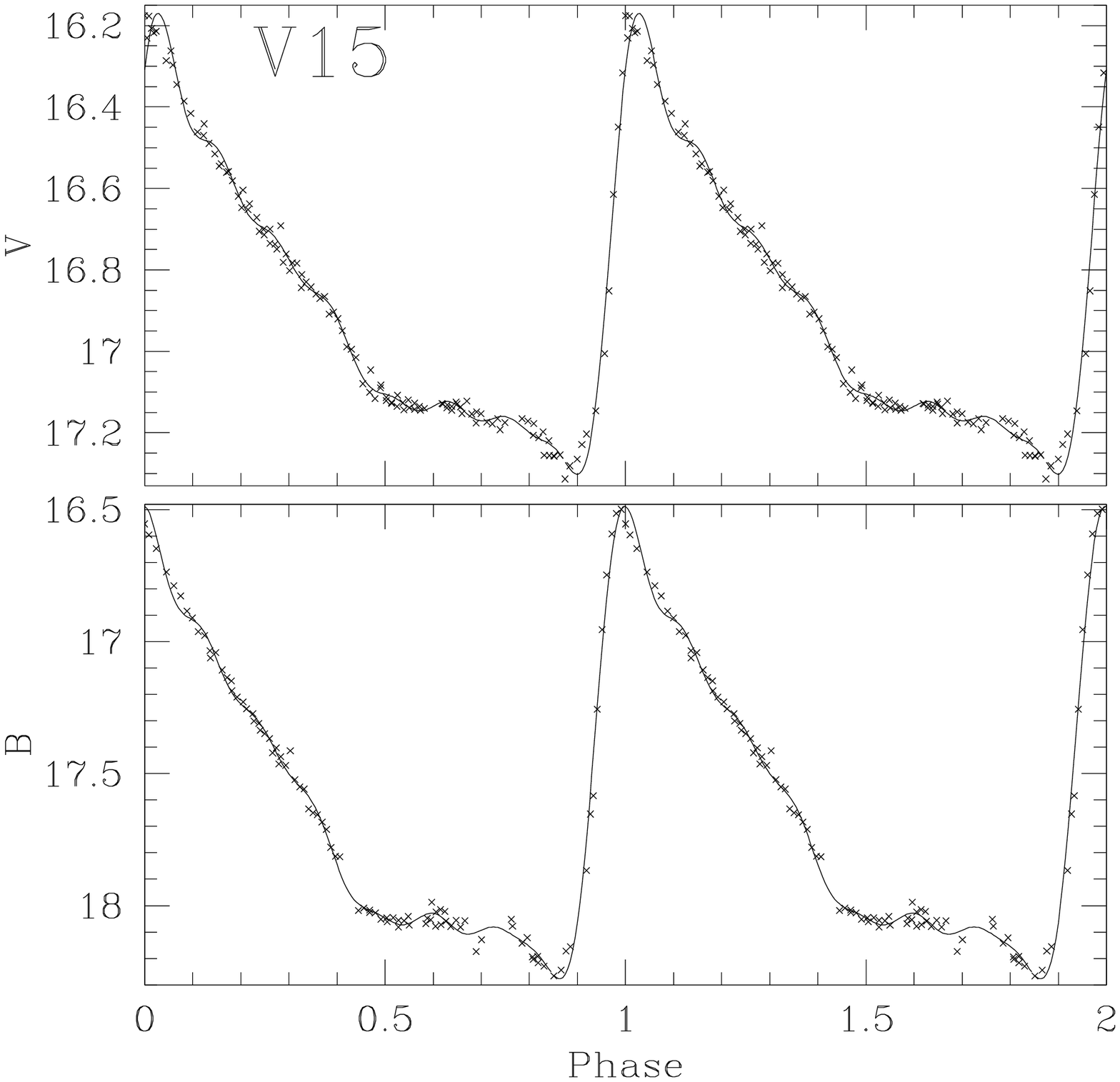}
\includegraphics[width=.28\textwidth]{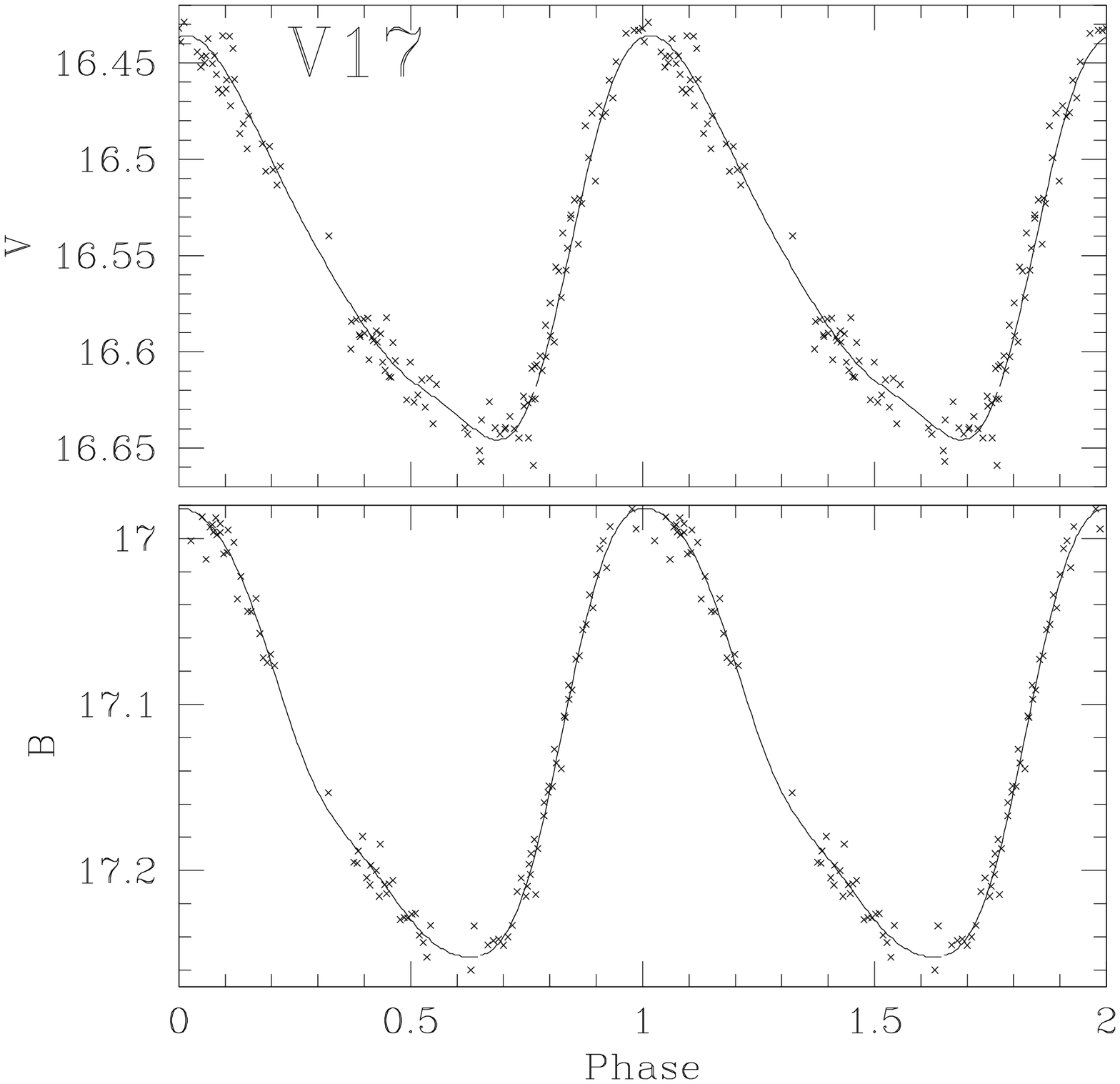}
\includegraphics[width=.28\textwidth]{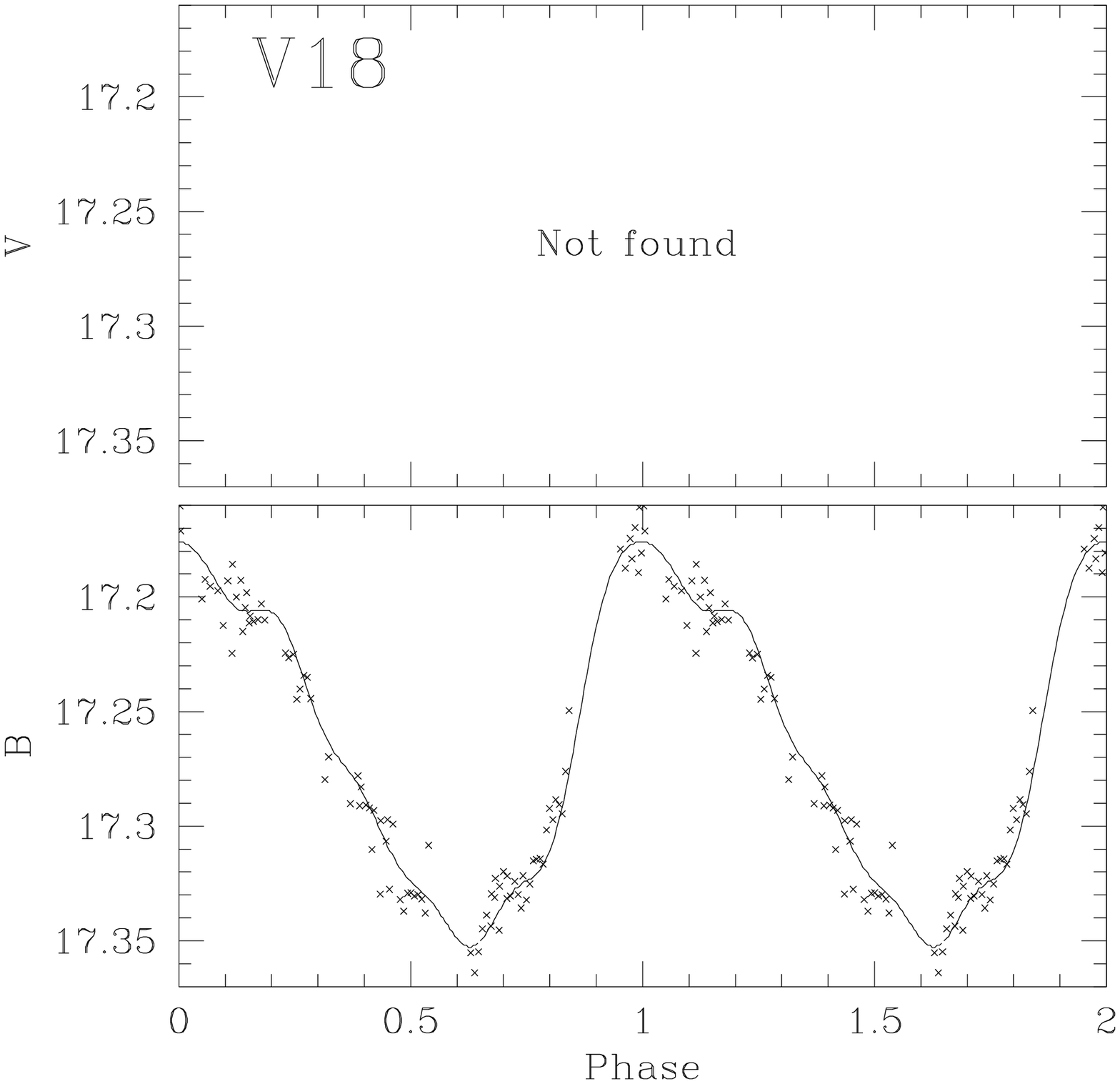}
\includegraphics[width=.28\textwidth]{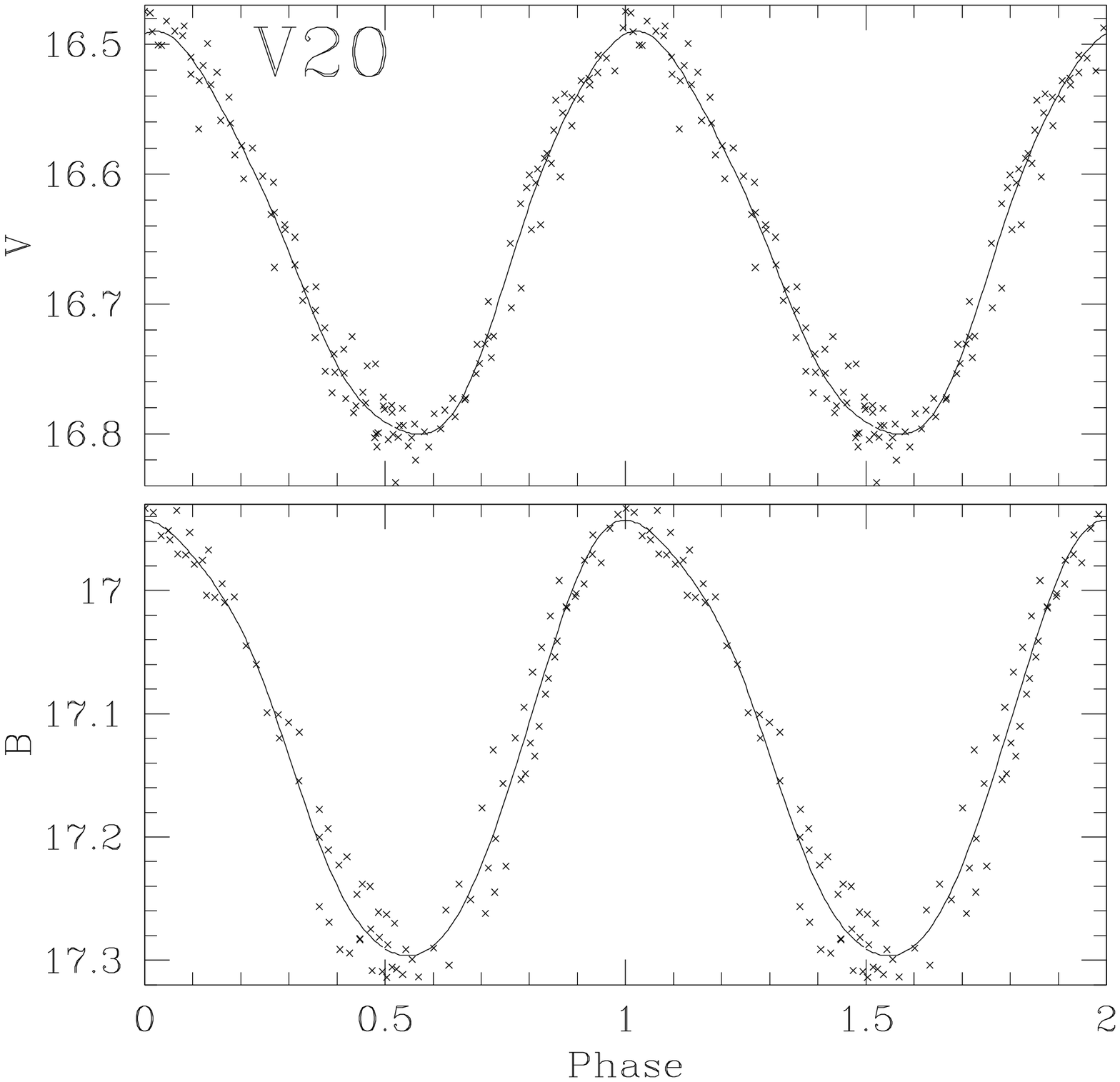}
\includegraphics[width=.28\textwidth]{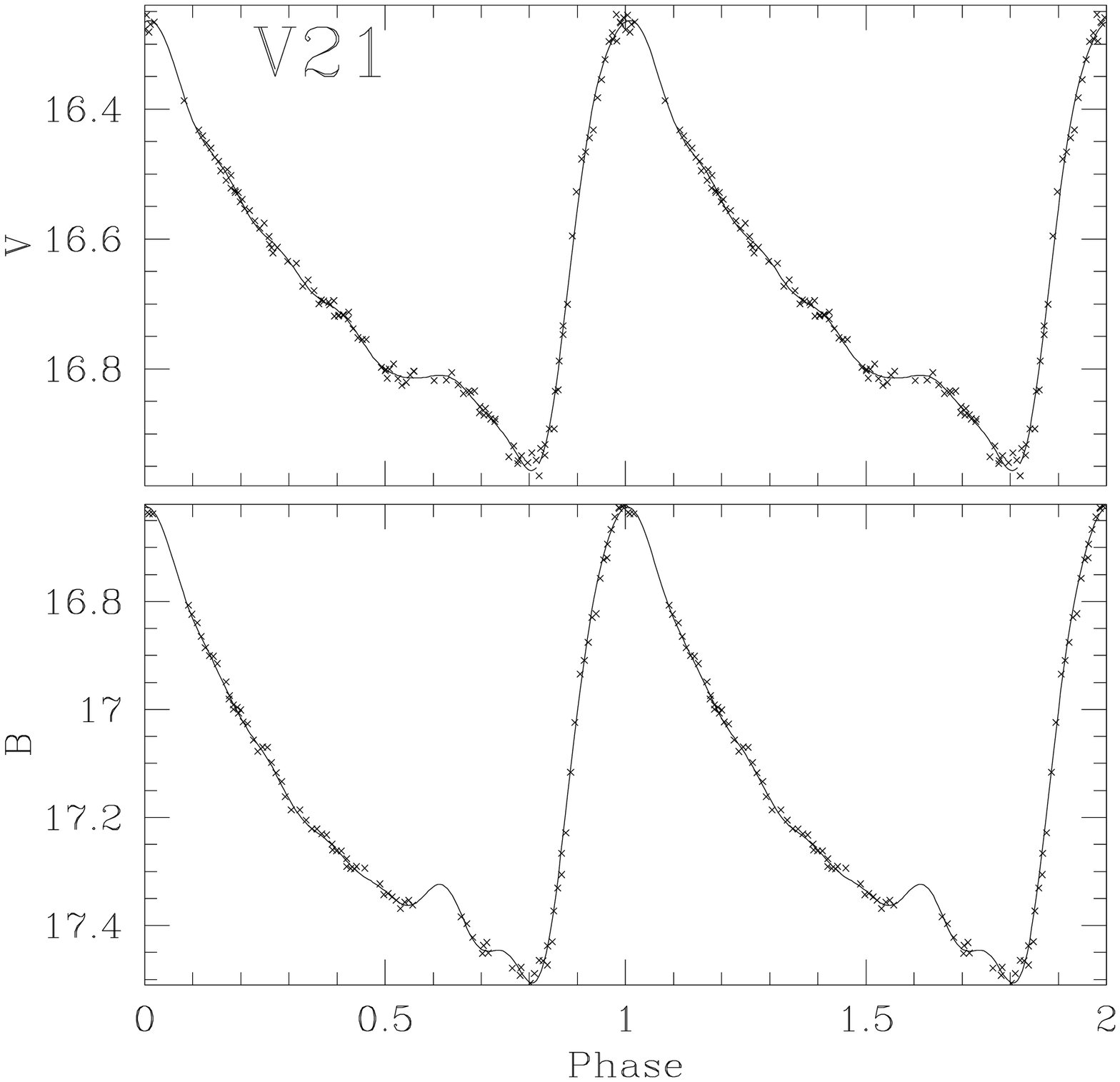}
\includegraphics[width=.28\textwidth]{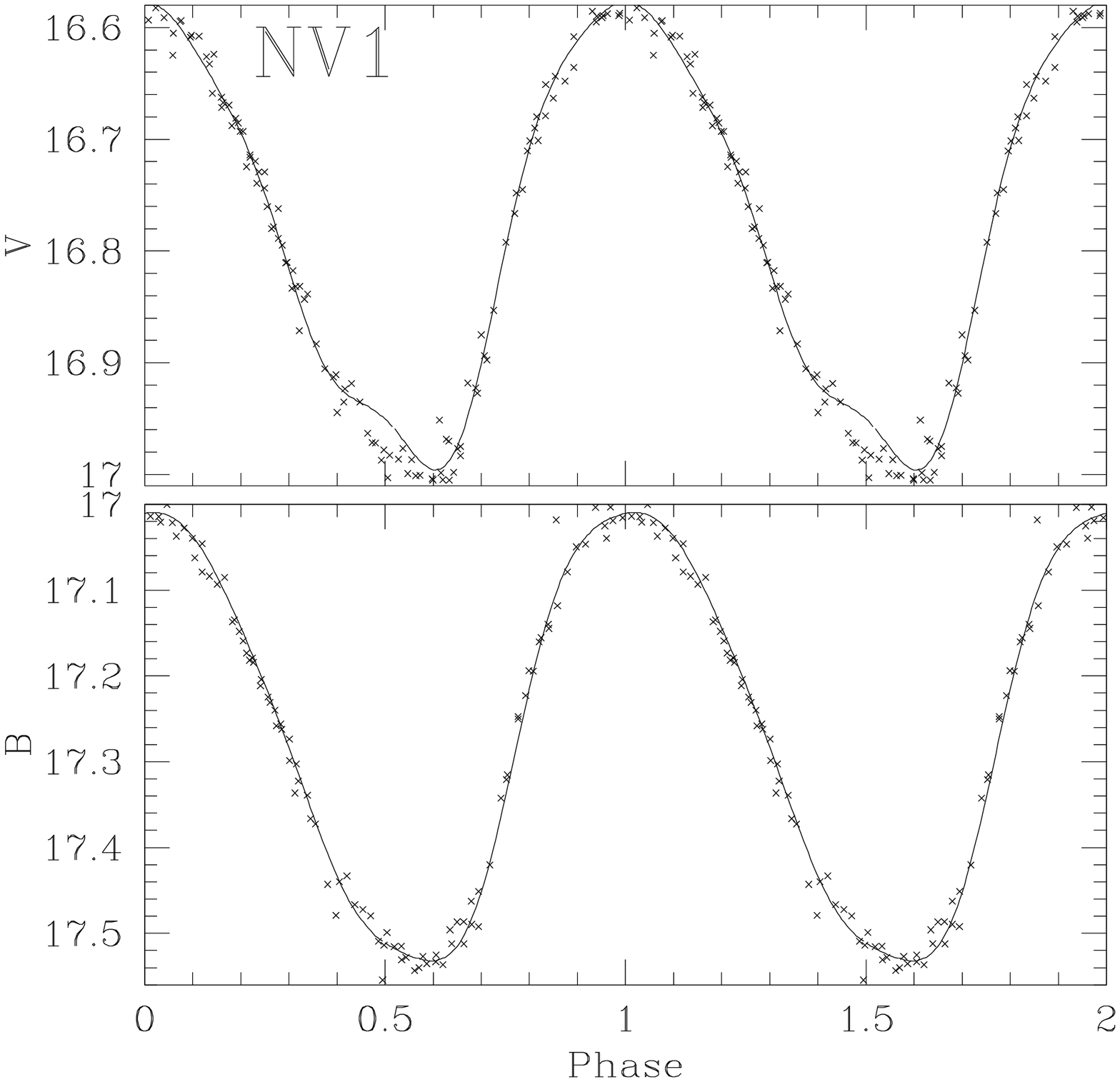}
\includegraphics[width=.28\textwidth]{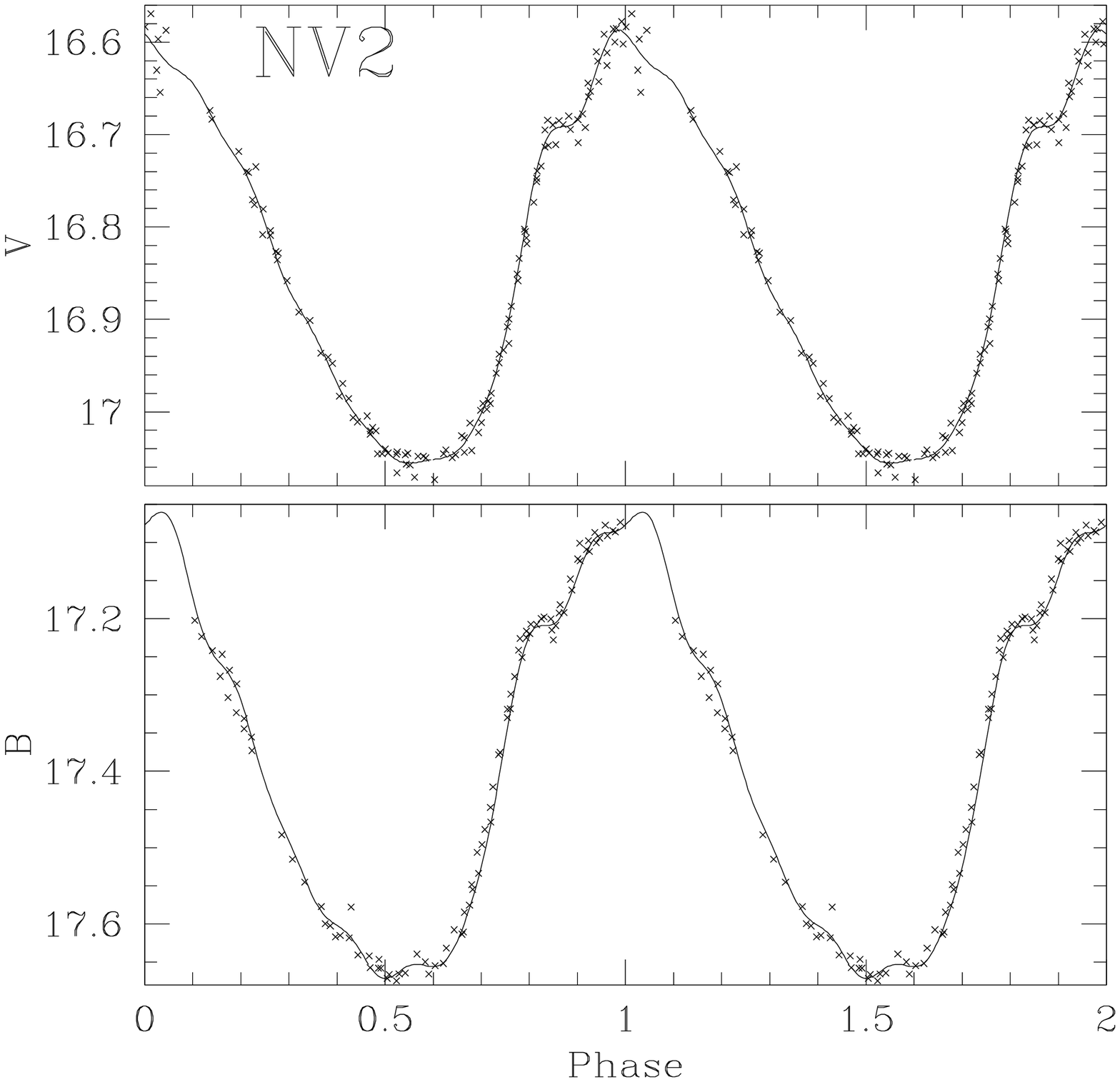}
\includegraphics[width=.28\textwidth]{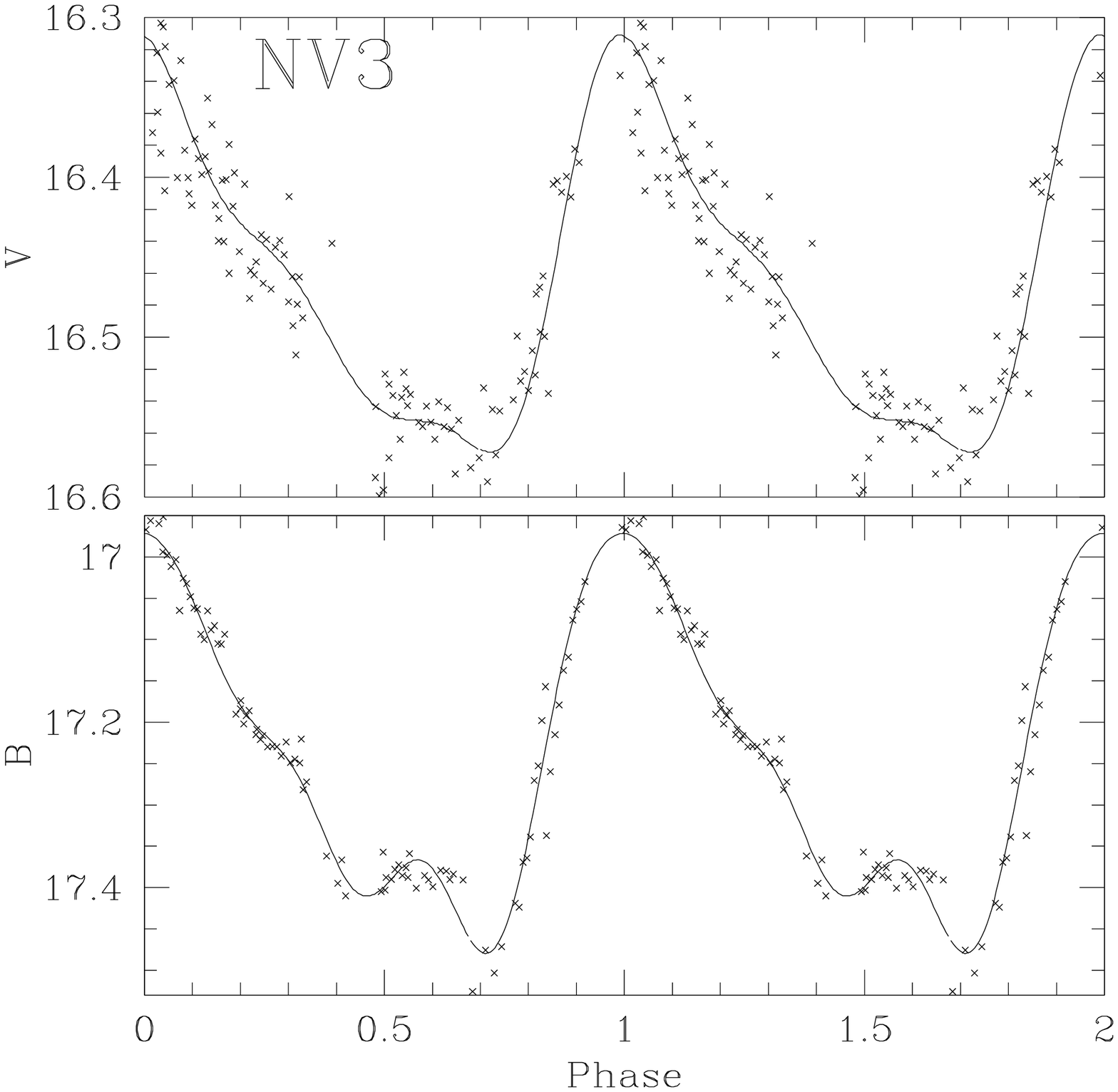}
\includegraphics[width=.28\textwidth]{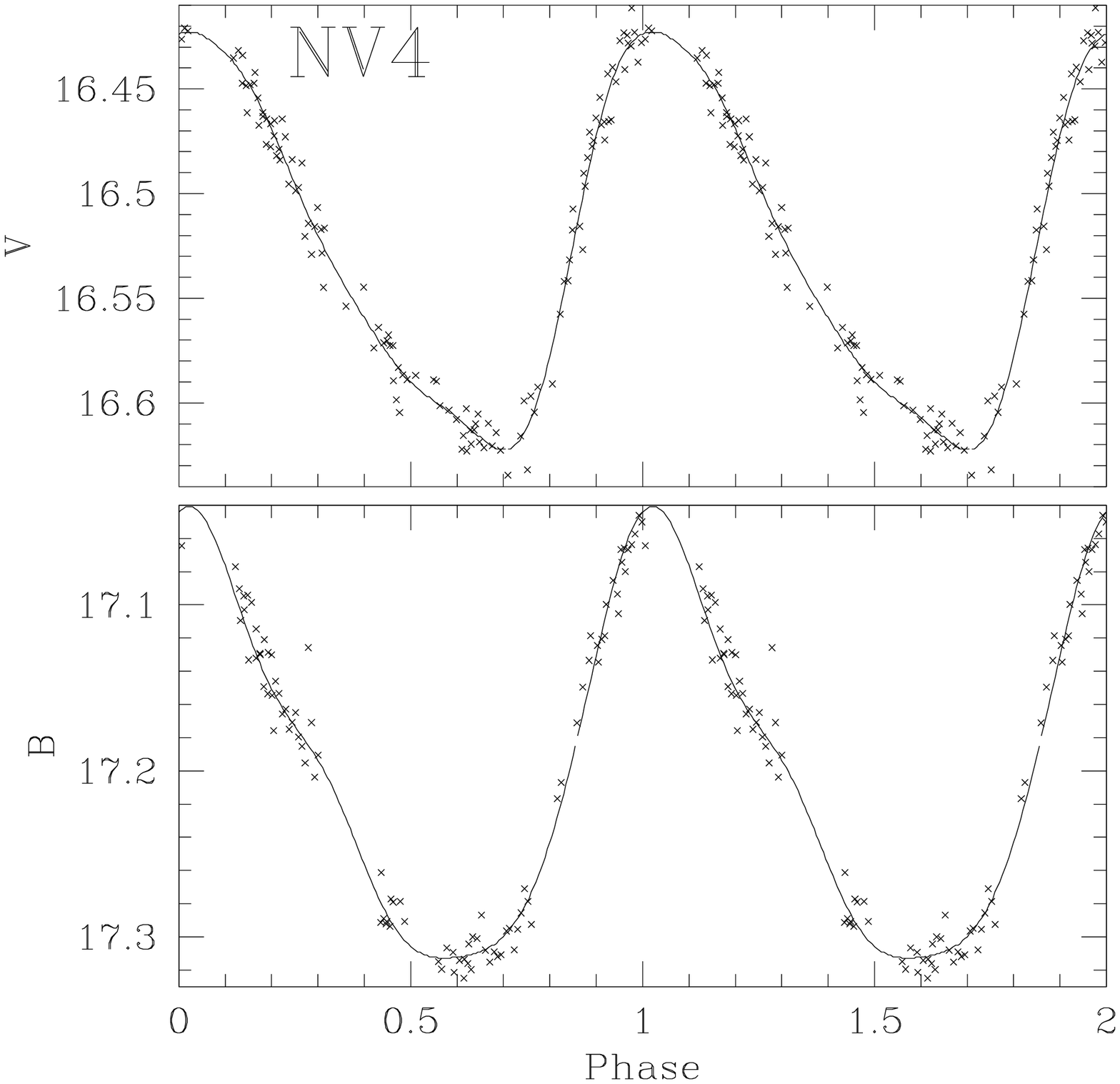}
\includegraphics[width=.28\textwidth]{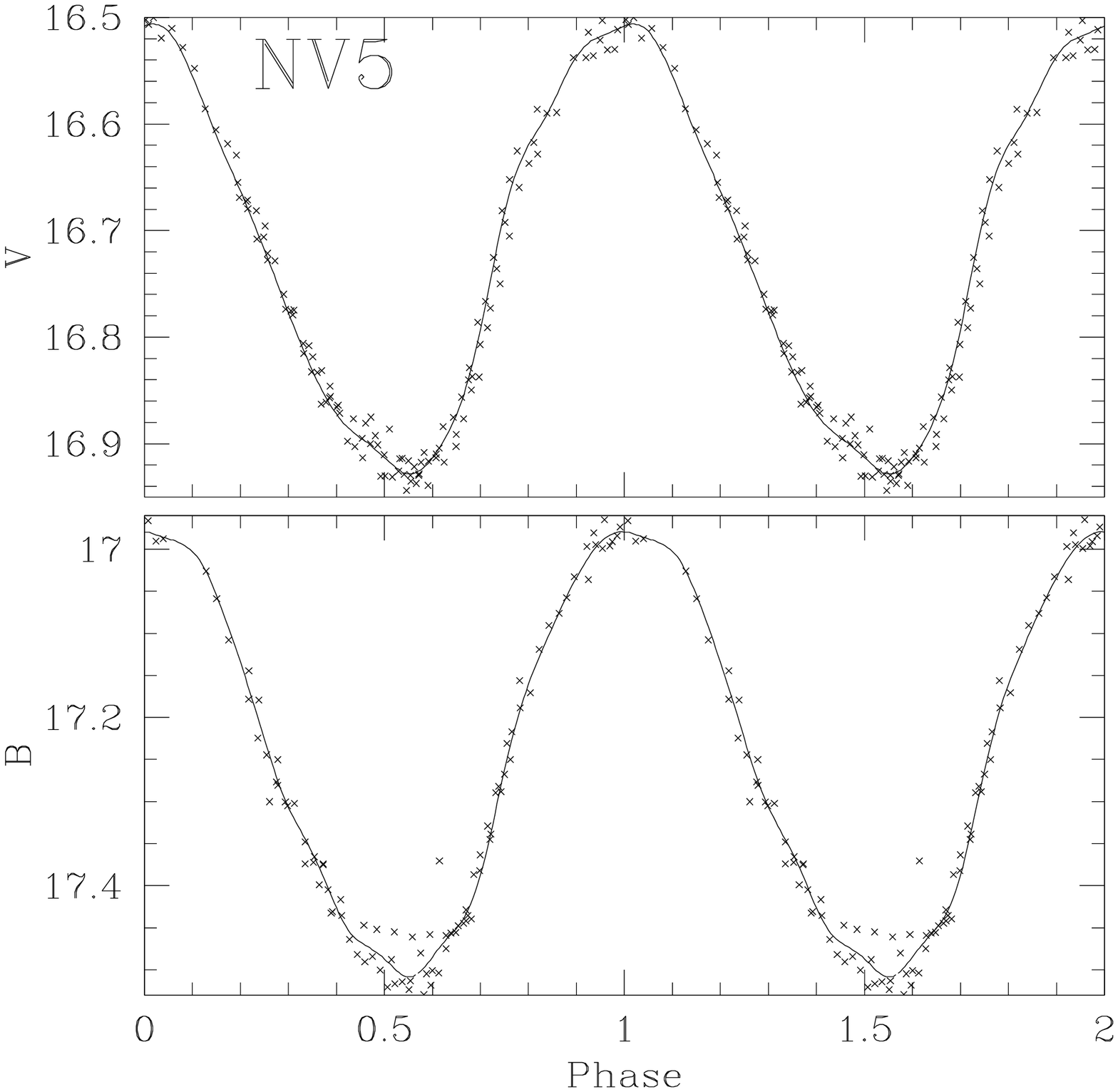}
  \end{center}
  \end{figure}

  \begin{figure}[t]
  \begin{center}
\includegraphics[width=.28\textwidth]{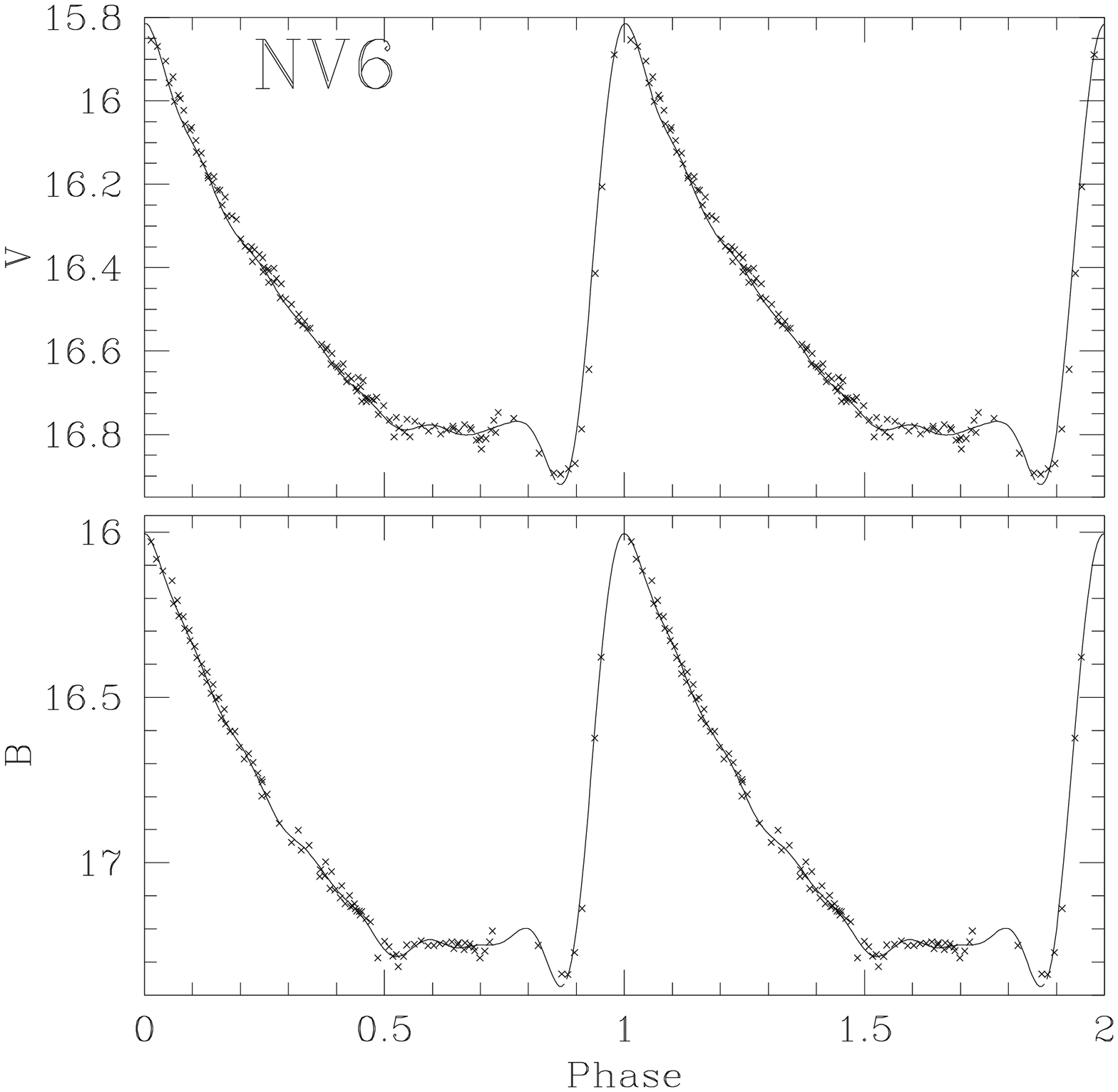}
\includegraphics[width=.28\textwidth]{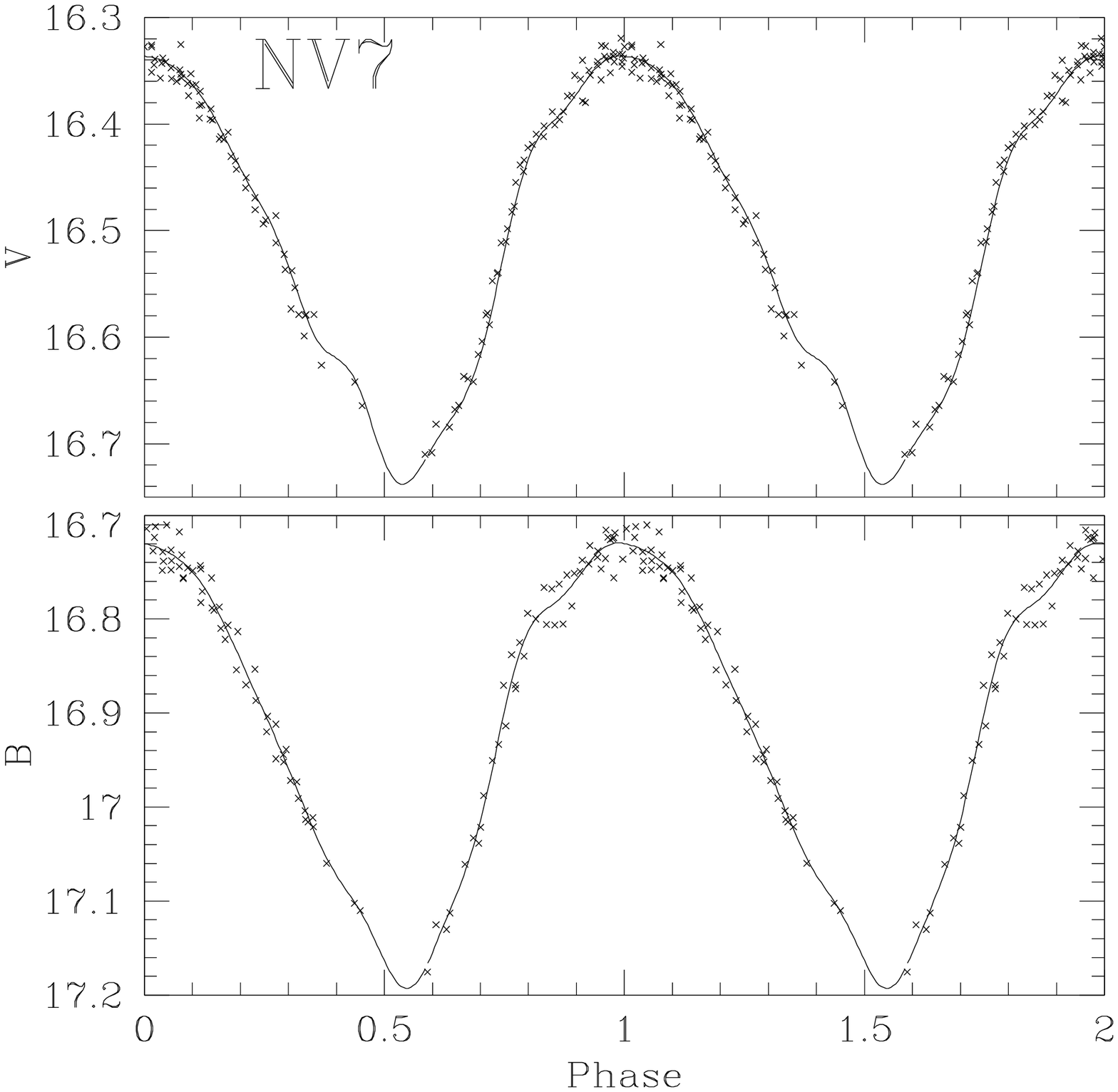}
\includegraphics[width=.28\textwidth]{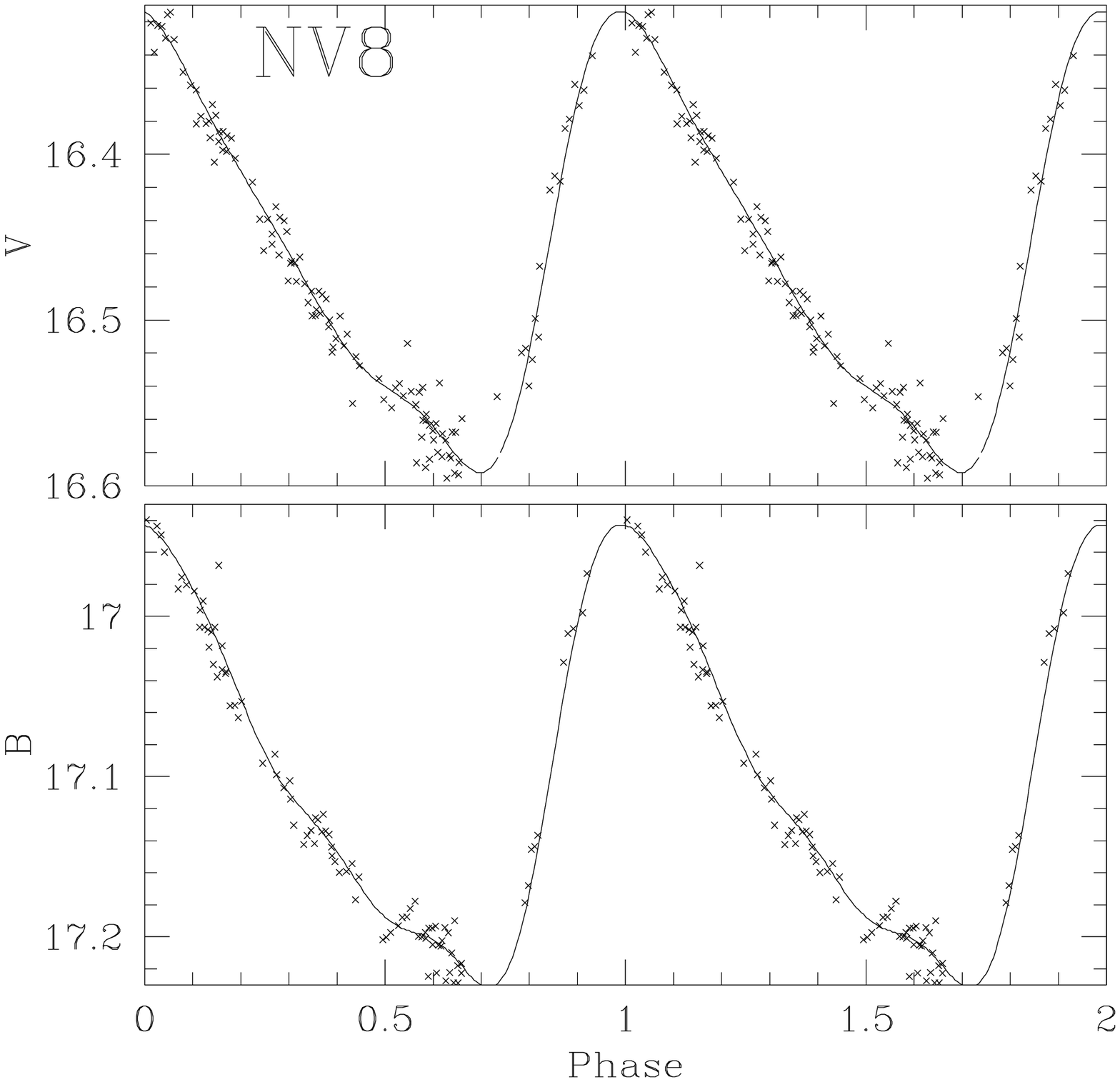}
\includegraphics[width=.28\textwidth]{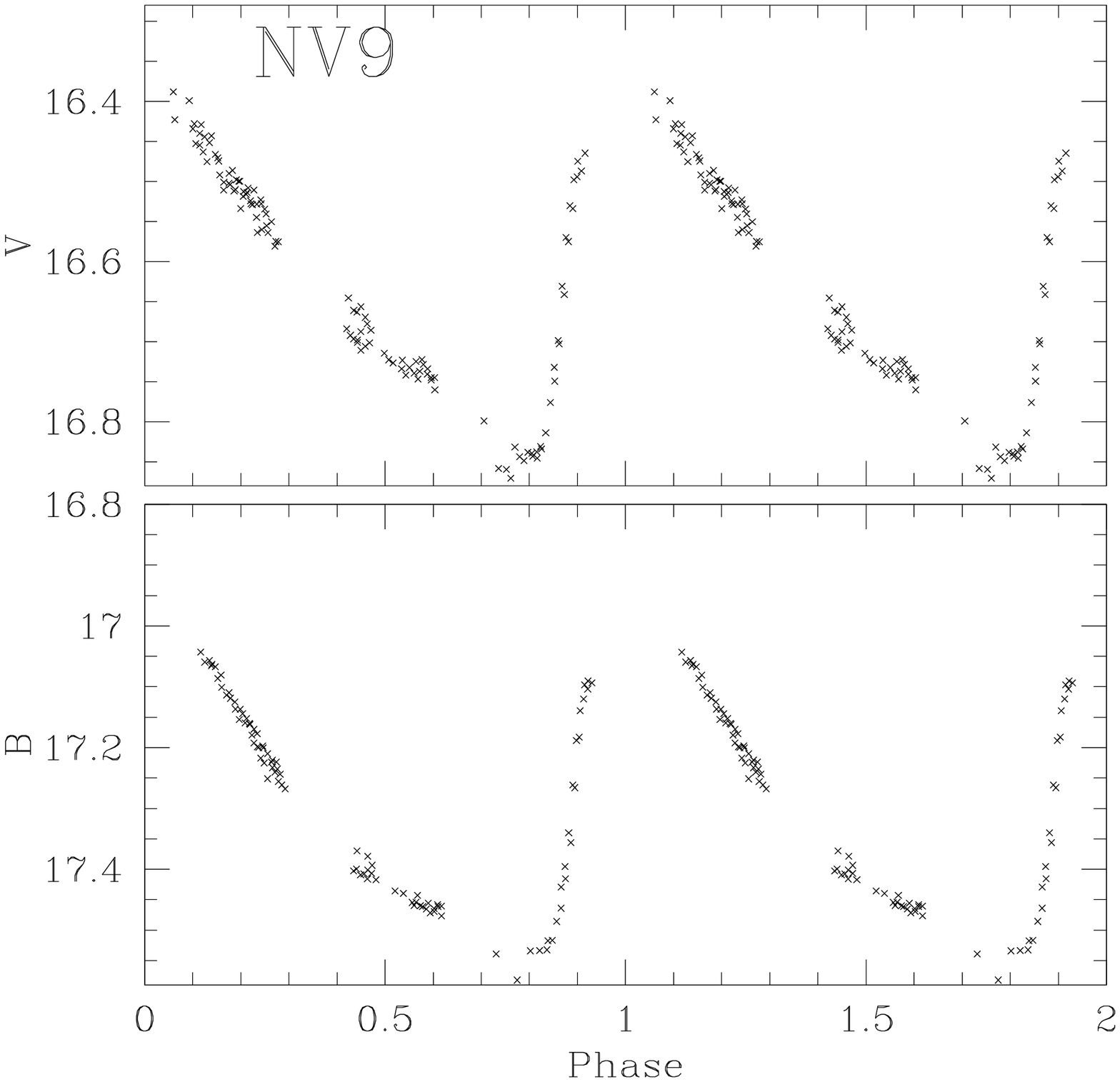}
\includegraphics[width=.28\textwidth]{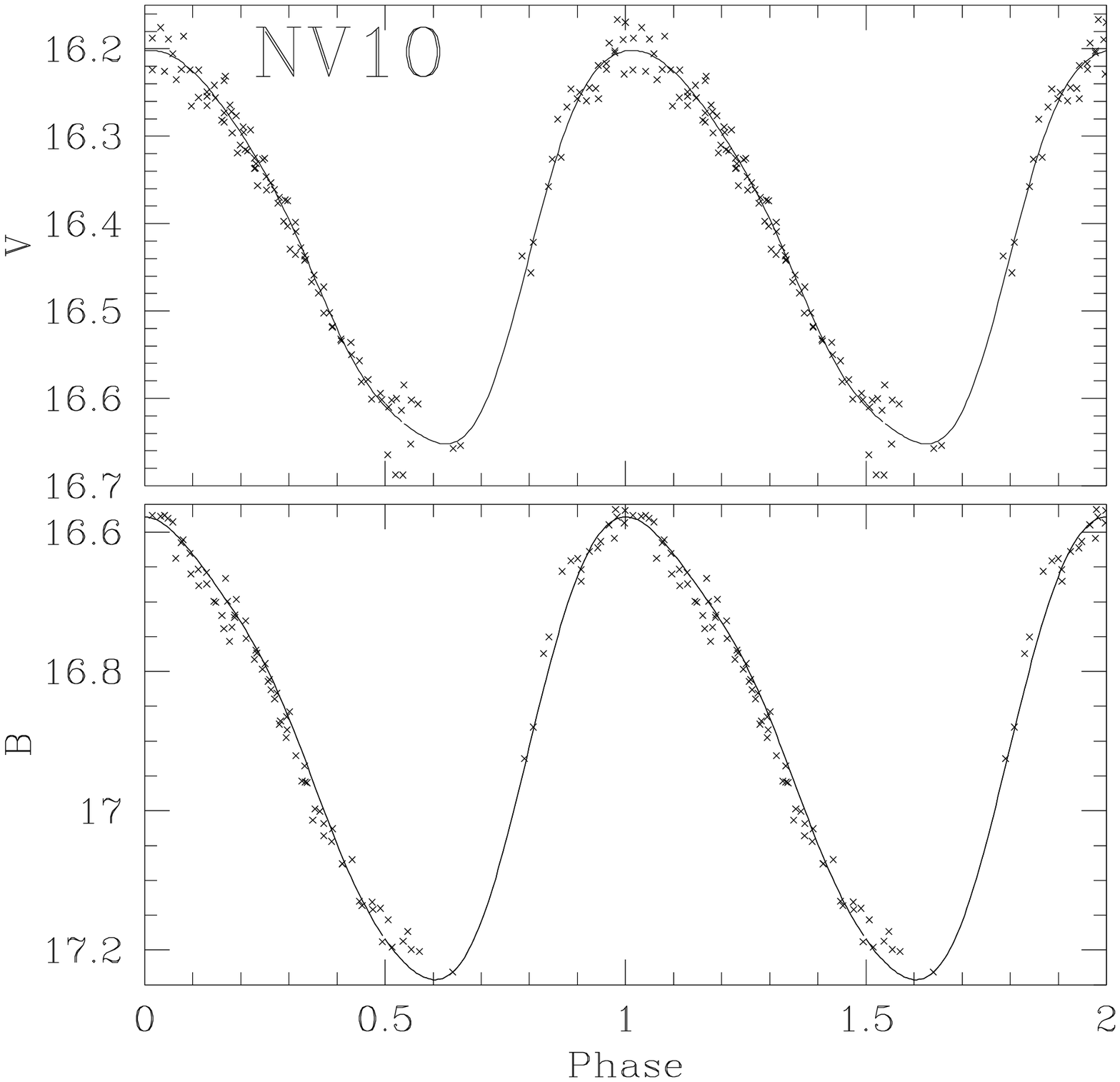}
\includegraphics[width=.28\textwidth]{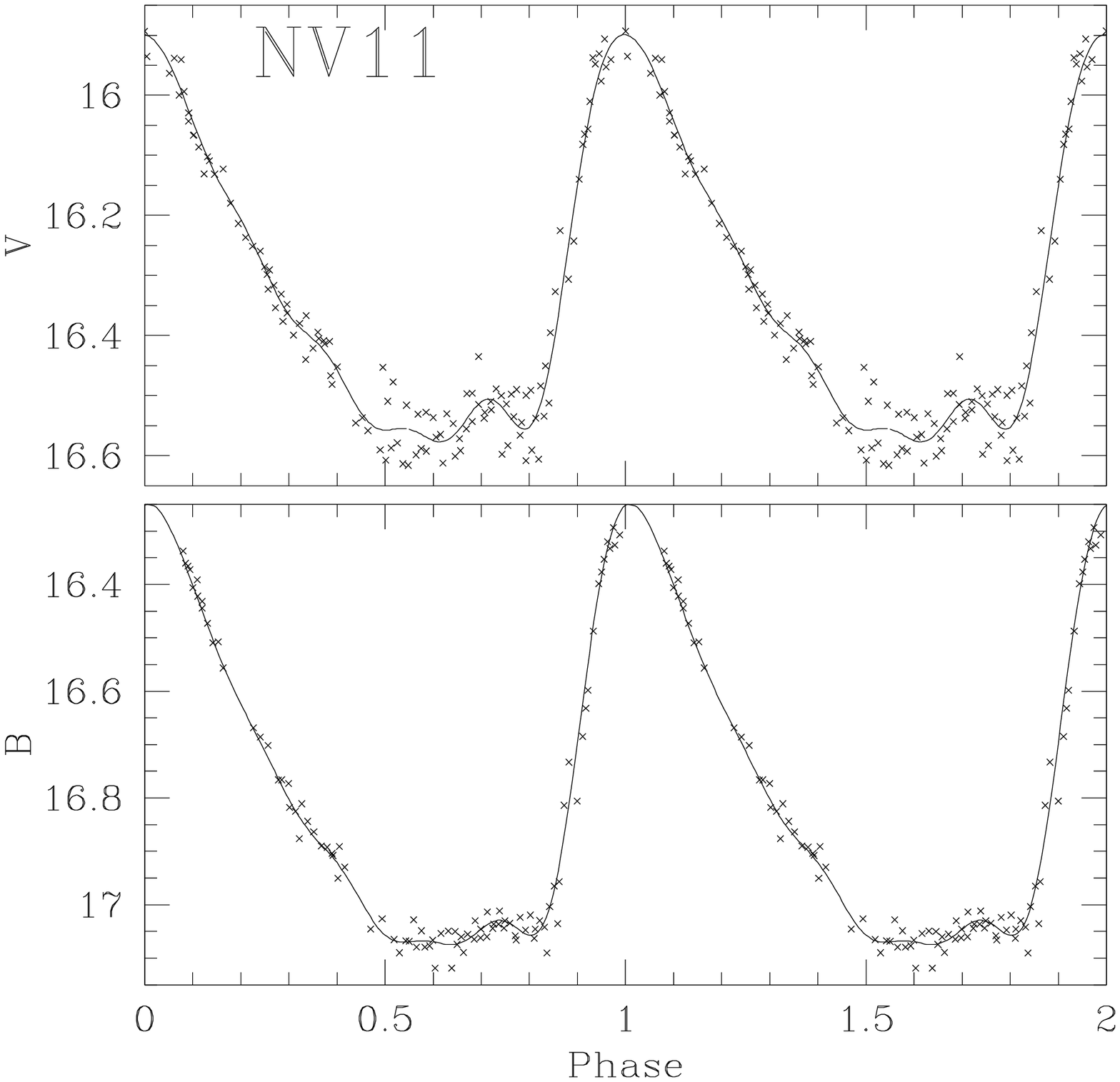}
\includegraphics[width=.28\textwidth]{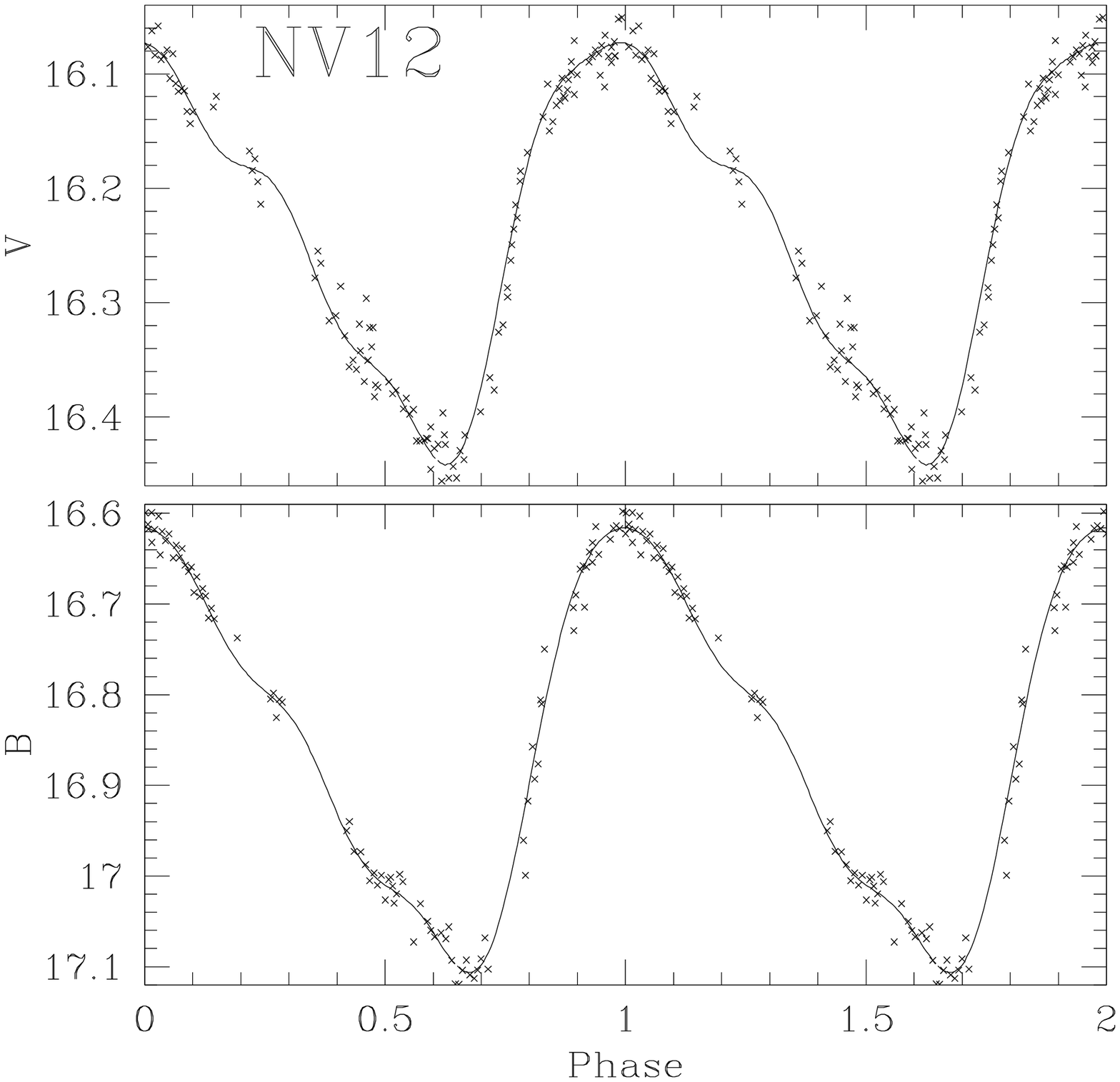}
\includegraphics[width=.28\textwidth]{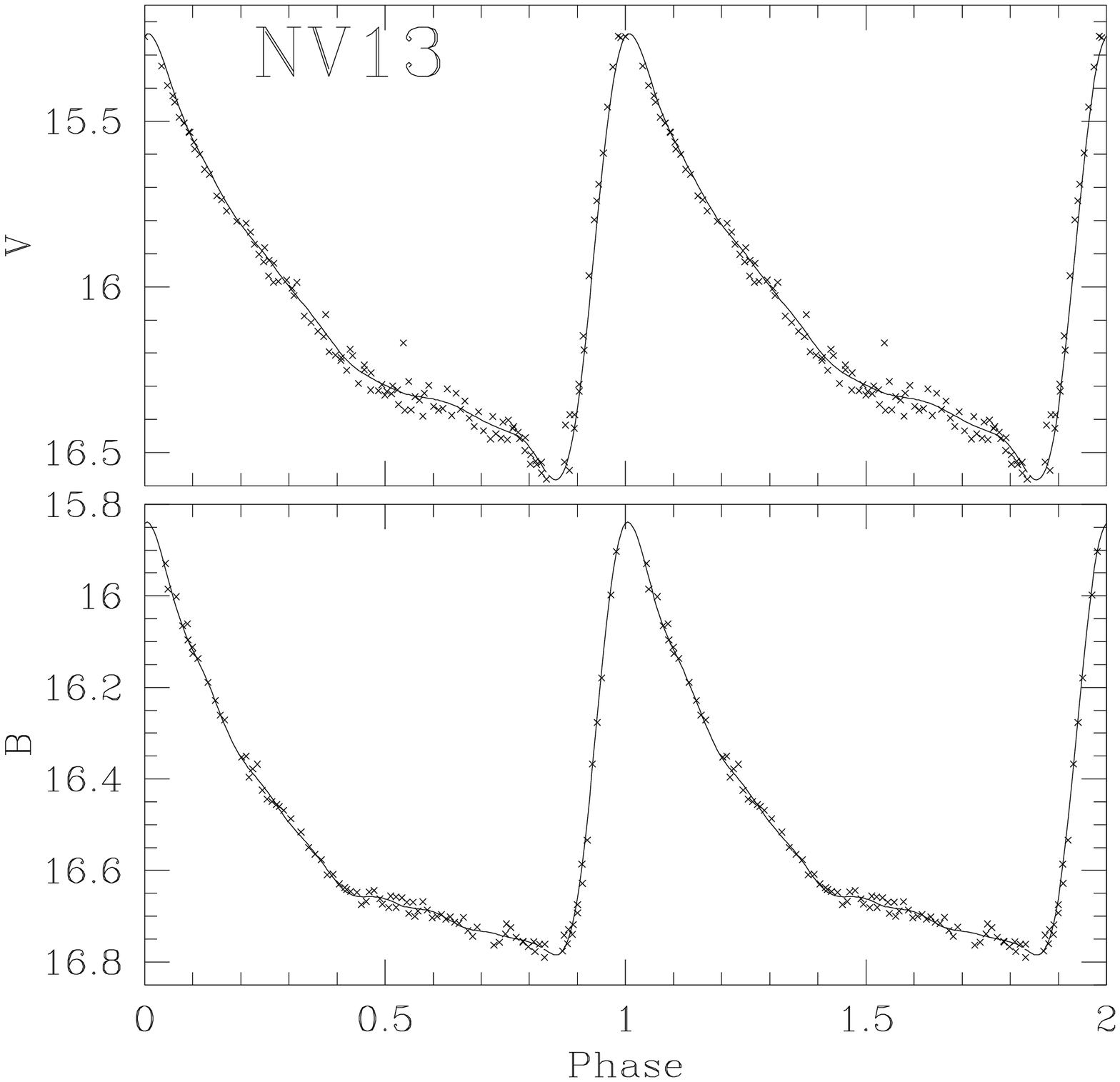}
\includegraphics[width=.28\textwidth]{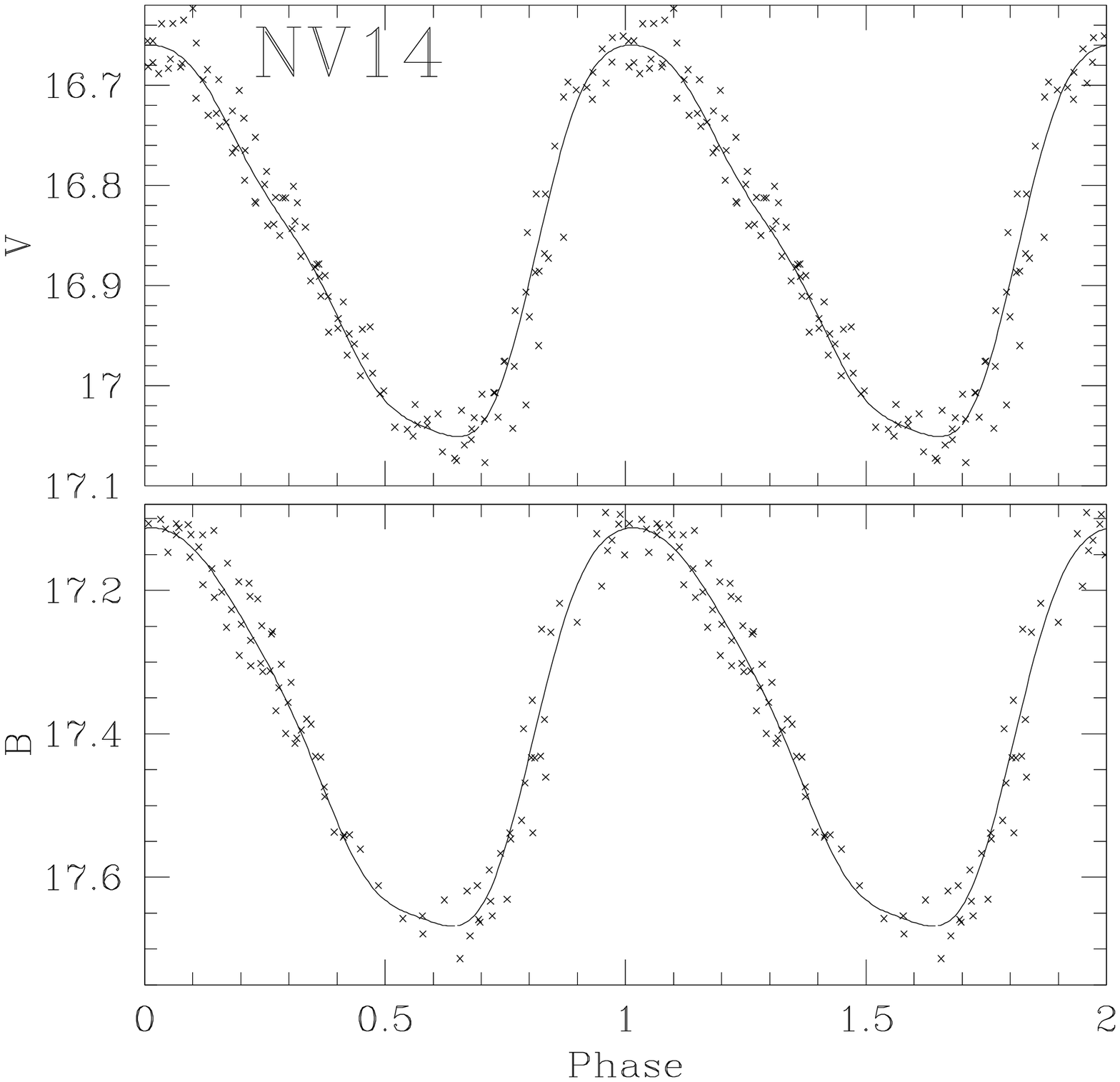}
\includegraphics[width=.28\textwidth]{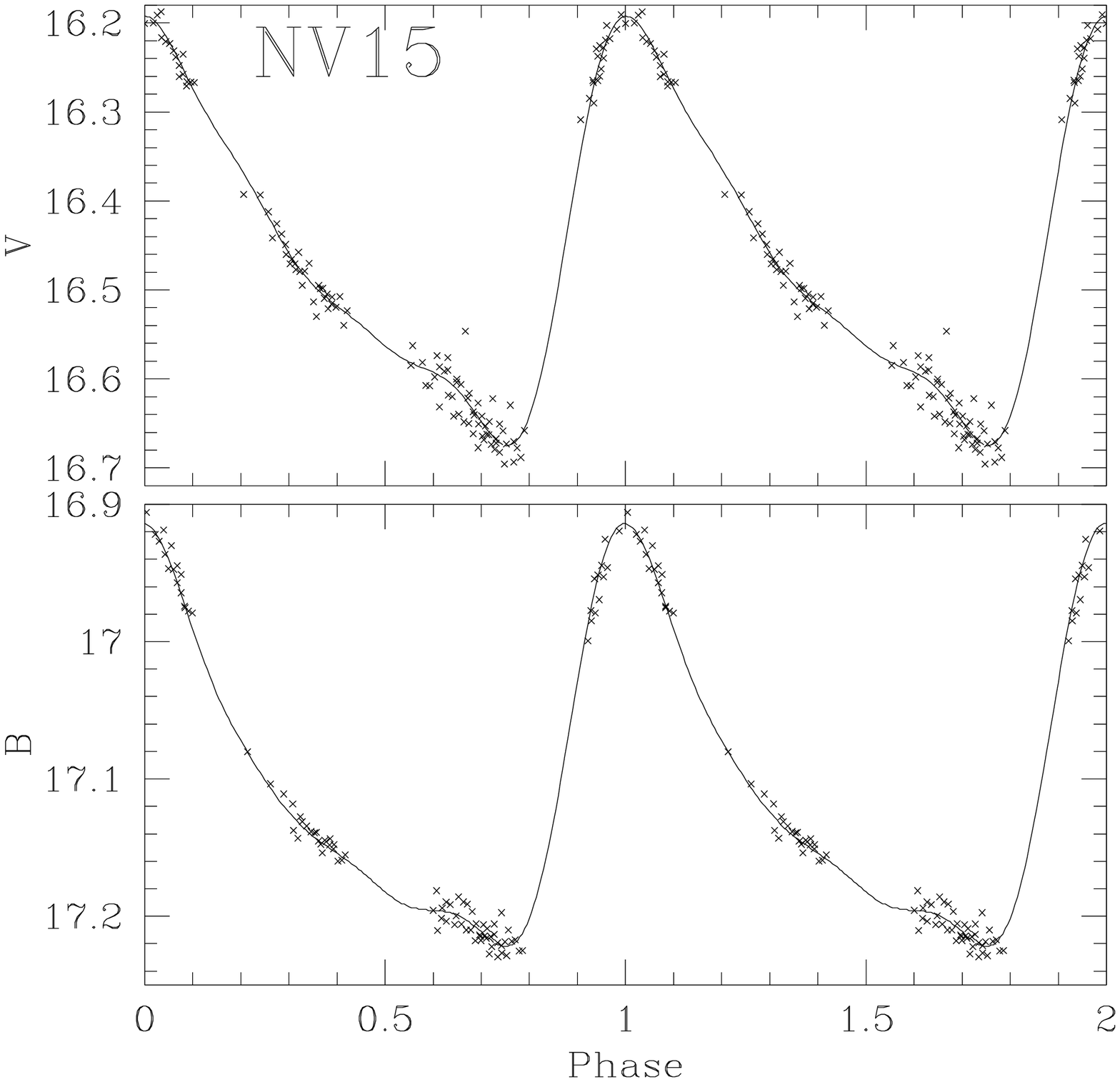}
\includegraphics[width=.28\textwidth]{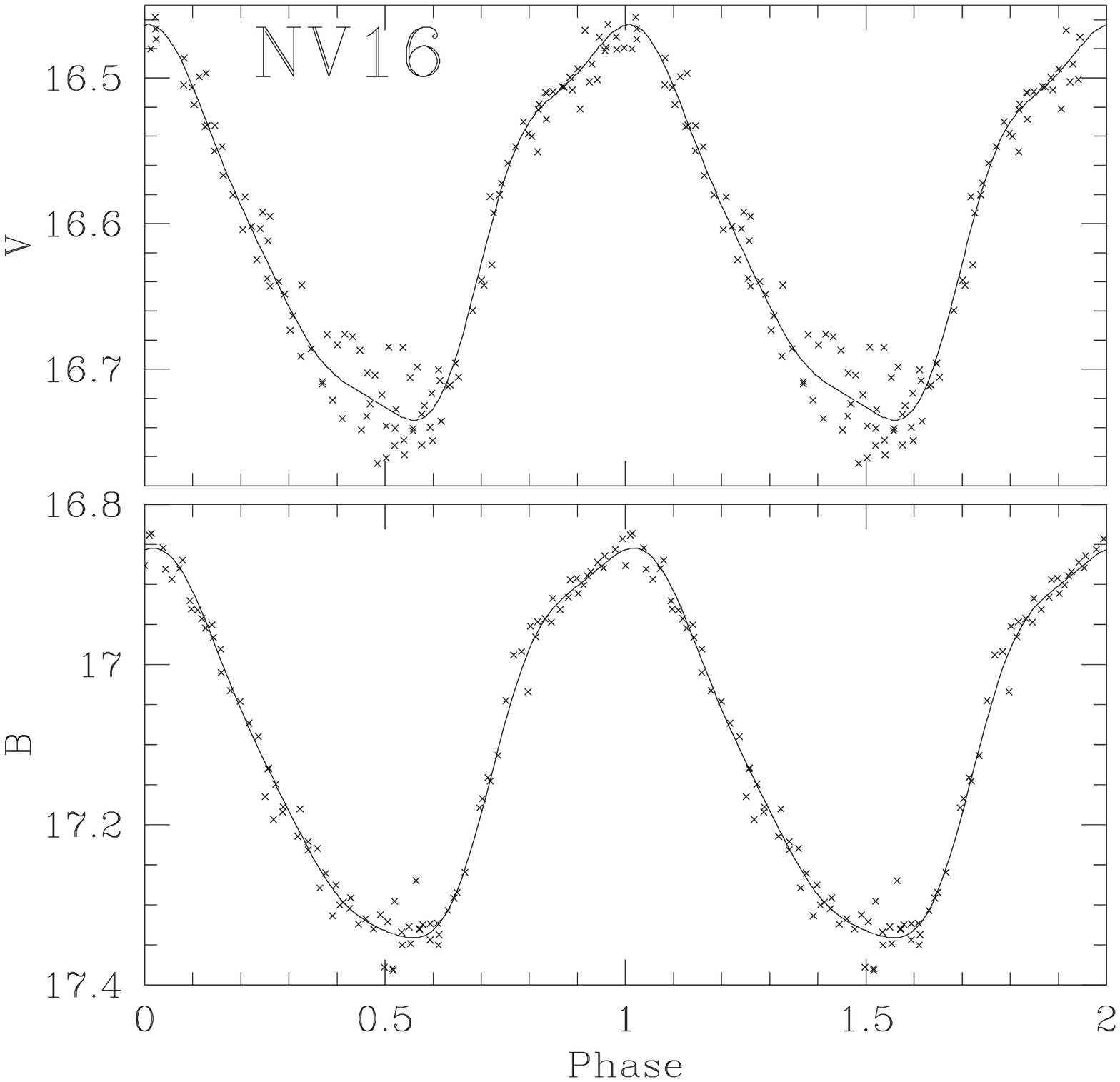}
\includegraphics[width=.28\textwidth]{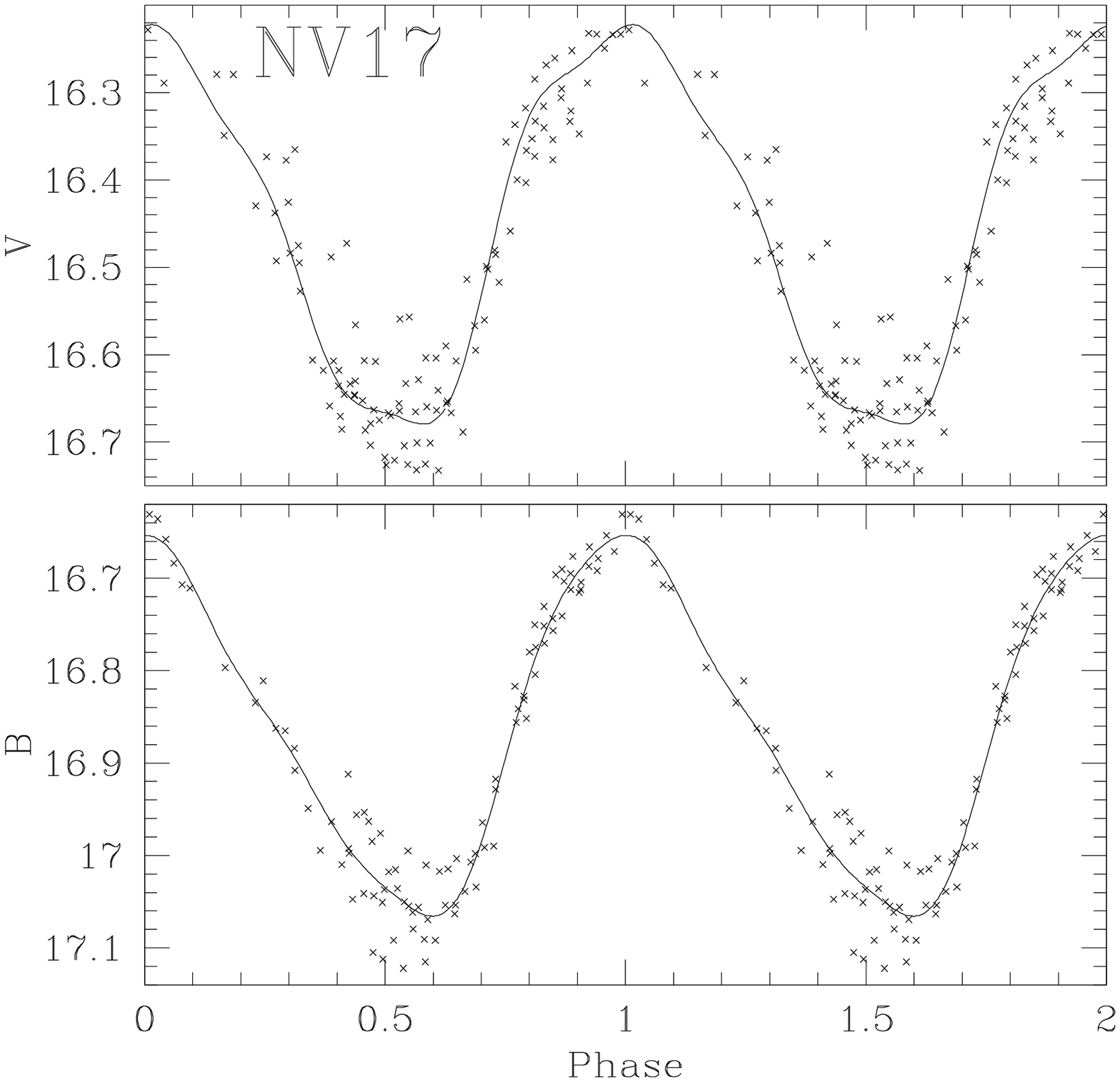}
  \end{center}
  \end{figure}
    
  \begin{figure}[t]
  \begin{center}
\includegraphics[width=.28\textwidth]{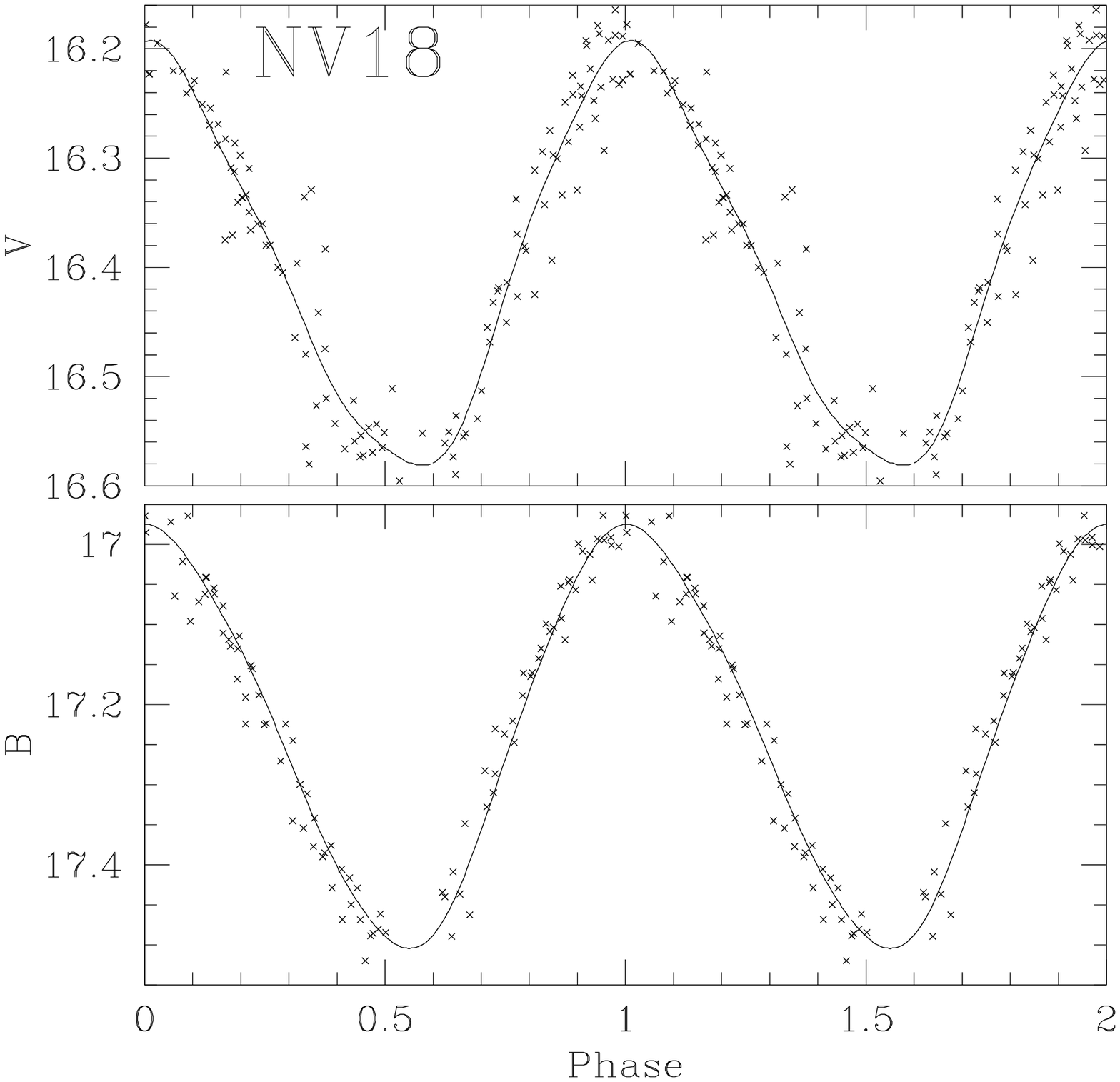}
\includegraphics[width=.28\textwidth]{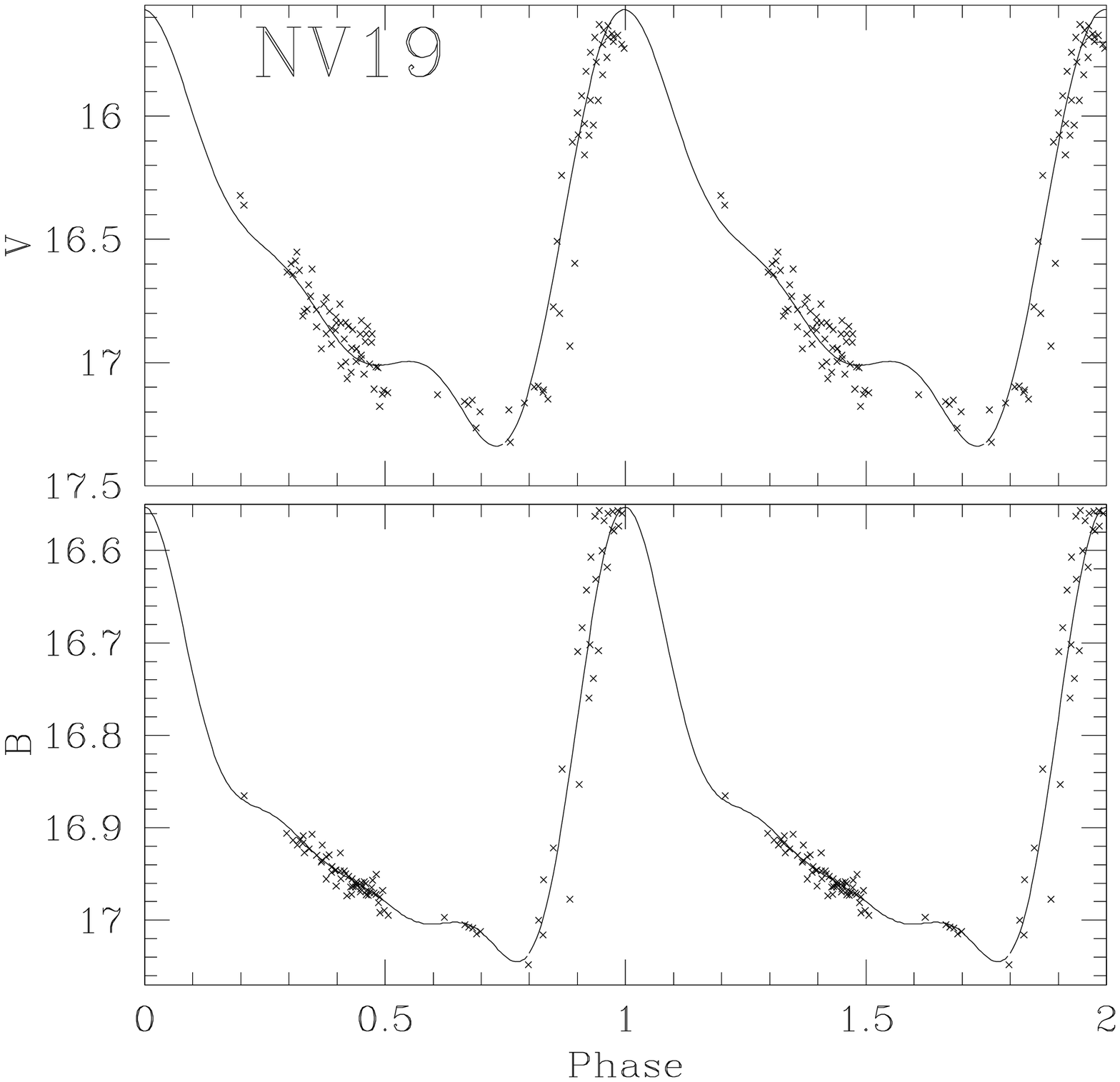}
\includegraphics[width=.28\textwidth]{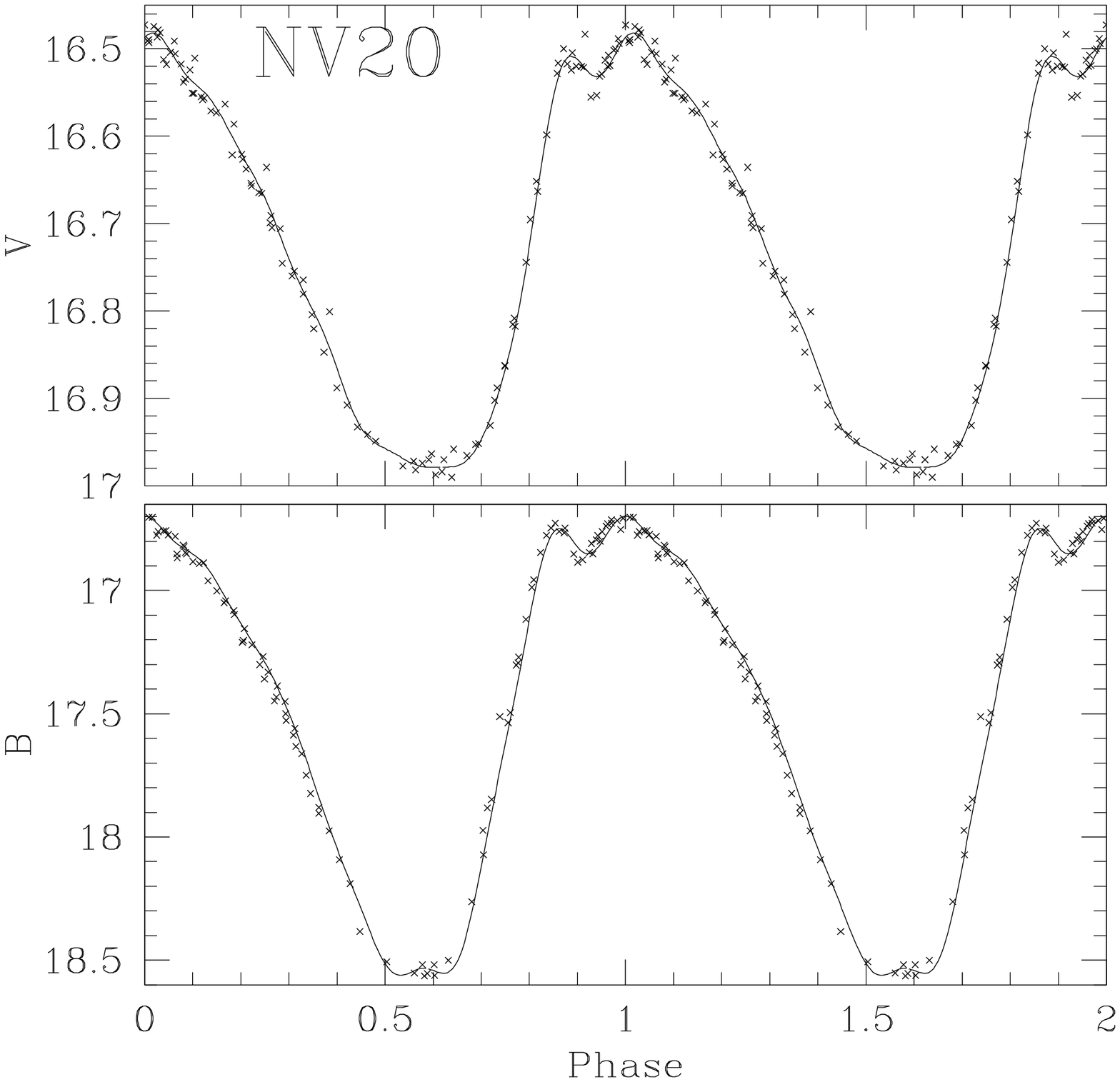}
\includegraphics[width=.28\textwidth]{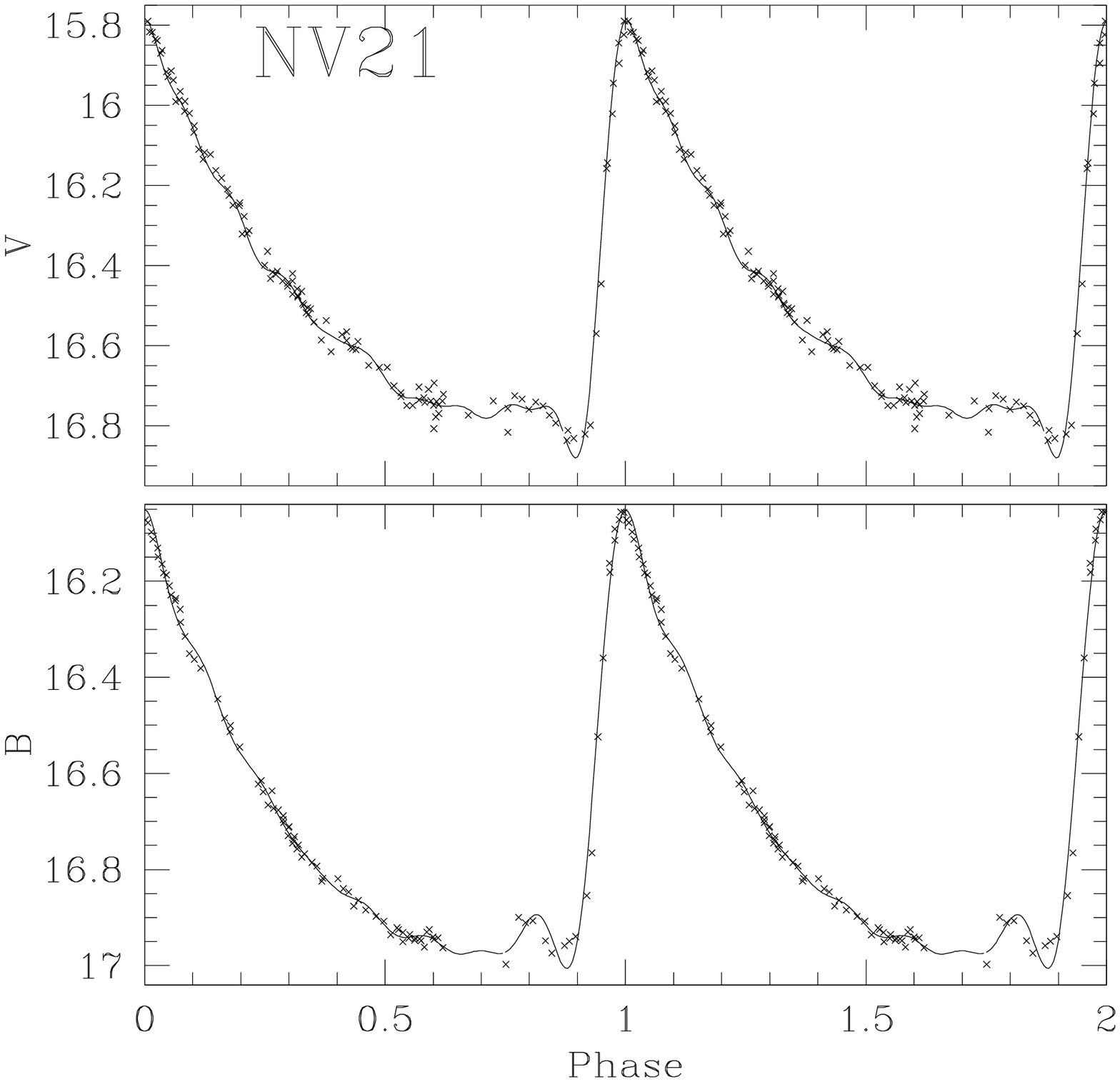}
\includegraphics[width=.28\textwidth]{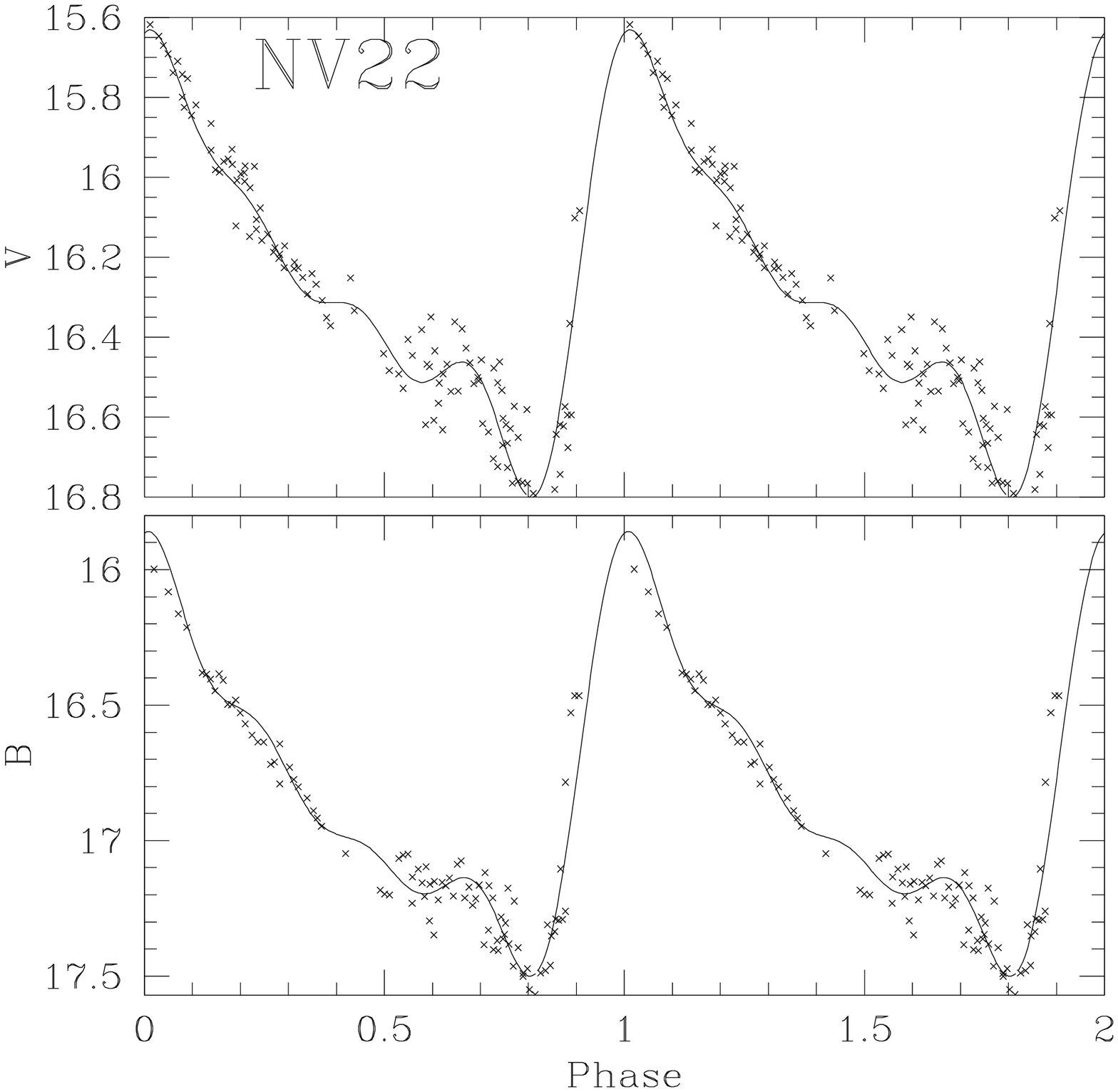}
\includegraphics[width=.28\textwidth]{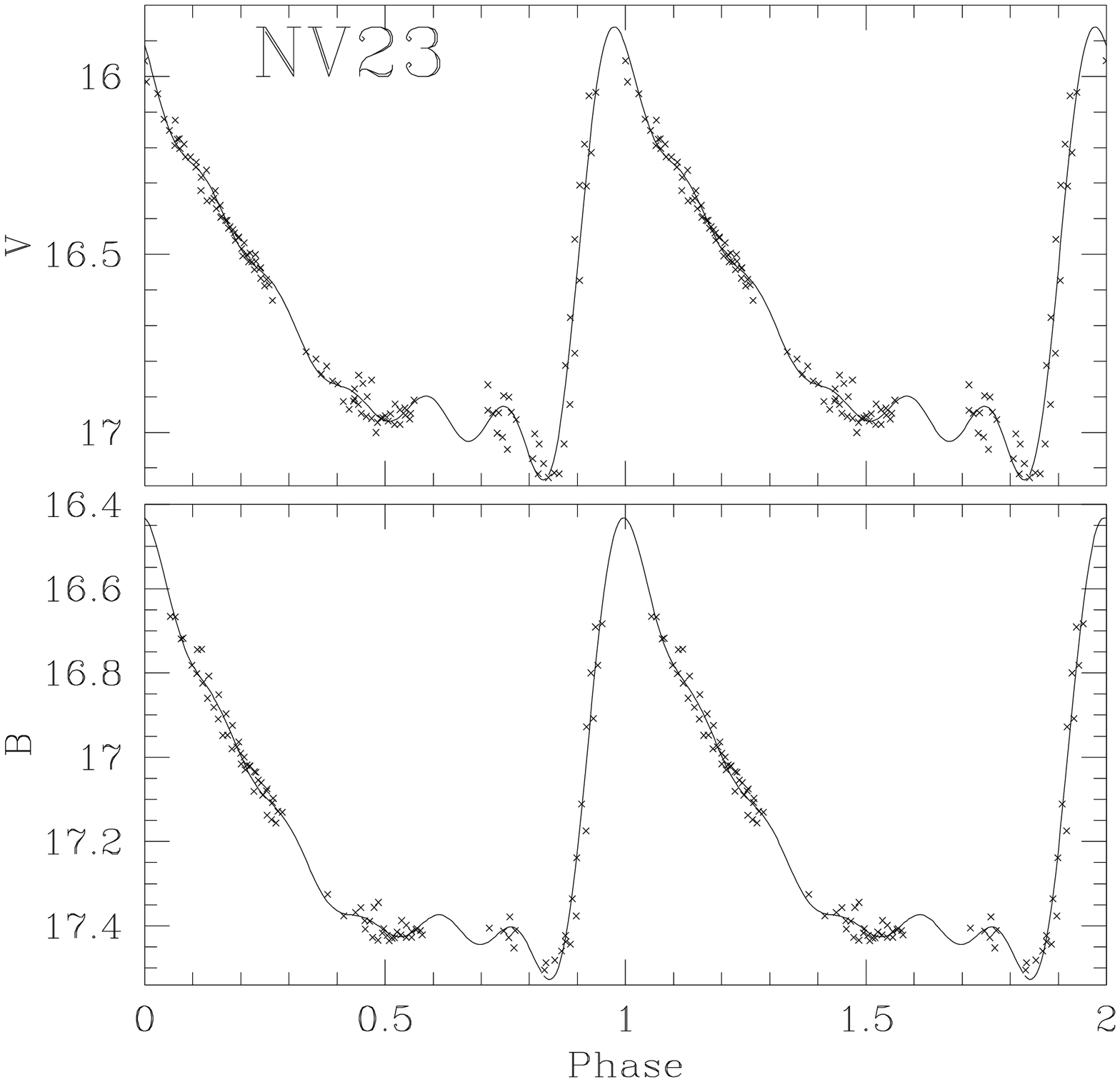}
\includegraphics[width=.28\textwidth]{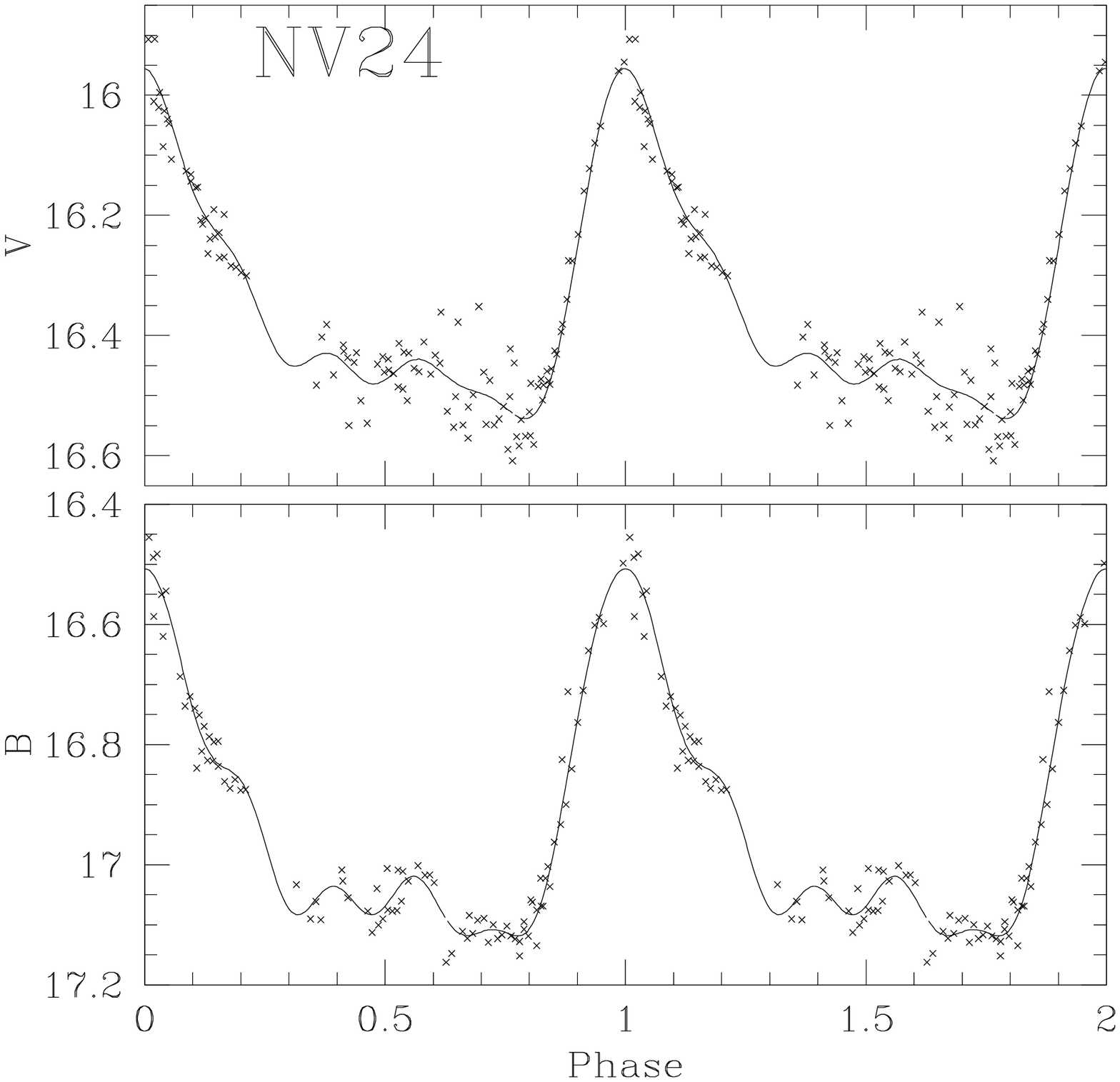}
\includegraphics[width=.28\textwidth]{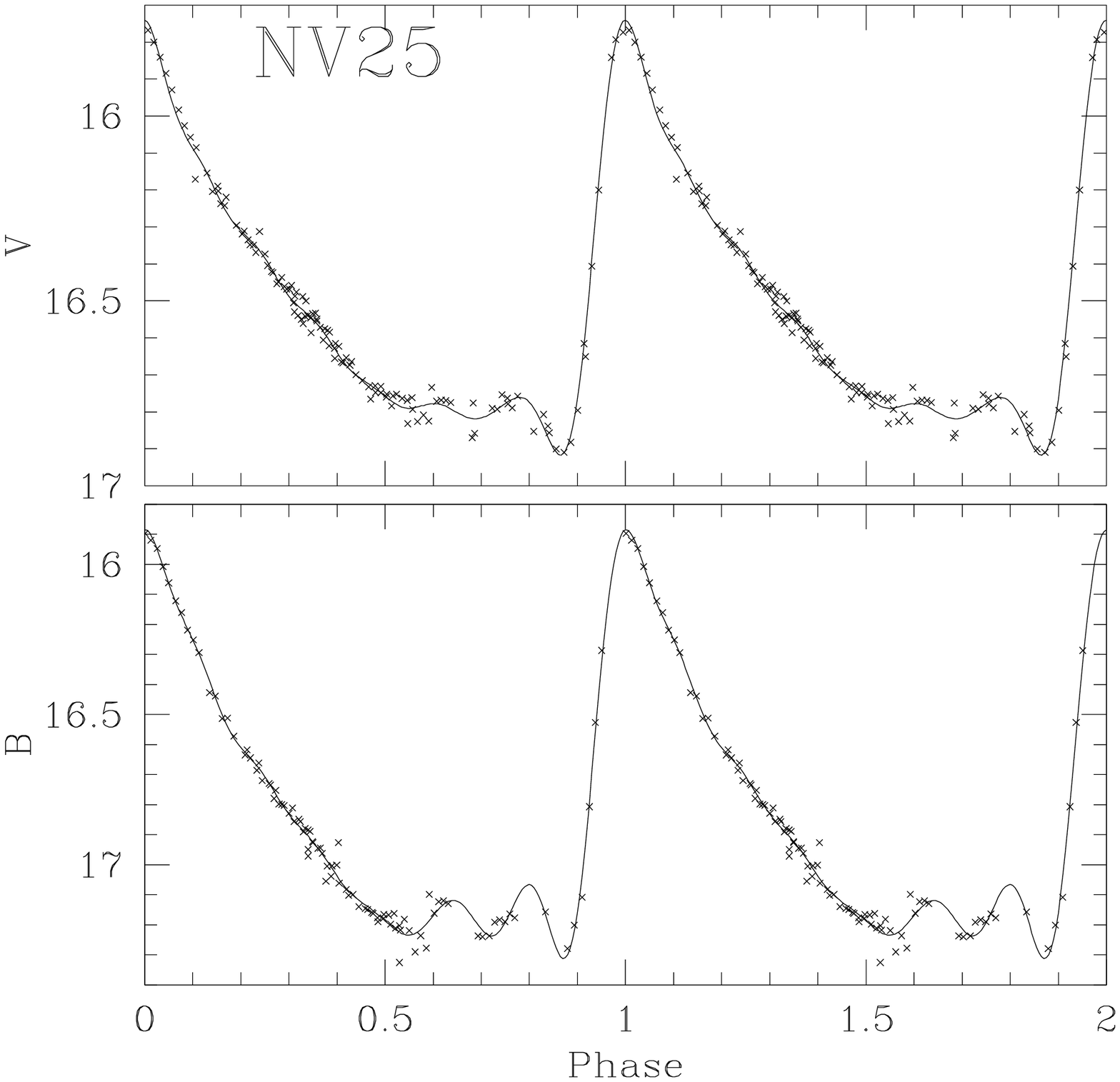}
\includegraphics[width=.28\textwidth]{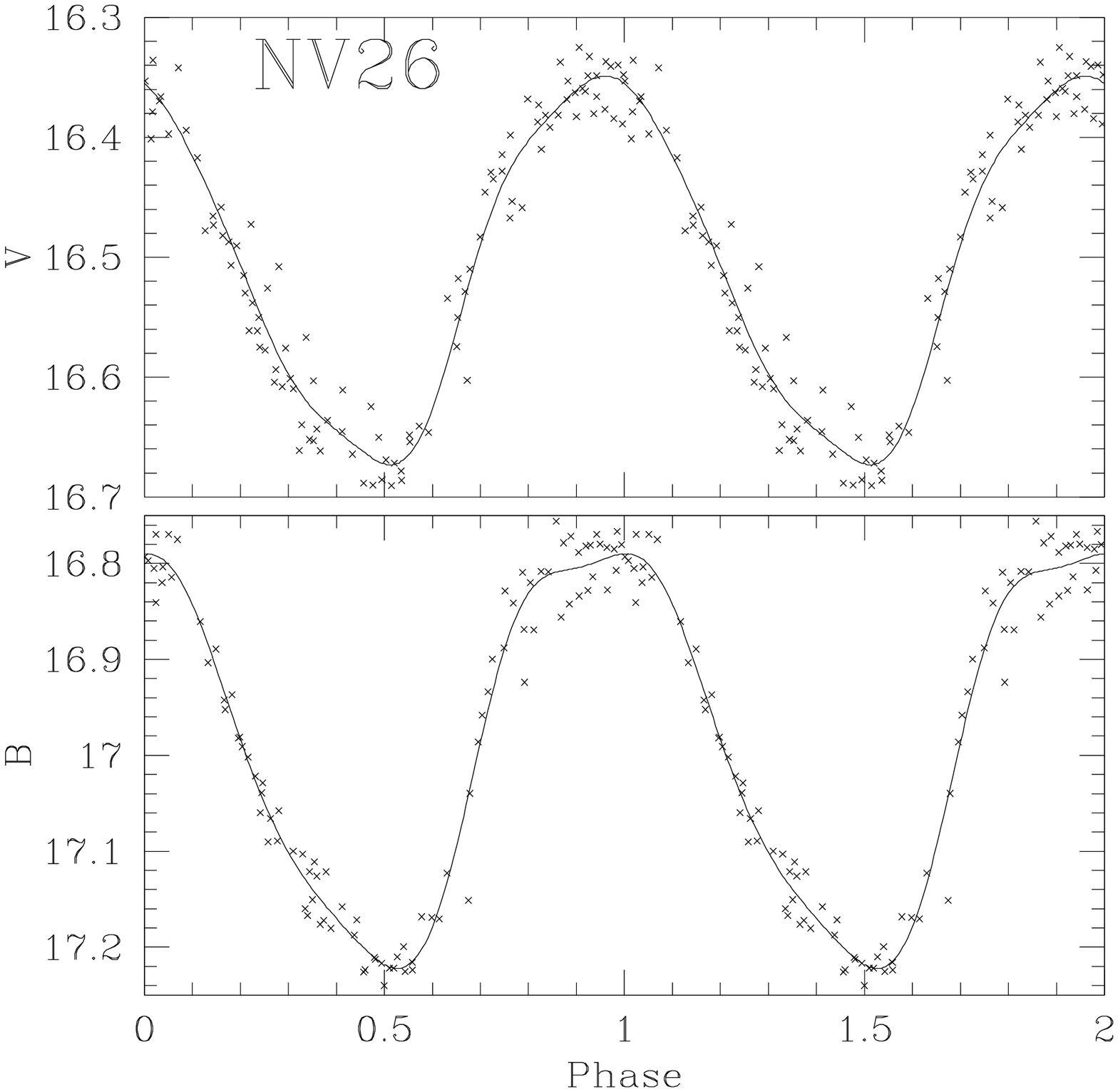}
\includegraphics[width=.28\textwidth]{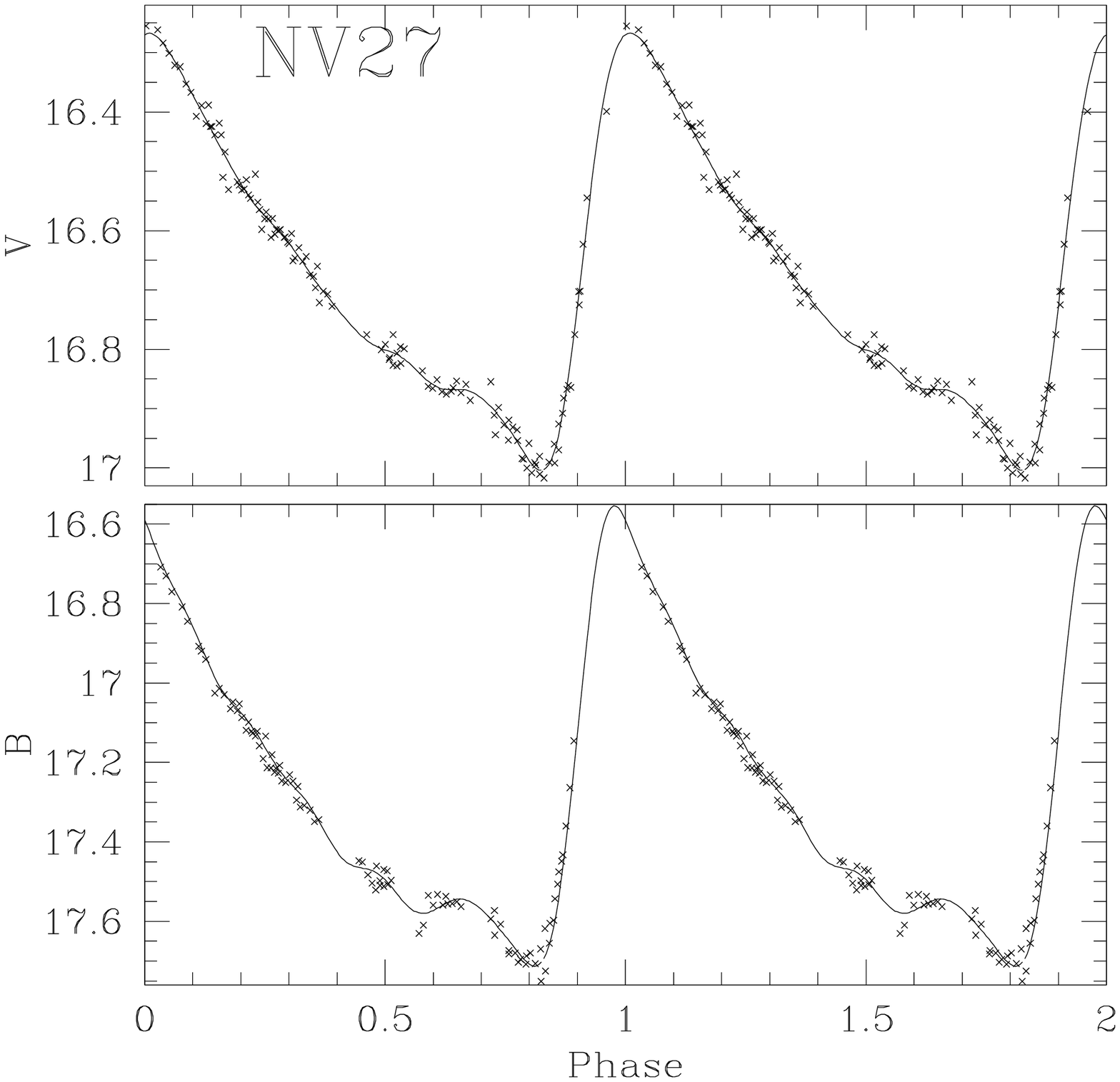}
\includegraphics[width=.28\textwidth]{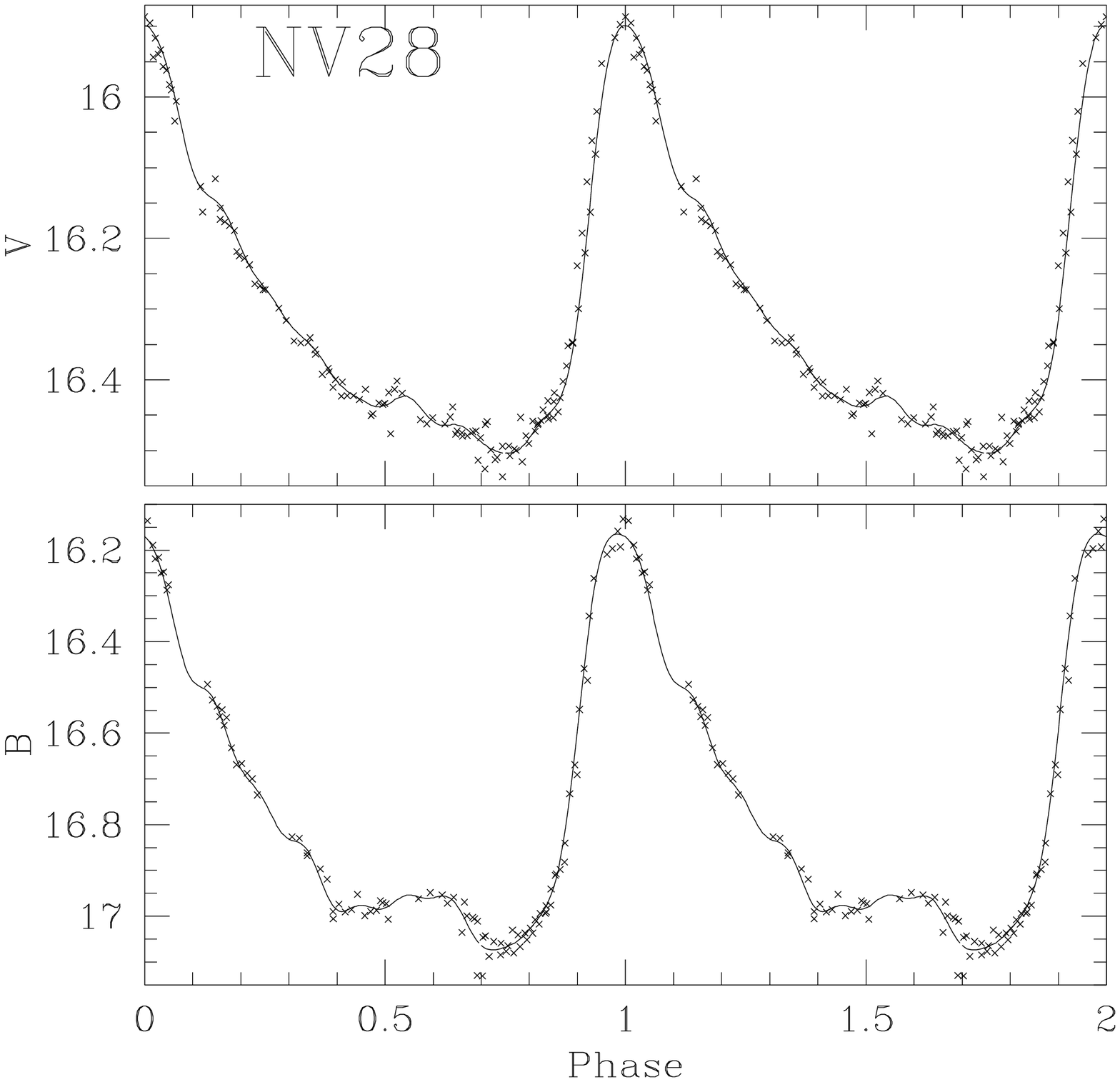}
\includegraphics[width=.28\textwidth]{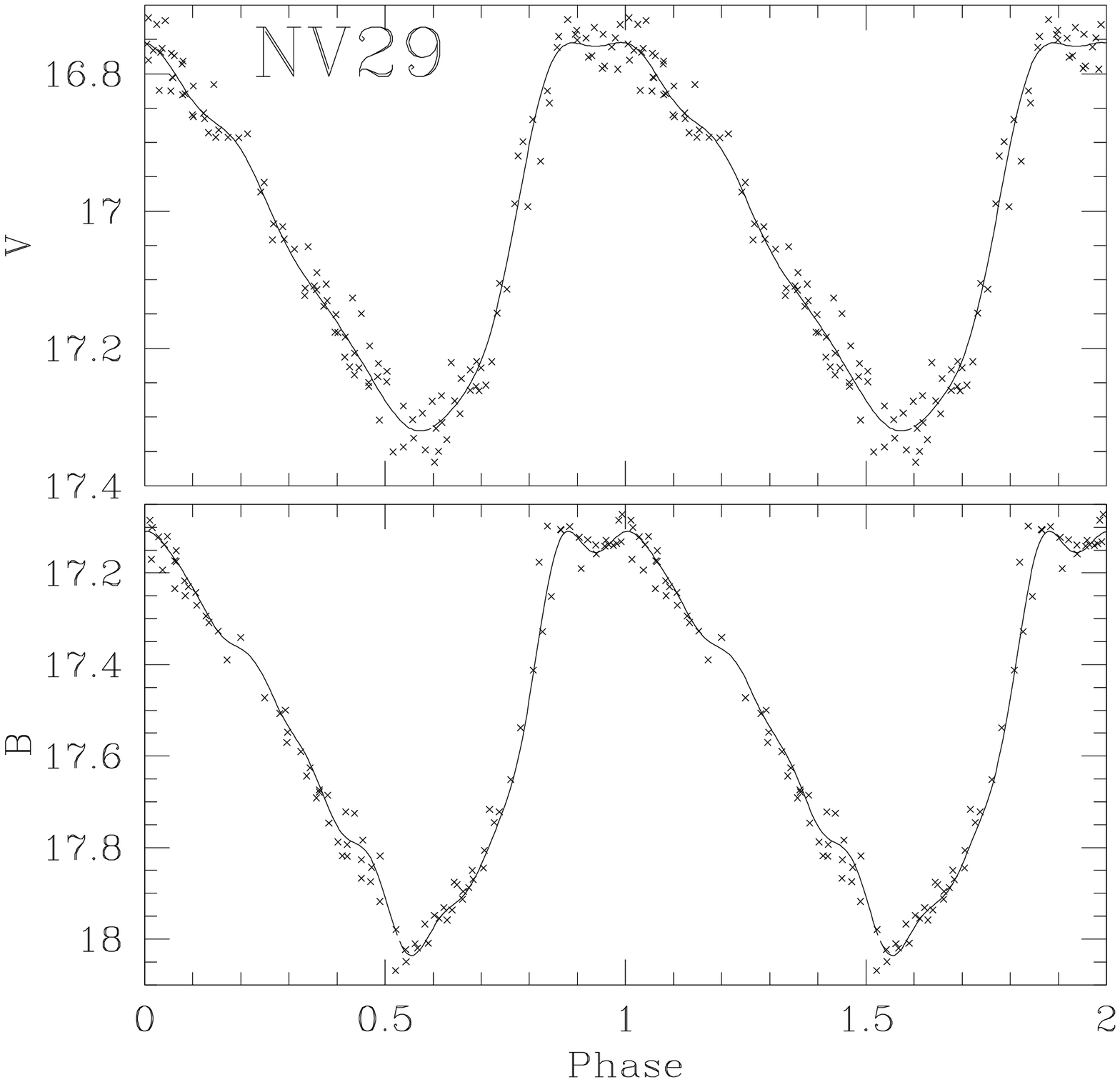}
  \end{center}
  \end{figure}
    
  \begin{figure}[t]
  \begin{center}
\includegraphics[width=.28\textwidth]{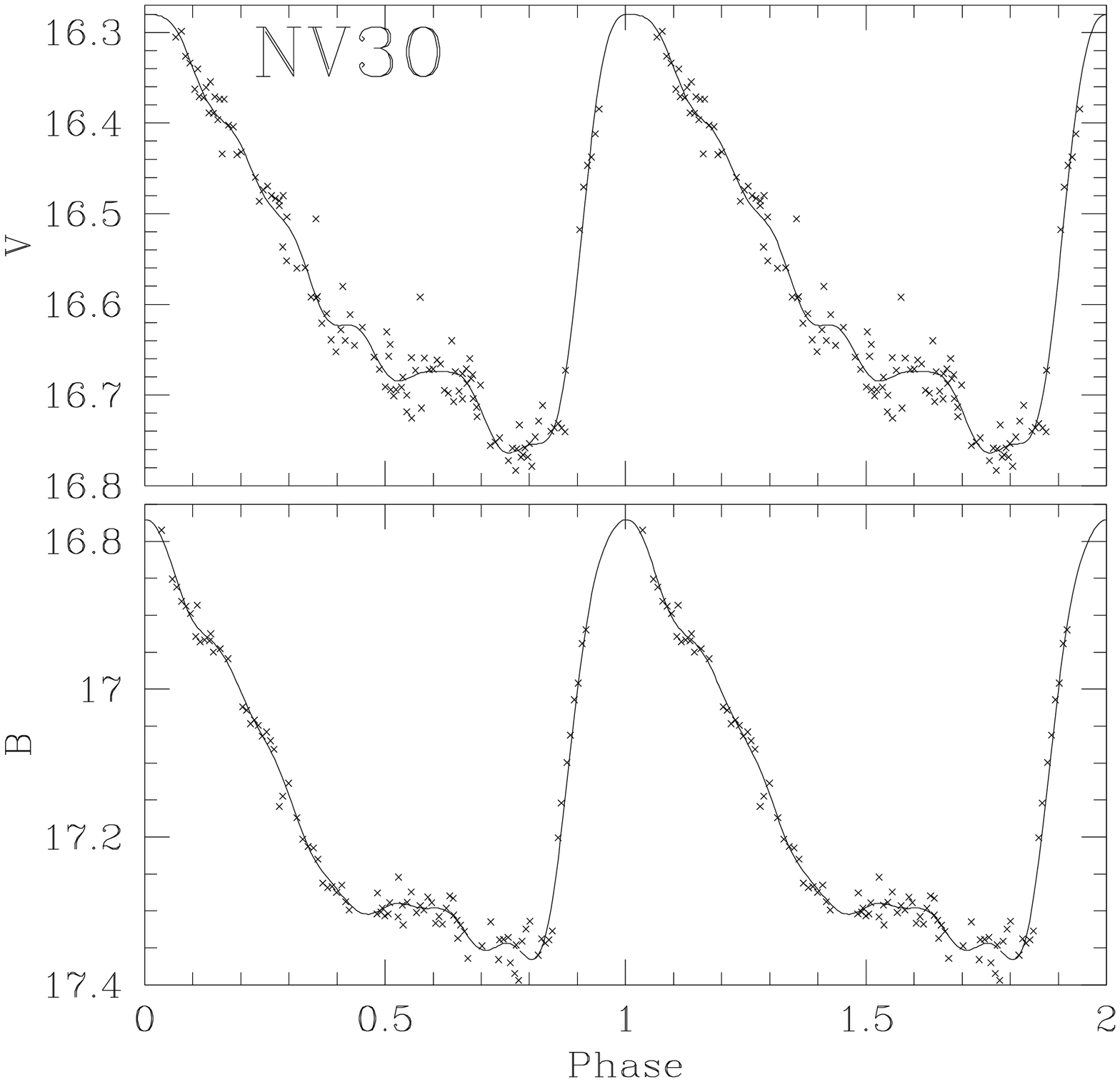}
\includegraphics[width=.28\textwidth]{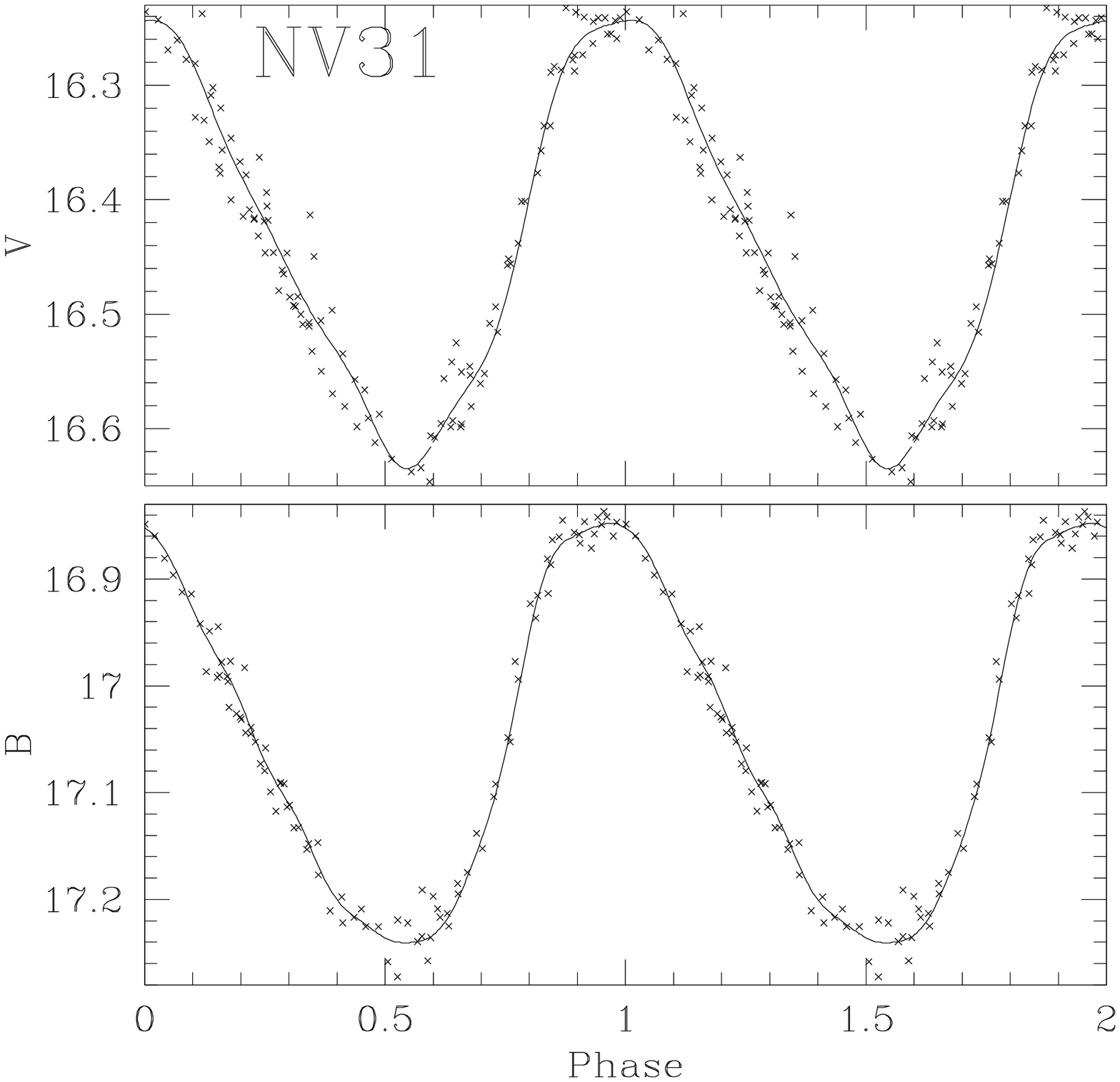}
\includegraphics[width=.28\textwidth]{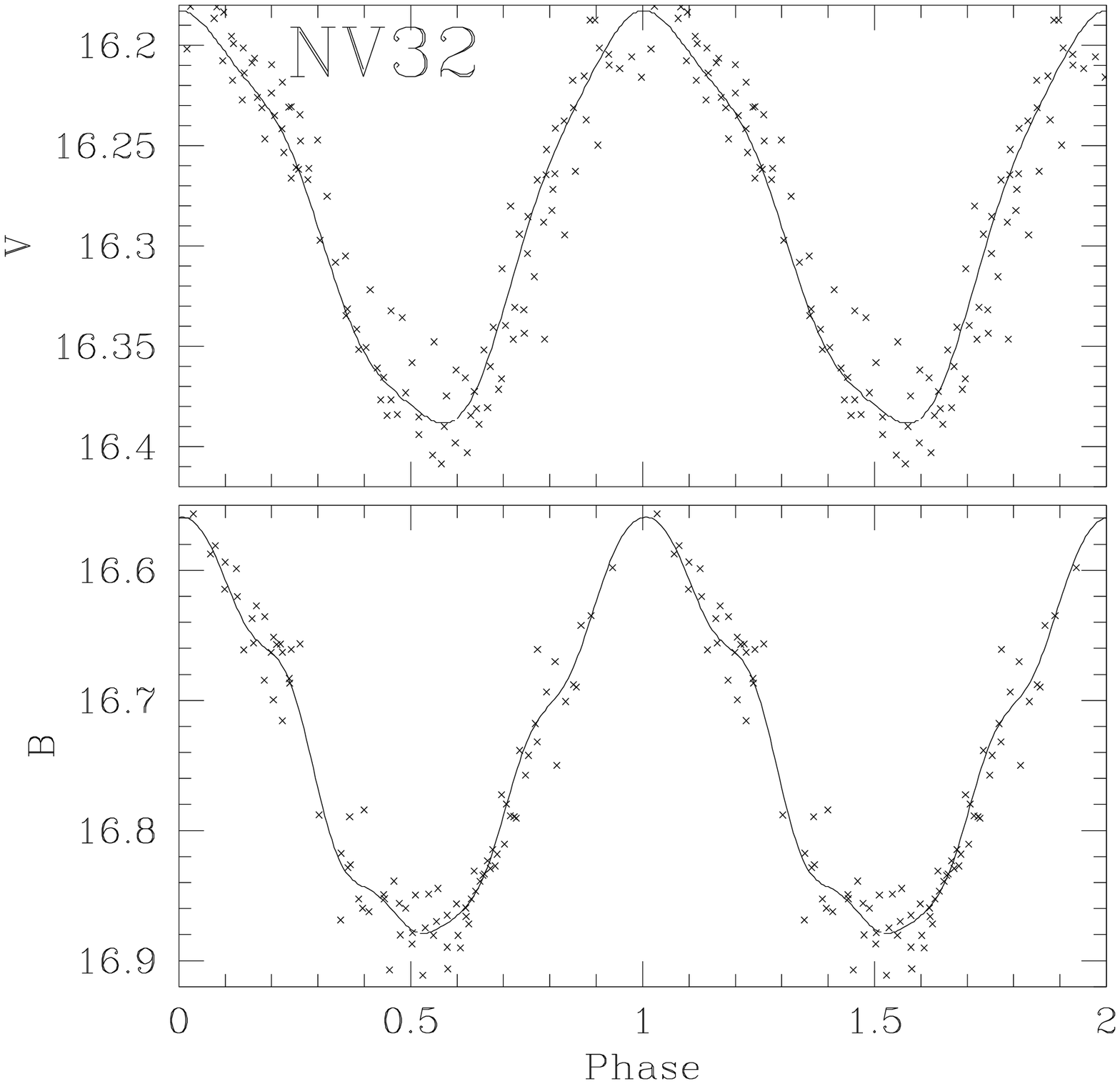}
\includegraphics[width=.28\textwidth]{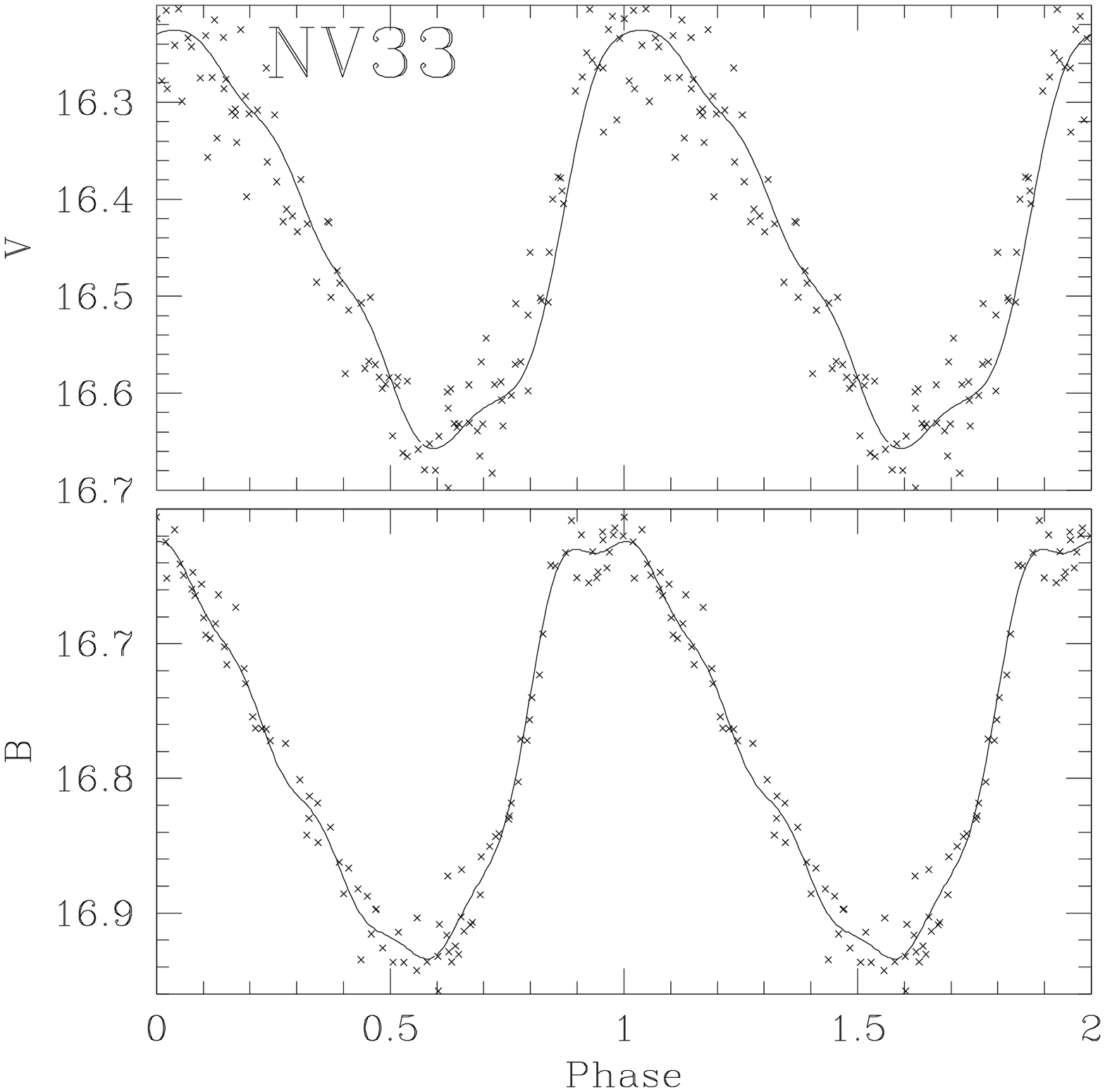}
\includegraphics[width=.28\textwidth]{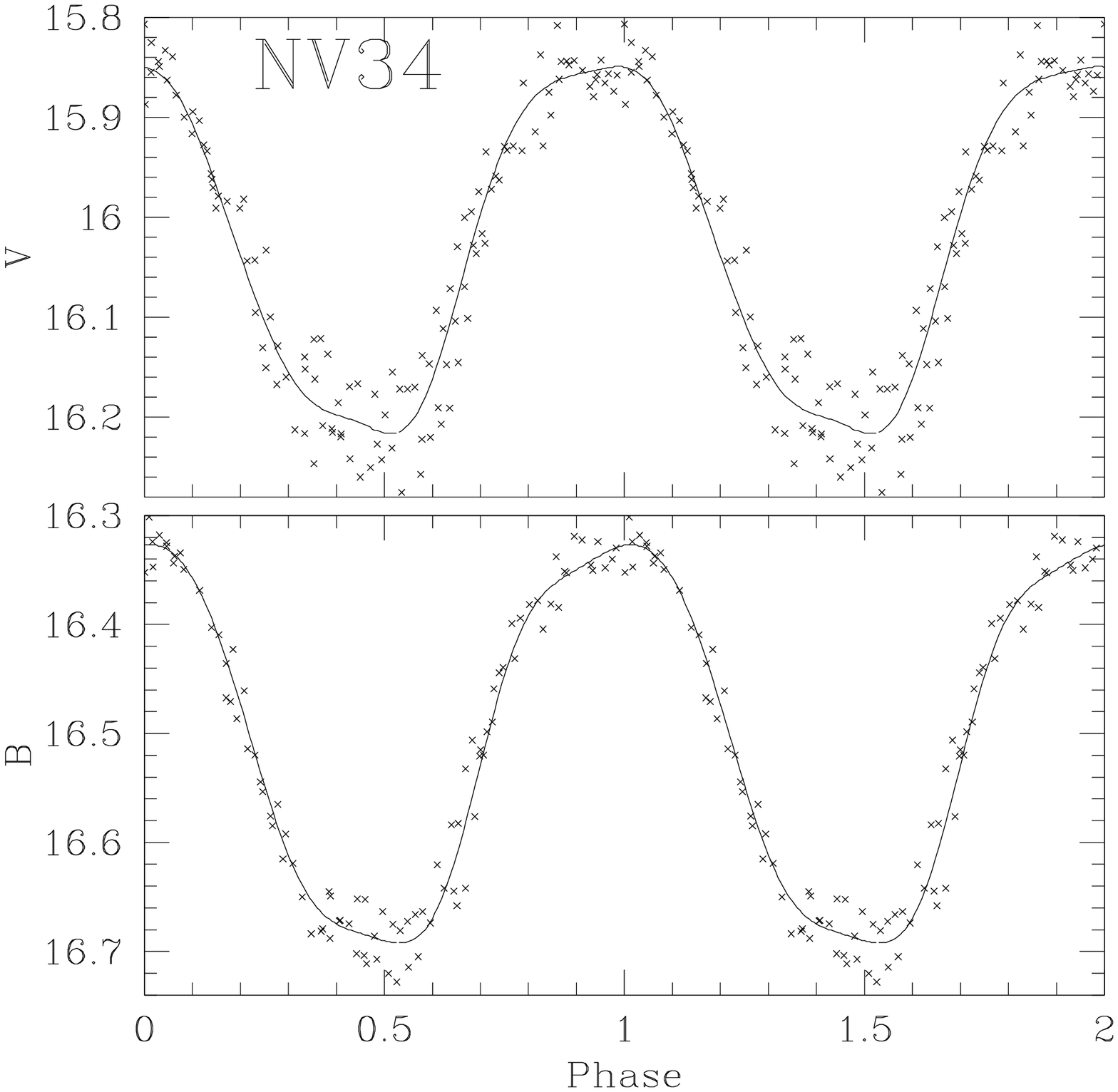}
\includegraphics[width=.28\textwidth]{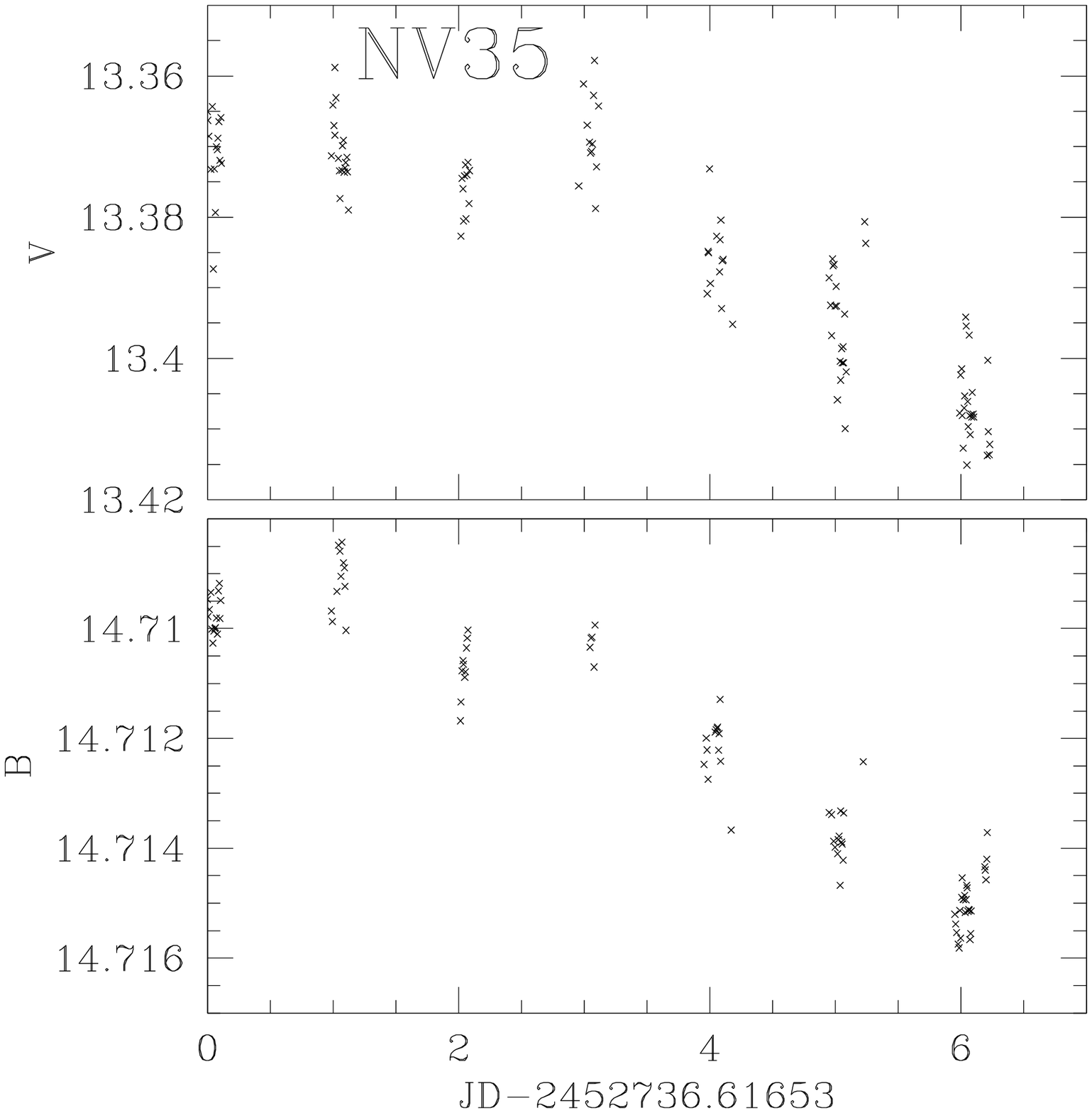}
\includegraphics[width=.28\textwidth]{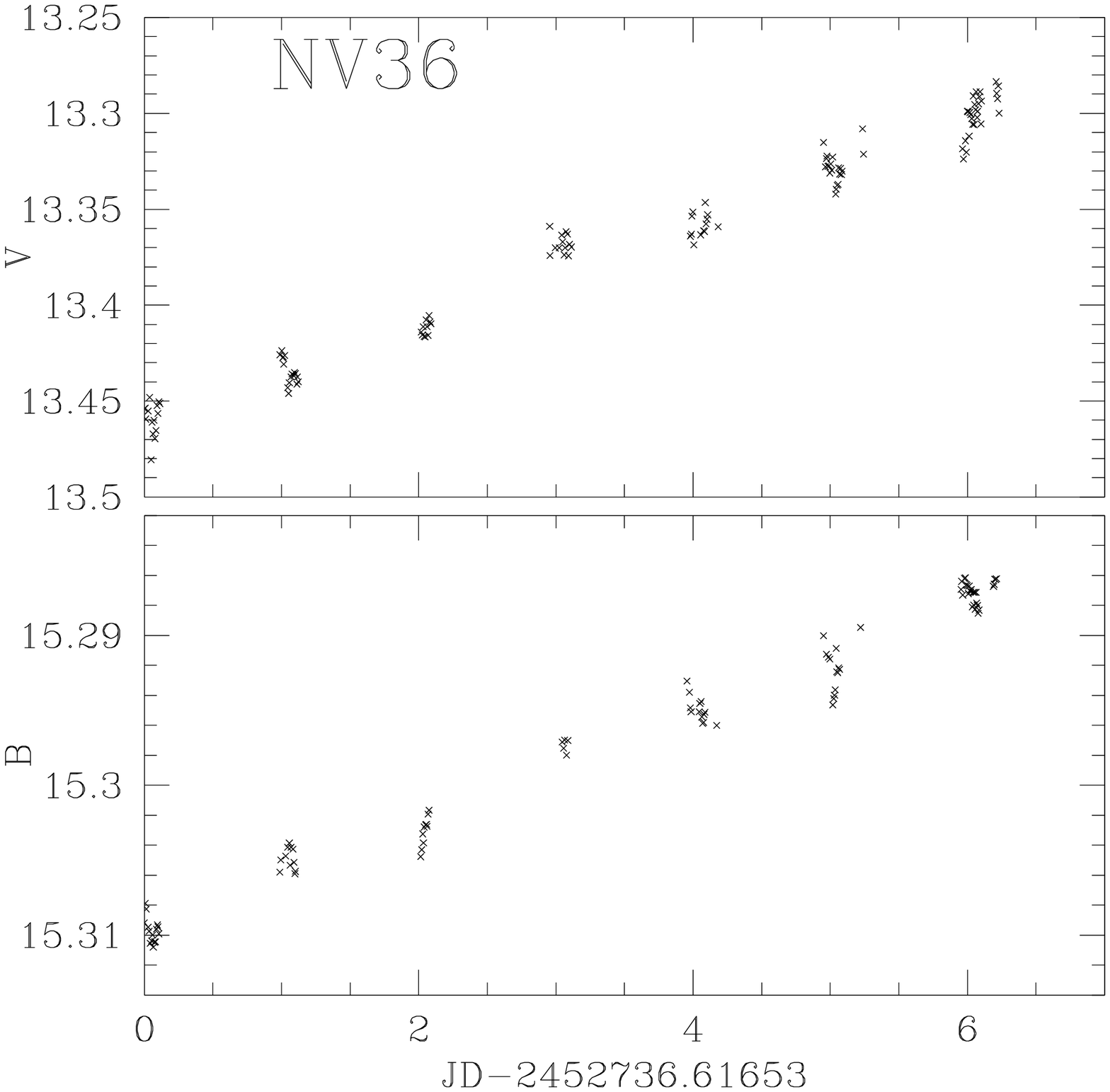}
\includegraphics[width=.28\textwidth]{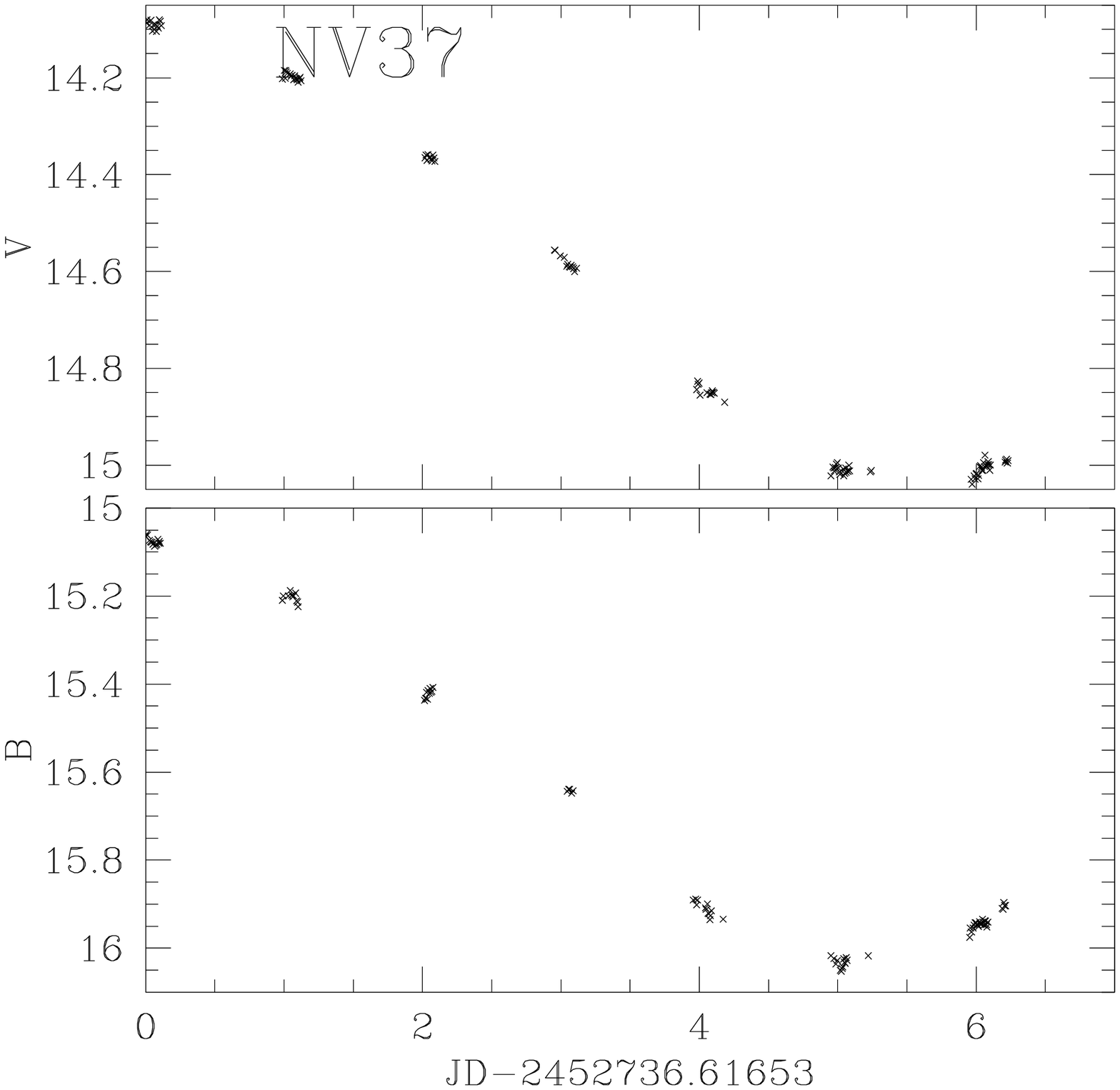}
\includegraphics[width=.28\textwidth]{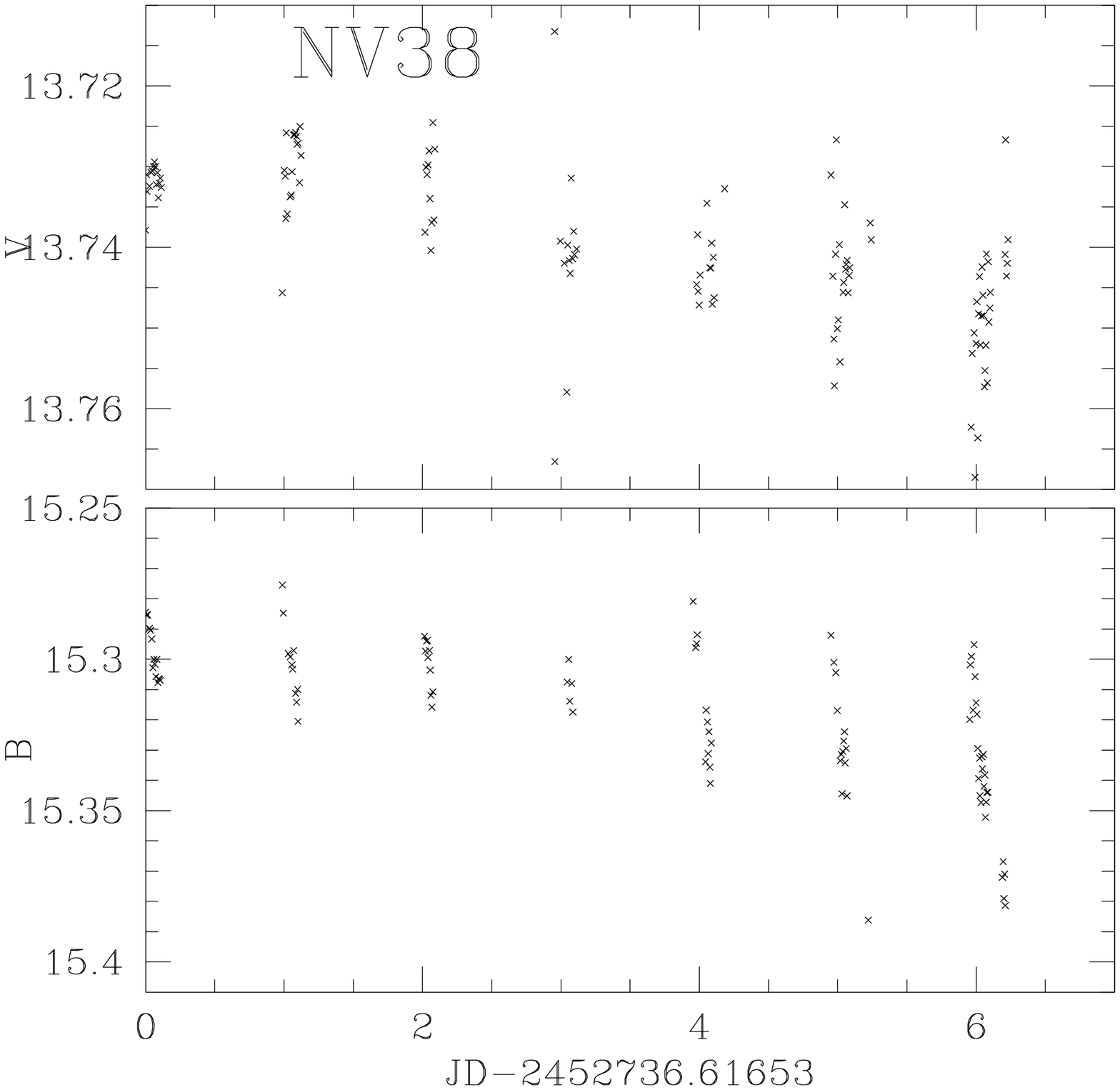}
  \end{center}
  \end{figure}

\end{document}